\documentclass{article}

\usepackage{jheppub}
\usepackage[T1]{fontenc}
\usepackage{breqn}

\usepackage{changepage}
\usepackage{multirow}
\usepackage{titlesec}
\allowdisplaybreaks[4]

\usepackage{shuffle}
\usepackage{mathtools}

\usepackage{array,longtable}
\newcolumntype{L}{>{$}l<{$}}  

\newenvironment{xsmallmatrix}[1]
  {\renewcommand\thickspace{\kern#1}\smallmatrix}
  {\endsmallmatrix}

\title{All orders structure and efficient computation of linearly reducible elliptic Feynman integrals}

\author[a,b]{Martijn Hidding,}
\author[c]{Francesco Moriello}

\affiliation[a]{Hamilton Mathematics Institute, Trinity College, Dublin 2, Ireland}

\affiliation[b]{School of Mathematics, Trinity College, Dublin 2, Ireland}

\affiliation[c]{ETH Z\"urich, Institut fur theoretische Physik, Wolfgang-Paulistr. 27, 8093, Z\"urich, Switzerland}

\emailAdd{hiddingm@tcd.ie}
\emailAdd{fmoriello@phys.ethz.ch}

\abstract{We define linearly reducible elliptic Feynman integrals, and we show that they can be algorithmically solved up to arbitrary order of the dimensional regulator in terms of a 1-dimensional integral over a polylogarithmic integrand, which we call the inner polylogarithmic part (IPP). The solution is obtained by direct integration of the Feynman parametric representation. When the IPP depends on one elliptic curve (and no other algebraic functions), this class of Feynman integrals can be algorithmically solved in terms of elliptic multiple polylogarithms (eMPLs) by using integration by parts identities. We then elaborate on the differential equations method. Specifically, we show that the IPP can be mapped to a generalized integral topology satisfying a set of differential equations in $\epsilon$-form. In the examples we consider the canonical differential equations can be directly solved in terms of eMPLs up to arbitrary order of the dimensional regulator. The remaining 1-dimensional integral may be performed to express such integrals completely in terms of eMPLs. We apply these methods to solve two- and three-points integrals in terms of eMPLs. We analytically continue these integrals to the physical region by using their 1-dimensional integral representation.}

\begin{document}

\maketitle

\section{Introduction}

The evaluation of Feynman diagrams is a crucial ingredient of particle physics calculations. Analytic evaluation is required up to high precision in order to make predictions for processes at the LHC, or other particle colliders. The evaluation of Feynman diagrams becomes increasingly complicated as the number of loops and scales increases. Many techniques have been developed to deal with this complexity. Modern methods focus on scalar Feynman diagrams, which span the diagrams for more general (gauge) theories. This result follows from applying Passarino-Veltman reduction~\cite{Passarino:1978jh} and integration-by-parts (IBP) identities~\cite{Tkachov:1981wb,Chetyrkin:1981qh,Laporta:1996mq,Laporta:2001dd} to rewrite non-scalar diagrams into scalar ones. In the remainder of the paper we will always let the term Feynman diagram refer to scalar Feynman diagrams, which we also call Feynman integrals.

Many Feynman integrals can be analytically computed in terms of special functions known as multiple polylogarithms (MPL)~\cite{Goncharov:1998kja,Remiddi:1999ew}. This could be considered the ``simplest'' class of functions that one may hope to consider, as for example the massless on-shell (1-loop) bubble, triangle, box and pentagon integrals are expressible in terms of MPLs up to all orders in the dimensional regulator $\epsilon$. Upon considering more complicated integrals, by adding external massive legs, internal massive lines, or considering higher loop diagrams, one quickly encounters integrals that are not expressible in terms of MPLs, and more general classes of functions need to be considered.

From the traditional viewpoint Feynman integrals are defined as momentum space integrals: one integrates over a 4-dimensional space for each internal momentum. Although many such integrals are divergent, one may adopt a dimensional regularization scheme such that the dimension is taken to be $d = 4 - 2\epsilon$, where $\epsilon$ is the dimensional regulator. Different integer dimensions than 4 may also be considered in this scheme. Unfortunately from the momentum-space viewpoint Feynman integrals are difficult to calculate both analytically and numerically. To get a better handle on them it is necessary to study things from a different viewpoint. Arguably the two most successful but different starting points are the differential equations method~\cite{Kotikov:1990kg,Kotikov:1991pm,Bern:1993kr,Remiddi:1997ny,Gehrmann:1999as} and the direct integration of the scalar parametrization of Feynman integrals~\cite{brownperiods, panzerthesis}. We give a short discussion of these two methods next.

The differential equations method has received significant attention in the last decade. In this method collections of Feynman integrals are considered which form a basis for the system of IBP relations that is associated with their topology.\footnote{Here the term topology approximately refers to the set of all Feynman integrals with the same propagators and subsets thereof, raised to arbitrary integer powers which may also be zero.} Such integrals are referred to as master integrals. The set of master integrals may  be minimized by using symmetry relations. For a basis of master integrals one may define a closed system of differential equations under differentiation with respect to external scales, like the Mandelstam variables and the masses. The study of such a system has been systematized by the introduction of the so-called canonical basis of master integrals \cite{Henn:2013pwa}, which is conjectured to exist for polylogarithmic integrals. For such a basis of master integrals, the differential equation matrix is $\epsilon$-factorized, where $\epsilon$ is the dimensional regulator, and the differential equation matrix is in $d\text{log}$-form. The general solution of such a system can be written as a path-ordered exponential, which gives the individual terms in the $\epsilon$-expansion as iterated integrals. Furthermore, if the arguments of the $d\text{log}$'s are rational, the resulting Chen-iterated integrals can be directly expressed in terms of multiple polylogarithms. The problem of finding the canonical basis - when it exists - has been automated in a number of software packages \cite{epsilon, fuchsia, canonica}. Another recurring challenge in the differential equations method is to perform the IBP reduction to obtain a closed form for the differential equations. Specialized software packages are available to perform these reductions \cite{litered, fire5, kira}, but all run into computational limits for sufficiently complicated topologies. 

The approach of directly integrating the scalar parametrization of Feynman integrals has seen major advancements by the works of (among others) Brown and Panzer, see for example \cite{brownperiods, panzerthesis}. In particular the Maple package HyperInt \cite{hyperint} automates the integration of multiple polylogarithmic functions. The direct integration method works as follows. First one computes the parametric (Feynman) representation of a Feynman integral, which will be discussed in section \ref{sec:parametric}. Starting from this representation one attempts to integrate one integration parameter at a time. In the ideal scenario there exists an integration sequence such that all integrations can be done in terms of multiple polylogarithms. In that case the Feynman integral is called ``linearly reducible'', as it requires a linear factorization of the set of polynomials occurring in each integration step. If instead a linear factorization can only be performed at the expense of introducing algebraic roots that contain remaining integration variables, one will generally not be able to perform these integrations in terms of MPLs. (Although in some cases a change of variables may be obtained to transform away the square roots.) The property of linear reducibility can be investigated without explicitly performing the integration, by considering if the compatibility graph associated to the parametric representation is linearly reducible \cite{brownperiods, panzerthesis}.

The next challenge in the computation of Feynman integrals is to get a good grasp on the structure of integrals that can't be expressed in terms of MPLs. It seems that the next non-trivial class are the so-called elliptic Feynman integrals, which have seen a lot of interest in the last years~\cite{Henn:2014qga,Caffo:1998du,Laporta:2004rb,Kniehl:2005bc,Kalmykov:2008ge,Bloch:2013tra,Bloch:2014qca,Adams:2014vja,Adams:2015gva,
Adams:2015ydq,Bloch:2016izu,Remiddi:2016gno,Adams:2016xah,Bonciani:2016qxi,Primo:2016ebd,vonManteuffel:2017hms,Ablinger:2017xml,Bogner:2017vim,Remiddi:2017har,Bourjaily:2017bsb,Chen:2017soz,Broedel:2017kkb,Broedel:2017siw,Adams:2018yfj, Adams:2018kez}. Elliptic integrals have traditionally been (loosely) identified as integrals for which the maximal cut, or the maximal cut of one of their subtopologies, is an elliptic integral.

Considerable progress has been made in the last years in the understanding of elliptic Feynman integrals. We discuss a few different starting points for solving elliptic Feynman integrals next. In \cite{vonManteuffel:2017hms} a non-planar triangle topology is considered for which the top sector is elliptic. In that case one may first find and solve a canonical $d\text{log}$ basis for the subtopologies, in terms of multiple polylogarithms. The solution for the elliptic Feynman integrals in the top sector may then be written down by the method of variation of parameters. This requires solving the homogeneous equations of the integrals in the top-sector. The problem of finding these homogeneous solutions can in turn be tackled by computing the maximal cuts of the integrals in the top sector. Since these maximally cut integrals have vanishing subtopologies, they solve the homogeneous part of the differential equation for the uncut integral (the inhomogeneous terms in the differential equation are subtopologies). The computation of maximally cut Feynman integrals has furthermore been aided by recent explorations of the Baikov parametrization \cite{Baikov1, Baikov2, Baikov3}. 

The equal-mass sunrise, which has tadpole integrals as subtopologies, has also been extensively studied in this way, see for example \cite{Laporta:2004rb, Bloch:2013tra, Adams:2014vja}. In more recent work it has been shown that an $\epsilon$-factorized form for the differential equations of the kite integral family, which contains the equal mass sunrise, may be obtained by allowing for non-algebraic integration kernels in the differential equations matrix. The resulting solution is then given in terms of iterated integrals of modular forms \cite{Adams:2018yfj}. A planar double box integral relevant to top-pair production with a closed top loop, involving several elliptic sub-sectors, has furthermore been computed by considering a differential equation linear in $\epsilon$ \cite{Adams:2018kez, Adams:2018bsn}.

On the more formal side significant progress has been made in the understanding of iterated integrals over an elliptic curve. More specifically, all iterated integrals over an elliptic curve can be expressed in terms of special functions called elliptic multiple polylogarithms (eMPLs) \cite{2011arXiv1110.6917B, Broedel:2014vla,Broedel:2017kkb}. Moreover, their functional identities can be systematically studied by using a suitable elliptic generalization of the symbol map \cite{Broedel:2018iwv}.  While this class of functions seems to be the natural candidate to express a large class of elliptic Feynman integrals, it is still unclear how and when this representation can be obtained.

In the first part of the  paper we investigate dimensionally regulated elliptic Feynman integrals by direct integration of the Feynman
parametric representation. In so doing we identify a class of Feynman integrals that we call \emph{linearly reducible elliptic Feynman integrals}. These integrals can be algorithmically solved up to arbitrary order of the dimensional regulator in terms of 1-dimensional integrals over a polylogarithmic expression, the so-called inner polylogarithmic part (IPP). More specifically, the direct integration approach requires that the polynomials in each integration step factor linearly with respect to the next integration variable (Feynman parameter). However, in the elliptic case, the factorization introduces irreducible square roots at some integration step and no further integration can be done thereafter within the class of multiple polylogarithms. Nevertheless we have observed empirically that in many cases the irreducible square roots appear only when the integration with respect to the last
parameter has to be performed. We call this class of elliptic Feynman integrals linearly reducible. When the IPP depends on one elliptic curve and no other algebraic functions this class of Feynman integrals can be algorithmically solved in terms of eMPLs by using, e.g., integration by parts identities \cite{Broedel:2017kkb}. 

While in many cases the direct integration approach is convenient as, e.g., no boundary conditions need to be computed and results at relatively low orders of the dimensional regulator can be easily obtained, it might be impractical to apply it when higher orders need to be considered. One reason is that the size of the expressions usually grows very rapidly and the algebraic manipulations involved become cumbersome. Moreover, the analytic properties of the answer to all orders are not immediately manifest in this approach. For these reasons we believe that a better understanding of the differential equations method applied to Feynman integrals beyond multiple polylogarithms is highly desirable. In the second part of the paper we show that the IPP of linearly reducible elliptic Feynman integrals can be mapped to a generalized integral topology satisfying a set of differential equations in $\epsilon$-form. The mapping is obtained by applying the Feynman trick to a suitably chosen pair of propagators in the momentum space representation of the Feynman integral under consideration. Remarkably, for all the examples we considered the canonical differential equations matrix can be directly expressed in terms of linear combinations of eMPL integration kernels, and the solution of these equations in terms of eMPLs becomes elementary. Once the IPP is expressed in terms of eMPLs the remaining 1-dimensional integral can be easily performed to express linearly reducible elliptic Feynman integrals completely in terms of eMPLs.

The remainder of the paper is organized as follows. In section \ref{sec:parametric} we review the properties of the parametric Feynman representation of Feynman diagrams. In particular we discuss the Cheng-Wu theorem, which is needed to find a linearly reducible integration order, and we review the direct integration algorithm and some of its properties. In section \ref{sec:ClassOfFunctions} we discuss linearly reducible elliptic Feynman integrals and some properties of the class of 1-fold integrals that arises in their computation. In section \ref{sec:IPPDifferentialEquations} we explain how to set up differential equations for the inner polylogarithmic part of a linearly reducible elliptic Feynman integral, which involves an application of the Feynman trick and is thereafter analogous to the polylogarithmic case. We then move on to section \ref{sec:examples}, where we treat a few examples of linearly reducible elliptic Feynman integrals. In section \ref{sec:OffshellUnequalMassSunrise} we discuss the unequal mass sunrise in depth. We study it by direct integration, and as a solution from the system of differential equations for its IPP. We also derive the analytic continuation of the first master integral to the physical region. A ``triangle with bubble'' integral is discussed in section \ref{sec:TriangleWithBubble} for which we perform a similar analysis. In section \ref{sec:non-planar triangle} we solve an elliptic non-planar triangle integral relevant for Higgs+jet production. This integral is linearly reducible, however the IPP depends on multiple algebraic curves and further development of the methods studied in this paper will be required. We end our discussion in section \ref{sec:conclusionsandoutlook}, where we reflect on the results obtained and give an outlook for the future.

\section{Parametric representation of Feynman integrals}
\label{sec:parametric}
The defining momentum space integral representation of scalar Feynman diagrams admits the form:
\begin{equation}
\label{eq:fidefmom}
I_{a_1,\ldots,a_n}(\{s\}) = N \int \left(\prod_{i=1}^l d^d k_i \right)\, \frac{1}{\prod_{i=1}^n D_i^{a_i}}\, , 
\end{equation}
where $\{s\} = \{p\} \cup \{m\}$ schematically denotes the dependence on the set of external momenta and masses, which is left implicit on the right hand side of the equation. Furthermore, $N$ is a normalization constant picked by convention, $l$ denotes the number of loops, $n$ denotes the number of propagators and $d$ is the dimension of the Minkowskian space-time. The convention $D_i = -q_i^2+m_i^2$ is used, where $q$ is the momentum flowing on the $i$-th propagator and $m$ is the mass on the propagator. We assume that the exponents $a_i$ of the propagators are positive integers.

A scalar parametrization can be found as follows. First one uses the Schwinger trick to rewrite every propagator as:\footnote{Convergence of the scalar integral is guaranteed by adding a small imaginary part to the propagators according to the Feynman prescription.}
\begin{equation}
\label{eq:SchwingerTrick}
\frac{1}{D^a} = \frac{i^a}{\Gamma(a)} \int_0^\infty d\alpha\, \alpha^{a-1} e^{-i D \alpha}\,.
\end{equation}
The momentum integrals of eq. (\ref{eq:fidefmom}) become Gaussian integrals which can be performed and result in the so-called Schwinger parametrization. From there, one may perform a change of variables $\alpha_i \rightarrow \eta \alpha_i$, under the constraint $\sum_{i=1}^n\alpha_i = 1$. The parameter $\eta$ can be integrated out, which yields the Feynman parametrization:
\begin{equation}
    \label{eq:fideffp}
    I_{a_1,\ldots,a_n}(\{s\}) = N \left(i\pi^{\frac{d}{2}}\right)^l \frac{\Gamma(a - \frac{ld}{2})}{\prod_{i=1}^n\Gamma(a_i)}\int_\Delta d^n\vec{\alpha}\, \left(\prod_{i=1}^n \alpha_i^{a_i-1}\right) \mathcal{U}^{a-\frac{d}{2}(l+1)}\mathcal{F}^{-a+\frac{l d}{2}}\, ,
\end{equation}
where $\mathcal{U}$ and $\mathcal{F}$ are called Symanzik polynomials, which are related to determinants taken during the Gaussian integration, but have an alternative interpretation from considering the set of spanning trees $T(G)$ and spanning 2-forests\footnote{A 2-forest is a disjoint union of 2 trees.} $F_2(G)$ of the Feynman diagram $G$:
\begin{align}
\mathcal{U} &= \sum_{T\in T(G)}\prod_{e_i \notin T}\alpha_i\,, &
\tilde{\mathcal{F}} &= \sum_{(T_1,T_2) \in F_2(G)}\left(\prod_{e_i \notin (T_1 \cup T_2)} \alpha_i\right) s_{(T_1,T_2)}\,, & \mathcal{F} &= -\tilde{\mathcal{F}} + \mathcal{U}\left(\sum \alpha_i m_i^2\right)\,,
\end{align} where $s_{(T_1,T_2)}$ is defined as the square of the external momentum travelling in between the components of the 2-forest. For a review of these concepts see for example \cite{graphpolsbogweinz, nakanishi1971graph}. The region $\Delta$ is defined as $\{\vec{\alpha}\,|\, \alpha_i > 0,\, \sum_{i=1}^n\alpha_i = 1\}$. The scalar parametrization of eq. (\ref{eq:fideffp}) is called the Feynman parametrization and the integration parameters are called Feynman parameters. We will also refer to it as the parametric representation. In the remaining sections we generally choose the normalization constant
\begin{equation}
\label{eq:normalizationchoice}
N=\left(i\pi^{\frac{d}{2}}\right)^{-l} \frac{\prod_{i=1}^n\Gamma(a_i)}{\Gamma(a - \frac{ld}{2})}\,,
\end{equation}
to remove the prefactor in the Feynman representation. We will often consider the external kinematics in the Euclidean region, which is determined by the condition that $\mathcal{F}\geq 0$ on the whole domain of integration.

\subsection{The Cheng-Wu theorem}
\label{sec:chengwu}
The Cheng-Wu theorem specifies various ways that one might perform the integration over the Feynman parametrization. Importantly, different applications of the theorem may be helpful in finding a linearly reducible integration order. The Cheng-Wu theorem states that a projective (Feynman) integral over $\Delta$ has the same value when integrated over the domain
\begin{equation}
    \Delta_S = \left\{\vec{\alpha}\,\bigg|\, \alpha_i \geq 0,\, \sum_{i\in S} \alpha_i = 1\right\}\, ,
\end{equation}
where $S\subseteq [1,n]$ is a nonempty set of integers \cite{chengwu}. Projectivity refers in this context to the property that the integrand should remain invariant under the rescaling:
\begin{align}
\label{eq:projrescl}
    \alpha_i \rightarrow \lambda \alpha_i,\,\, d\alpha_i \rightarrow \lambda d\alpha_i\,.
\end{align}
If one starts with an integral over $\Delta$ which does not remain invariant under eq. (\ref{eq:projrescl}), the following projective transform can be performed to obtain such an integral
\begin{align}
    \alpha_i \rightarrow \frac{\alpha_i'}{\sum_{j=1}^n \alpha_j'}\, ,
\end{align}
which has Jacobian $(\sum_{j=1}^n \alpha_j')^{-n}$ \cite{hooftveltmanprojectivetransform}. Note that this change of variables keeps the integration domain $\Delta$ invariant and is in fact only a proper change of variables because we integrate over $\Delta$, as it manifestly sets the sum of the integration parameters to 1. As an illustrative example consider the beta function:
\begin{align}
    \textrm{B}(x,y) &= \int_0^1 d\alpha_1\, \alpha_1^{x-1}(1-\alpha_1)^{y-1} = \int_\Delta d^2\vec{\alpha}\,\alpha_1^{x-1} \alpha_2^{y-1} \nonumber\\\label{eq:bbbb}&\stackrel{\left(\alpha_i\rightarrow \frac{\alpha_i'}{\alpha_1'+\alpha_2'}\right)}{=} \int_\Delta d^2\vec{\alpha}'\,\frac{(\alpha_1')^{x-1} (\alpha_2')^{y-1}}{(\alpha_1'+\alpha_2')^{x+y}} = \int_0^\infty d\alpha_1'\,\frac{(\alpha_1')^{x-1}}{(\alpha_1'+1)^{x+y}}\, .
\end{align}
Eq. (\ref{eq:bbbb}) starts with the usual definition of the beta function, upgrades it to a 2-dimensional integral over $\Delta$, projectivizes the integrand, and lastly uses Cheng-Wu to achieve a different representation of the integral. (Of course in this trivial example the same result would have easily been obtained with the M{\" o}bius transformation $\alpha_1\rightarrow \alpha_1'/(\alpha_1'+1)$.)

To integrate the scalar parametrization one seeks a linearly reducible integration order, which is discussed in the upcoming section. As a rule of thumb it is often sufficient to apply Cheng-Wu before performing any integrations, and to let one of the Feynman parameters, say $\alpha_i$, go to 1 by picking $S = \{i\}$. The other parameters are then integrated from 0 to infinity. It is sometimes convenient to factor out the Cheng-Wu parameter:
\begin{equation}
    I_{a_1,\ldots,a_n}(\{s\}) = \int d\alpha_i\, \delta(1-\alpha_i)\left[\int_0^\infty d\alpha_1\ldots \hat{d\alpha_i}\ldots d\alpha_n\, \left(\prod_{i=1}^n \alpha_i^{a_i-1}\right) \mathcal{U}^{a-\frac{d}{2}(l+1)}\mathcal{F}^{-a+\frac{l d}{2}}\right]\, .
\end{equation}
One can then focus on integrating out Feynman parameters successively in the bracketed part. Projectivity of the integrand is usually manifestly preserved after performing these integrations. Suppose one ends up with the following expression after performing a number of integrations:
\begin{equation}
    \label{eq:fiinterm}
    I_{a_1,\ldots,a_n}(\{s\}) = \int d\alpha_i\, \delta(1-\alpha_i)\left[\int_0^\infty d^k \vec{\alpha}_{S'}\,\textrm{MPL}(\vec{\alpha}_{S'\cup\{i\}})\right]\, ,
\end{equation}
where there are $k<n-1$ non-trivial Feynman parameters remaining, labeled by a set $S'$, and where the integrand is projective. By the Cheng-Wu theorem we can turn this into:
\begin{equation}
    \label{eq:fiinterm2}
    I_{a_1,\ldots,a_n}(\{s\}) = \int_{\Delta_{S''}} d^{k+1} \vec{\alpha}_{S'\cup\{i\}}\,\textrm{MPL}(\vec{\alpha}_{S'\cup\{i\}})\, ,
\end{equation}
where $S''\subseteq S' \cup \{i\}$.

\subsection{Direct integration, linear reducibility and all orders statements}
\label{sec:integration}
First we remind the reader of the definition of (Goncharov) multiple polylogarithms (MPLs). They are the following recursively defined functions:
\begin{align}
    \label{eq:Gdef}
    G(\vec{a}_n;\, z) = G(a_1,\ldots,a_n;\, z) = \int_0^z \frac{dt}{t-a_1}G(a_2,\ldots,a_n;\, t)\, ,
\end{align}
where $a_1,\ldots,a_n, z$ are complex variables. The recursion is ended at $n=0$ where one lets by convention:
\begin{align}
    G(;\,z) \equiv 1\, .
\end{align}
As a general feature of iterated integrals, MPLs obey the shuffle product:
\begin{align}
G(\vec{a}_n;\, z)G(\vec{b}_m;\, z) = \sum_{\vec{c}_{n+m}\,\in\, \vec{a}_n\, \shuffle\, \vec{b}_m}G(\vec{c}_{n+m};\, z)\,,
\end{align}
where the set of shuffles of $\vec{a}_n$ and $\vec{b}_m$ denoted $\vec{a}_n\, \shuffle\, \vec{b}_m$ may be understood to contain all permutations of the sequence $(\vec{a},\vec{b})$ that preserve the ordering of the individual vectors.

A (small) technical complication in the definition of multiple polylogarithms is that a divergence at the basepoint 0 occurs when $a_n = 0$. One may adopt the definition:
\begin{align}
    \label{eq:G0def}
    G(\vec{0};\,z) \equiv \frac{1}{n!}\log(z)^n\, ,
\end{align}
to deal with the divergent case where all $n$ parameters $a_i$ are equal to zero. Cases with $a_n = 0$ and at least one $a_i \neq 0$ can then be dealt with in a consistent manner by rearranging parameters using the shuffle product, and using eq. (\ref{eq:G0def}). A pedagogical review of multiple polylogarithms and their functional identities is given in \cite{Duhr:2014woa}.

The parametric representation discussed in the previous sections can be integrated, e.g., by using the computer program HyperInt \cite{hyperint}. We sketch a few ideas underlying the algorithm next, in order to illustrate the concept of linear reducibility. First one performs a series expansion of the integrand of eq. (\ref{eq:fideffp}) on the dimensional regulator. Assuming that the Feynman integral is finite in the integer dimension $\tilde{d}$, we let $d=\tilde{d}-2\epsilon$ and find:
\begin{align}
    I_{a_1,\ldots,a_n}(\{s\}) = \sum_{k=0}^\infty I_{a_1,\ldots,a_n}^{(k)}(\{s\}) \epsilon^k\,,
\end{align}
where the coefficients are:
\begin{align}
    I_{a_1,\ldots,a_n}^{(k)}(\{s\}) = \frac{1}{k!}\int_\Delta d^n\vec{\alpha}\, \left(\prod_{i=1}^n \alpha_i^{a_i-1}\right) \mathcal{U}^{a-\frac{\tilde{d}}{2}(l+1)}\mathcal{F}^{-a+\frac{l \tilde{d}}{2}}\left((1+l)\log(\mathcal{U})-l\log(\mathcal{F})\right)^k\,.
\end{align}
It is clear that for even dimensions $\tilde{d}$ and integer powers of the propagators the resulting integrand is a polylogarithmic expression without algebraic terms. The remaining integrations to be performed, at each order in $\epsilon$, take the schematic form:
\begin{equation}
\label{eq:alphaint}
f_k(\alpha_{k+1},\dots, \alpha_{n})=\int_0^\infty d\alpha_k\, f_{k-1}(\alpha_{k},\dots, \alpha_{n})\,,
\end{equation}
where  $k\in \{1,\dots,n\}$. We aim to perform these integrations in such a way that at each integration step the integrand $f_{k-1}(\alpha_{k},\dots, \alpha_{n})$ is a polylogarithmic expression. More precisely, we require that the integrand at each integration step is a combination of multiple polylogarithms with prefactors and arguments that are rational functions of the remaining integration parameters. 

Now suppose that $f_{k-1}(\alpha_{k},\ldots, \alpha_{n})$ is a polylogarithmic expression. Then it depends on a set of irreducible polynomials in the remaining integration parameters which we denote by $\vec{P}^{(k-1)}(\alpha_{k},\ldots, \alpha_{n})$. A requirement for the integral in eq. (\ref{eq:alphaint}) to be polylogarithmic again is that all polynomials in $\vec{P}^{(k-1)}(\alpha_{k},\ldots, \alpha_{n})$ are linear in $\alpha_k$. If at each step of the integration the set of polynomials $\vec{P}^{(k-1)}(\alpha_{k},\ldots, \alpha_{n})$ is linear in $\alpha_{k}$, we have found that $\alpha_1,\ldots,\alpha_n$ is a so-called \textit{linearly reducible integration order}.

Each integration can then be performed in terms of multiple polylogarithms along the following lines:
\begin{enumerate}
    \item express $f_{k-1}(\alpha_{k},\ldots, \alpha_{n})$ as a combination of multiple polylogarithms of argument $\alpha_k$,
    \item find a primitive $F_{k}(\alpha_{k},\dots, \alpha_{n})$ such that $\partial_{\alpha_k} F_{k}(\alpha_{k},\dots, \alpha_{n})=f_{k-1}(\alpha_{k},\dots, \alpha_{n})$,
    \item compute the limit $f_k(\alpha_{k+1},\dots, \alpha_{n})=\lim_{\alpha_{k}\rightarrow \infty}F_{k}(\alpha_{k},\dots, \alpha_{n})-\lim_{\alpha_{k}\rightarrow 0}F_{k}(\alpha_{k},\dots, \alpha_{n})$.
\end{enumerate}
To search for a linearly reducible integration order one can enumerate over all possible integration sequences. Luckily the set of polynomials $\vec{P}^{(k)}(\alpha_{k+1},\dots, \alpha_{n})$ at each integration step can be exposed from a so-called \textit{compatibility graph} without performing the actual integration, as introduced in \cite{brownperiods}. While we refer the reader to that reference for further details, an important consequence of the compatibility graph method is that the polynomials $\vec{P}^{(k)}(\alpha_{k+1},\dots, \alpha_{n})$, $k\in\{1,\dots,n\}$ are independent of the order in $\epsilon$ we are considering. It should however be noted that at leading order in $\epsilon$ the exponent of one of the Symanzik polynomials may become 0 for special configurations of the dimension and powers of the propagators, and hence one may find an integration sequence that does not work at higher orders in $\epsilon$. In section \ref{sec:examples} we solve a number of finite linearly reducible elliptic Feynman integrals up to and including the order $\epsilon^1$, and it is understood that this yields linear reducibility up to all orders.

In some cases one finds that applying the Cheng-Wu theorem with one Feynman parameter set to 1 does not lead to a linearly reducible integration order. A nontrivial application of the Cheng-Wu theorem may then occasionally lead to a linearly reducible integration order. This is the case for the non-planar triangle integral in section \ref{sec:non-planar triangle}. In other cases the situation may be worse, and a change of variables is needed at some point during the integration. Some more discussion on this topic, and explicit examples of changes of variables are discussed in \cite{panzerthesis, analyticregularization}.

The previous story applies when the Feynman integral is expressible in terms of multiple polylogarithms. To our knowledge there is no known example of an elliptic Feynman integral which can be expressed in terms of MPLs, and it is believed that such a representation is not possible for elliptic Feynman integrals in general. That means in particular that no linearly reducible integration order exists for these integrals. We therefore define \emph{linearly reducible elliptic Feynman integrals} as elliptic Feynman integrals that are linearly reducible if one excludes the last integration. In particular, we allow the next-to-last integration to be performed at the expense of introducing algebraic terms of the last integration parameter. The final expression of a linearly reducible elliptic Feynman integral reads, at a generic $\epsilon$-order:
\begin{equation}
    f_n=\int_0^\infty d\alpha_n\, f_{n-1}(\alpha_n)\,,
\end{equation}
where $f_{n-1}(\alpha_n)$ is a polylogarithmic expression with algebraic coefficients and arguments, depending on the elliptic curves of the problem.
In the framework of the direct parametric integration, linearly reducible elliptic Feynman integrals are the simplest instance of elliptic Feynman integrals.

\section{Structure of linearly reducible elliptic Feynman integrals}
\label{sec:ClassOfFunctions}
In the previous section linearly reducible elliptic Feynman integrals have been introduced which are, to all orders in $\epsilon$, expressible as 1-fold integrals. These 1-fold integrals take the following schematic form:
\begin{equation}
    \label{eq:schem}
    \int_0^\infty dx\, \sum_i A_i\left(x \right) \textrm{MPL}_i\left(x\right) ,
\end{equation}
where the sum over $i$ denotes a generic collection of terms grouped by factors $A_i$ which are algebraic functions in $x$, and $\textrm{MPL}_i\left(x\right)$ denotes a polylogarithmic term with algebraic arguments. Here we investigate some properties of these 1-fold integral, and sketch some general strategies that may be employed to write eq. (\ref{eq:schem}) in terms of a minimal class of integrals. 

From the direct integration point of view the algebraic dependence in $A_i$ arises from forcing a linear factorization of the polynomials in the previous integration step. For the upcoming examples of the unequal mass sunrise, and the triangle with bubble, we find that the only algebraic dependence of the inner polylogarithmic part is on 1 elliptic curve. In that case one may write:
\begin{align}
    A_i\left(x \right) = R_i(x, y(x))\,,
\end{align}
such that $y(x)^2 = P(x)$ defines an elliptic curve, i.e. $P(x)$ is an irreducible cubic or quartic polynomial, and where $R_i$ is a rational function in its arguments. One may furthermore factorize the dependence on the elliptic curve such that one has:
\begin{align}
    \label{eq:RationalEllipticCurveFactorized}
    R_i(x, y(x)) = R_{i,1}(x) + \frac{1}{y(x)}R_{i,2}(x)\,,
\end{align}
where $R_{i,1}(x)$ and $R_{i,2}(x)$ are rational functions in $x$. This can be achieved in the following manner. Firstly one may absorb any even power of $y(x)$ in the rational part. Furthermore, for denominators of the type $1/(S_1(x) + S_2(x)y(x))^k$, where $S_1(x)$ and $S_2(x)$ are some polynomials in $x$ and $k$ is some positive integer, one may multiply by the conjugate
\begin{align}
    \frac{1}{(S_1(x)+S_2(x)y(x))^k}=\frac{(S_1(x)-S_2(x)y(x))^k}{\left(S_1(x)^2-S_2(x)^2P(x)\right)^k}\,,
\end{align}
observe that the new denominator is a polynomial, and expand out the numerator, again absorbing all even powers of $y(x)$ in the rational part in $x$. Lastly one may use the relation $y(x) = P(x)/y(x)$ to obtain a representation of the form of eq. (\ref{eq:RationalEllipticCurveFactorized}). One may furthermore partial fraction a rational term $R(x)$ such that:
\begin{align}
    \label{eq:partialfractioning}
     R(x) = \frac{N(x)}{\prod_{i=1}^k (x-b_{i})^{p_i}} = \sum_{i=1}^k \sum_{j=1}^{p_i}\frac{ N_{i,j}}{(x-b_{i})^{j}} + \sum_{j=0}^{\text{deg}(N(x)) - p} M_j x^j ,
\end{align}
where $p_i \in \mathbb{N}_{+}$, $p = \sum_{i=1}^k p_i$, $N(x)$ is a polynomial in $x$, and $N_{i,j}$ and $M_j$ are complex coefficients that do not depend on $x$. From now on we will shorten the notation $y(x)$ to $y$. By the previous arguments we may reduce the integrand to the following cases:
\begin{align}
    \label{eq:IntegrationKernelsBeforeIBP}
     \frac{dx}{(x-\beta)^k y}\textrm{MPL}(x,y)\,, &&  \frac{dx}{y}\textrm{MPL}(x,y)\,,& &&   \frac{x^k\,dx}{y}\textrm{MPL}(x,y)\,, \nonumber\\  dx x^k\textrm{MPL}(x,y)\,, &&
     \frac{dx}{(x-\beta)^k}\textrm{MPL}(x,y)\,,&
\end{align}
where $k$ is a positive integer, $\beta$ is a constant with respect to $x$, and $\text{MPL}(x, y)$ is a polylogarithmic expression. We will refer to the algebraic factors, without the polylogarithmic term $\textrm{MPL}(x,y)$, as integration kernels. Note that splitting up the integral in eq. (\ref{eq:schem}) in terms of 1-fold integrals of the type of eq. (\ref{eq:IntegrationKernelsBeforeIBP}) requires that the individual contributions are finite. We will ignore the issue of regularization of the individual contributions for now, and provide results in terms of a single one fold integral that contains all contributions. Furthermore, to save space we will use the shorthand subscript notation:
\begin{align}
    G(a_1,\ldots,a_n;\,1) = G_{a_1,\ldots,a_n}\,,
\end{align}
later on in the text to present the polylogarithmic terms. In the upcoming examples we will find that we are able to pick a basis of master integrals such that we only encounter kernels with $k = 1$ and with an elliptic curve that is quartic. Nonetheless, in general one might expect other integration kernels to show up as well, and we show next that it is possible to reduce kernels of the type with k > 1 to the case with k = 1 by employing IBP relations. This is similar to the treatment in \cite{Broedel:2017kkb}, where these kinds of IBP identities are considered to provide an integration algorithm for elliptic polylogarithms multiplied by rational functions. In our treatment we keep the factor multiplying the algebraic integration kernel explicitly polylogarithmic, and we will work with a quartic elliptic curve whose roots will be denoted by $a_1,\ldots, a_4$. 

Besides reducing the integration kernels to the case with $k=1$, it is possible to relate kernels of the form $dx/((x-a_i) y)$. There are 4 such kernels, and one may trade them using IBP identities for one of the following:
\begin{align}
    &\frac{dx}{(x-a_1)y}\,, && \frac{dx}{(x-a_2)y}\,, && \frac{dx}{(x-a_3)y}\,,  \nonumber\\& \frac{dx}{(x-a_4)y}\,,&& \frac{dx\,x^2}{y}\,,
\end{align}
where we note that the polylogarithmic part $\textrm{MPL}(x,y)$ that multiplies the kernels is affected by the IBP relations. The final result that is obtained after reducing the set of integration kernels is not necessarily unique. For example, the inner polylogarithmic part is still subject to the usual functional identities between multiple polylogarithms. Furthermore, a kernel of the type $dx/(x-\beta)$ may be exchanged for different ones using the IBP identity:
\begin{align}
    \int\frac{\text{\footnotesize MPL}(x, y)}{x-c}dx=\text{\footnotesize MPL}(x, y)\log(x-c)-\int\log(x-c)\text{\footnotesize MPL}'(x, y)dx\,,
\end{align}
where the new kernels depend on the precise form of $\text{\footnotesize MPL}'(x, y)$. 

Next we provide the explicit IBP relations that may be used to reduce the set of integration kernels to the cases with $k=1$. First we consider the following relation for $k>1$:
\begin{align}
    \label{eq:IBPRelationxk}
    \int\frac{x^k \text{\footnotesize MPL}(x, y)}{y}dx=\frac{x^{k+1}\text{\footnotesize MPL}(x, y)}{(k-1)y}+\frac{1}{2(k-1)}\int\left(\text{\footnotesize MPL}(x, y)\left(\frac{a_1x^k}{y\left(x-a_1\right)}+\frac{a_2x^k}{y\left(x-a_2\right)}+\right.\right.\nonumber\\\left.\left.\frac{a_3x^k}{y\left(x-a_3\right)}+\frac{a_4x^k}{y\left(x-a_4\right)}\right)-\frac{2x^{k+1}}{y}\text{\footnotesize MPL}'(x, y)\right)dx\,,
\end{align}
where we note that:
\begin{align}
    \frac{x^k}{x-a_i}=a_i^{k-1} \left(\sum_{i=0}^{k-1}\left(\frac{x}{a_i}\right)^i+\frac{a_i}{x-a_i}\right)\,,
\end{align}
from which it is clear that the power of $x$ in the numerator is reduced for each term in the indefinite integral on the right hand side that carries a factor $\text{MPL}(x)$. Note that there is a term carrying a factor $x^{k+1}$, but this is multiplied by the derivative of $\text{MPL}(x)$, which has its weight reduced by 1. This allows for an inductive procedure that terminates at weight $0$. Similarly we may derive the following relation for $k>1$:
\begin{align}
    \label{eq:IBPRelationxmck}
    \int&\frac{\text{\footnotesize MPL}(x, y)}{y(x-c)^{k}}dx=-\frac{(x-c)^{1-k}\text{\footnotesize MPL}(x, y)}{(k-1)y}-\frac{1}{2(k-1)}\int\left(\text{\footnotesize MPL}(x, y)\left(\frac{(x-c)^{1-k}}{y\left(x-a_1\right)}+\right.\right.\nonumber\\&\left.\left.\frac{(x-c)^{1-k}}{y\left(x-a_2\right)}+\frac{(x-c)^{1-k}}{y\left(x-a_3\right)}+\frac{(x-c)^{1-k}}{y\left(x-a_4\right)}\right)+\left(\frac{2c}{y(x-c)^{k}}-\frac{2x}{y(x-c)^{k}}\right)\text{\footnotesize MPL}'(x, y)\right)dx\,.
\end{align}
We remark that partial fractioning a term of the form $(x-c)^{1-k}/\left(x-a_i\right)$ decomposes it into pieces that carry a factor $(x-c)^{-j}$, where $1\leq j\leq k-1$, and a piece that carries a factor $1/(x-a_i)$, see eq. (\ref{eq:partialfractioning}). Hence we may safely iterate eq. (\ref{eq:IBPRelationxmck}) to reduce the power of $(x-c)$ in the denominator, up to polylogarithmic terms of lower weight. Lastly, we provide the following relation that may be used to trade the kernel $dx/(y(x-a_1))$ for $x^2\,dx/y$, up to polylogarithmic terms of lower weight:
{\small
\begin{align}
    \label{eq:InadmissableKernelIBP}
    \int&\frac{\text{\footnotesize MPL}(x, y)}{y\left(x-a_1\right)}dx=-\frac{2\left(x-a_2\right)\left(x-a_3\right)\left(x-a_4\right)\text{\footnotesize MPL}(x, y)}{a_{12}a_{13}a_{14}\,y}+\frac{1}{a_{12}a_{13}a_{14}}\int\left(\text{\footnotesize MPL}(x, y)\left(\frac{2x^2}{y}+\right.\right.\nonumber\\&\left.\left.\frac{\left(-a_1-a_2-a_3-a_4\right)x}{y}+\frac{a_1\left(-a_1+a_2+a_3+a_4\right)}{y}\right)+\left(\frac{2x^3}{y}-\frac{2\left(a_2+a_3+a_4\right)x^2}{y}+\right.\right.\nonumber\\&\left.\left.\frac{2\left(a_3a_4+a_2\left(a_3+a_4\right)\right)x}{y}-\frac{2a_2a_3a_4}{y}\right)\text{\footnotesize MPL}'(x, y)\right)dx\,,
\end{align}}
where we used the notation $a_{ij} = a_i - a_j$. One may obtain similar relations for the kernels $dx/(y(x-a_j))$, $j = 2,3,4$, by cyclically permuting the labels of the roots: $a_i\rightarrow a_{i+1}$. This way one may remove every kernel of the type $dx/(y(x-a_i))$, at the expense of introducing a kernel $x^2\,dx/y$. One may rearrange eq. (\ref{eq:InadmissableKernelIBP}) afterwards to obtain an expression that contains just the kernel $dx/(y(x-a_1))$. Lastly we have the relations:
\begin{align}
    \int\frac{\text{\footnotesize MPL}(x, y)}{(x-c)^{k}}dx&=\frac{\text{\footnotesize MPL}(x, y)}{(1-k)(x-c)^{k-1}}-\int\frac{\text{\footnotesize MPL}'(x, y)}{(1-k)(x-c)^{k-1}}dx\,, \\
    \int x^k \text{\footnotesize MPL}(x, y)dx&=\frac{x^{k+1}\text{\footnotesize MPL}(x, y)}{k+1}-\int\frac{x^{k+1}\text{\footnotesize MPL}'(x, y)}{k+1}dx\,,
\end{align}
of which the right hand side in both cases involves a piece that has been integrated, and an indefinite integral that contains terms of lower weight.

\section{Differential equations for the inner polylogarithmic part}
\label{sec:IPPDifferentialEquations}
In the previous sections we have considered linearly reducible elliptic Feynman integrals from the viewpoint of the direct integration method. In section \ref{sec:FeynmanTrick} we will show that the inner polylogarithmic part of these integrals can be mapped to a (generalized) Feynman integral topology that arises from an application of the Feynman trick to two propagators. This topology can be studied in momentum space, where it is easy to derive IBP relations and setup a system of differential equations for its master integrals. We then review the differential equations method in section \ref{sec:DifferentialEquationsGen}. In particular, we discuss the canonical basis of differential equations and how, in the upcoming examples, it can be used to algorithmically solve the IPP in terms of eMPLs. The full integral is then solved in terms of eMPLs by performing the remaining 1-fold integral over the IPP (see section \ref{sec:examples}). The resulting approach essentially bridges a gap between the direct integration method and the differential equations method. One can either find the IPP by direct integration of the Feynman parametrization, or alternatively one can find it by solving a canonical system of differential equations for its corresponding topology. However, in the direct integration approach the solution of a given integral in terms of eMPLs involves first integrating the IPP in terms of MPLs, and then iteratively writing the polylogarithms as an integral over their derivative. The complexity of this approach usually grows quickly with the order of the dimensional regulator.  On the other hand in section \ref{sec:examples} we show  that by using the canonical differential equations method for the IPP it is possible, in some cases, to solve the full integrals up to arbitrary order of the dimensional regulator in a fully algebraic manner, since the relevant integration kernels coincide with the ones defining the eMPLs.

\subsection{The Feynman trick}
\label{sec:FeynmanTrick}
In the following treatment we consider the inner polylogarithmic part with respect to the last integration parameter $\alpha_{n-1}$, which is a generic choice as we have the freedom to relabel variables. We start by considering a general topology $I_{a_1,\ldots,a_n}$. First write down the Feynman parametrization, and apply the Cheng-Wu theorem to put $\alpha_n = 1$:
\begin{align}
    \label{eq:FeynmanParametrizationCW}
    I_{a_1,\ldots,a_n} &\equiv N \int \left(\prod_{i=1}^l d^d k_i \right)\, \frac{1}{\prod_{i=1}^{n} D_i^{a_i}} \nonumber\\
    &= \left(\prod_{i=1}^{n-1}\int_0^\infty d\alpha_i\,  \alpha_i^{a_i-1}\right) \left.\left(\mathcal{U}^{a-\frac{d}{2}(l+1)}\mathcal{F}^{-a+\frac{l d}{2}}\right)\right|_{\alpha_n = 1} \nonumber \\
    &= \int_0^\infty d\alpha_{n-1}\,\text{IPP}^{(n-1)}\,,
\end{align}
where we denote the inner polylogarithmic part with respect to the last integration on $\alpha_{n-1}$ as $\text{IPP}^{(n-1)}$. The Feynman trick tells us that:
\begin{align}
\label{eq:GeneralFeynmanTrick}
\frac{1}{D_{n-1}^{a_{n-1}} D_n^{a_n}} &= \frac{\Gamma \left(a _{n-1}+a _n\right)}{\Gamma \left(a _{n-1}\right) \Gamma \left(a _n\right)}\int_0^\infty\frac{\alpha_{n-1}^{a_{n-1}-1}}{(\alpha_{n-1} D_{n-1} + D_n )^{a _{n-1}+a_n}}d\alpha_{n-1}\,.
\end{align}
Inspired by this, we consider a new topology that contains a generalized propagator of the form $\alpha_{n-1} D_{n-1} + D_n$. In this topology $\alpha_{n-1}$ is an external scale, and to avoid confusion we'll denote the Feynman parameters of this topology with a tilde ($\tilde{\alpha}_i$). We define:
\begin{align}
    \label{eq:FeynmanParametrizationFeynmanTrick}
    \tilde{I}_{a_1,\ldots,a_{n-2},a_{n-1}+a_n} &\equiv \tilde{N} \int \left(\prod_{i=1}^l d^d k_i \right)\, \frac{1}{\left(\prod_{i=1}^{n-2} D_i^{a_i}\right) \left(\alpha_{n-1} D_{n-1}+D_n\right)^{a_{n-1}+a_n} }\,.
\end{align}
Like before a normalization factor $\tilde{N}$ is included to remove an overall prefactor from the Feynman parametrization: 
\begin{align}
    \tilde{N} = \left(i\pi^{\frac{d}{2}}\right)^{-l} \frac{\left(\prod_{i=1}^{n-2}\Gamma(a_i)\right)\Gamma(a_{n-1}+a_n)}{\Gamma(a - \frac{ld}{2})}\,.
\end{align}
From eq. (\ref{eq:GeneralFeynmanTrick}) it is clear that this yields:
\begin{align}
    \label{eq:IPPFundamentalRelation}
    I_{a_1,\ldots,a_n} = \int_0^\infty d\alpha_{n-1} \,\alpha_{n-1}^{a_{n-1}-1} \tilde{I}_{a_1,\ldots,a_{n-2},a_{n-1}+a_n}\,.
\end{align}
Next we show explicitly that:
\begin{align}
    \label{eq:IPPRelationTildeTopology}
    \text{IPP}^{(n-1)} = \alpha_{n-1}^{a_{n-1}-1} \tilde{I}_{a_1,\ldots,a_{n-2},a_{n-1}+a_n}\,.
\end{align}
Setting up the Feynman parametrization yields:
\begin{align}
    \tilde{I}_{a_1,\ldots,a_{n-2},a_{n-1}+a_n} &= \left(\prod_{i=1}^{n-2}\int_0^\infty d\tilde{\alpha}_i\, \tilde{\alpha}_i^{a_i-1}\right)\, \left.\left(\tilde{\mathcal{U}}^{a-\frac{d}{2}(l+1)}\tilde{\mathcal{F}}^{-a+\frac{l d}{2}}\right)\right|_{\tilde{\alpha}_{n-1}=1}\,,
\end{align}
where the Cheng-Wu theorem has been applied to put the Feynman parameter $\tilde{\alpha}_{n-1}$ (which corresponds to the generalized propagator) to 1. Next let $\mathcal{U}(\alpha_1,\ldots,\alpha_n)$ and $\tilde{\mathcal{U}}(\tilde{\alpha}_1,\ldots,\tilde{\alpha}_{n-1})$ explicitly denote the dependence of the Symanzik polynomials of the two topologies on their Feynman parameters. One may show that they are related by:\footnote{The relation between the Symanzik polynomials of both topologies can be read off by comparing the argument of the exponent that is obtained from applying the Schwinger trick to each propagator, which is proportional to $\sum_i \alpha_i D_i$, for a topology with propagators $\{D_i\}$ and corresponding integration parameters $\{\alpha_i\}$.}
\begin{align}
    \tilde{\mathcal{U}}(\tilde{\alpha}_1,\ldots,\tilde{\alpha}_{n-1}) &= \mathcal{U}(\tilde{\alpha}_1,\ldots,\tilde{\alpha}_{n-2},\tilde{\alpha}_{n-1} \alpha_{n-1},\tilde{\alpha}_{n-1})\,, \nonumber \\
    \tilde{\mathcal{F}}(\tilde{\alpha}_1,\ldots,\tilde{\alpha}_{n-1}) &= \mathcal{F}(\tilde{\alpha}_1,\ldots,\tilde{\alpha}_{n-2},\tilde{\alpha}_{n-1} \alpha_{n-1},\tilde{\alpha}_{n-1})\,.
\end{align}
Putting $\tilde{\alpha}_{n-1} = 1$ in correspondence with the choice of the Cheng-Wu theorem in eq. (\ref{eq:FeynmanParametrizationFeynmanTrick}), and relabeling $\tilde{\alpha}_i$ to $\alpha_i$ for $i = 1,\ldots,n-2,$ without ambiguity, yields the special case:
\begin{align}
    \label{eq:SymanzikRelnsCW}
    \tilde{\mathcal{U}}(\alpha_1,\ldots,\alpha_{n-2},1) &= \mathcal{U}(\alpha_1,\ldots,\alpha_{n-2},\alpha_{n-1},1)\,, \nonumber \\
    \tilde{\mathcal{F}}(\alpha_1,\ldots,\alpha_{n-2},1) &= \mathcal{F}(\alpha_1,\ldots,\alpha_{n-2},\alpha_{n-1},1)\,.
\end{align}
In other words, the Symanzik polynomials of both topologies match if we use the Cheng-Wu theorem to put $\alpha_n = 1$ for the original topology and to put $\tilde{\alpha}_{n-1} = 1$ for the topology with 2 combined propagators. Comparing eqs. (\ref{eq:FeynmanParametrizationCW}), (\ref{eq:FeynmanParametrizationFeynmanTrick}), and (\ref{eq:SymanzikRelnsCW}) we conclude that eq. (\ref{eq:IPPRelationTildeTopology}) holds. Hence the topology of eq. (\ref{eq:FeynmanParametrizationFeynmanTrick}) may be used to represent the IPP of a linearly reducible elliptic Feynman integral using eq. (\ref{eq:IPPRelationTildeTopology}).

\subsection{The differential equations method}
\label{sec:DifferentialEquationsGen}
Next we remind the reader of some points that are relevant for the differential equations method. Firstly one requires a reduction of the integrals in the topology in terms of a set of master integrals, which we denote by $\vec{B} = (B_1,\ldots,B_k)$. The master integrals are independent with respect to all IBP relations. (Such IBP relations are most easily generated from the momentum space picture of the integrals.) Furthermore, generally one also takes into account symmetry relations. A set of master integrals, and the reduction of the remaining integrals in the topology - up to some finite bound on the propagator exponents - may be obtained using programs such as LiteRed, FIRE and  KIRA \cite{litered, fire5, kira}. We will make use of the C++ version of FIRE5 which seemingly has no problem in dealing with combined propagators that are obtained from the Feynman trick.

One may write down a closed form system of differential equations for $\vec{B}$ with respect to each external scale $s_j$:
\begin{align}
    \label{eq:difeqmain}
    \frac{d}{ds_j}\vec{B}=\tilde{\mathbf{A}}_j\vec{B}\,,
\end{align}
where $\tilde{\mathbf{A}}_j$ is a matrix whose elements depend on the external scales and the dimension. For polylogarithmic topologies a basis can be found in a ``canonical'' $d\log$ $\epsilon$-factorized form \cite{Henn:2013pwa} where the matrices satisfy:
\begin{align}
    \label{eq:hennmatrix}
    \tilde{\mathbf{A}}_j = \epsilon \frac{d}{d s_j}\mathbf{A} = \epsilon \sum_{l\in \mathcal{A}} \mathbf{A}_l \frac{d\log(l)}{ds_j}\,,
\end{align}
such that $\mathbf{A}$ has no more dependence on the dimension $d = \tilde{d} - 2\epsilon$, where $\tilde{d}$ is an integer and the dimensional regulator is $\epsilon$. The set of ``letters'' $\mathcal{A}$ consists of rational or algebraic functions of the external scales. Lastly, $\mathbf{A}_l$ is a matrix with integer coefficients. The differential equations for each $s_i$ may now be combined:
\begin{align}
    d \vec{B} = \epsilon\, \left(d\mathbf{A}\right)\vec{B} = \epsilon \sum_{l\in \mathcal{A}} d\log(l)\,\mathbf{A}_l \vec{I}\,.
\end{align}
Differential equations in canonical form have two important properties. Firstly, since $\epsilon$ is factored out one may write the general solution of the equation in terms of a path-ordered exponential:
\begin{align}
    \vec{B} = \mathbb{P} \exp\left[\epsilon \int_\gamma d\mathbf{A}\right]\vec{B}_{\text{boundary}}\,,
\end{align}
which order-by-order in $\epsilon$ expresses the result in terms of iterated integrals:
\begin{align}
    \label{eq:PathOrderedExponentialExpanded}
    \vec{B} = \vec{B}^{(0)}_{\text{boundary}} + \sum_{k\geq 1}\epsilon^k \sum_{j=1}^k \int_0^1 \gamma^*(d\mathbf{A})(t_1) \int_0^{t_1} \gamma^*(d\mathbf{A})(t_2) \ldots \int_0^{t_{j-1}} \gamma^*(d\mathbf{A})(t_j) \,\vec{B}^{(k-j)}_{\text{boundary}}\,,
\end{align}
where $\gamma$ is a path with domain $[0,1]$ in the space of external invariants, and where we have a boundary term $\vec{B}_{\text{boundary}}$, which is $\vec{B}$ evaluated at the point in kinematic space given by $\gamma(0)$. We furthermore denote the $\epsilon$-expansion of the boundary term by:
\begin{align}
    \vec{B}_{\text{boundary}} = \sum_{k\geq 0}\vec{B}^{(k)}_{\text{boundary}} \epsilon^k\,,
\end{align}
which we assume to be finite. Note that one may obtain a set of master integrals that is finite as $\epsilon\rightarrow 0$ by multiplying all the master integrals by a power of $\epsilon$. If the letters are rational functions, and one is able to find a boundary term, eq. (\ref{eq:PathOrderedExponentialExpanded}) allows us to directly write down the master integrals in terms of MPLs order by order in $\epsilon$.

The second important result is that a canonical form differential equation provides the symbol of the master integrals, given that they are uniformly transcendental and that we have the leading coefficient of $\vec{B}$ in $\epsilon$. In particular, one finds that:
\begin{align}
	\mathcal{S}(\vec{B}^{(k)})=\epsilon^k\left(R\left(\mathbf{A}^{\otimes^k} \vec{B}^{(0)}_{\text{boundary}}\right)\right)\,,
\end{align}
where $\mathbf{A}^{\otimes^2}_{ij} =\mathbf{A}_{ik}\otimes \mathbf{A}_{kj}$, etc., and where $R$ is an operator that reverses the ordering of the tensor product: $R(a \otimes b \otimes c) = c \otimes b \otimes a$.

In the upcoming sections we will also use the differential equations method to find results in terms of $\text{E}_4$-functions \cite{Broedel:2017kkb, Broedel:2017siw}, for examples that depend on a single quartic elliptic curve. This will be done by rescaling the last integration parameter, so that it runs from 0 to 1. We will denote the rescaled parameter as $x'$, and consider a system of differential equations with respect to $x'$:
\begin{align}
    \frac{\partial}{\partial x'}\vec{B} = \epsilon \frac{\partial\mathbf{A}}{\partial x'} \vec{B}\,,
\end{align}
where $\vec{B}$ will be a canonical basis for the inner polylogarithmic part. The solution in terms of $\text{E}_4$-functions will follow because the entries of the matrix $\partial\mathbf{A}/\partial x'$ will turn out to correspond to linear combinations of integration kernels of $\text{E}_4$-functions. The particular kernels that show up in the upcoming sections are presented here for completeness:
\begin{align}
    \label{eq:E4IntegrationKernels}
    \psi_0(0,x)=\frac{c_4}{y}\,,&&\psi_{-1}(\infty,x)=\frac{x}{y}\,,&&\psi_{-1}(c,x) = \frac{y_c}{(x-c)y}-\frac{\delta_{c0}}{x}\,, && \psi_1(c,x) = \frac{1}{x-c}\,.
\end{align}
For the definitions of the other integration kernels we refer to \cite{Broedel:2017kkb}. We note that $c_4 \equiv \frac{1}{2}\sqrt{a_{13}a_{24}}$, where $a_{ij} = a_i - a_j$, and where $a_1,a_2,a_3$ and $a_4$ are roots of the elliptic curve: $y^2 = (x-a_1)(x-a_2)(x-a_3)(x-a_4)$. 

\section{Analytic continuation}
\label{sec:continuation}
In this section we describe how to perform the analytic continuation to the physical region of linearly reducible elliptic Feynman integrals in a form that is suitable for fast and reliable numerical evaluations. Feynman integrals are analytically continued to the physical region by using the Feynman prescription, which is implemented by shifting the external invariants by a vanishing positive imaginary part $i\delta$. Our starting point will be the one-fold integral representation of eq. (\ref{eq:schem}) obtained from direct integration. Moreover we will assume that the integrand, at every $\epsilon$ order, is a pure polylogarithmic function of fixed transcendental weight multiplied by an overall algebraic function, of the form:
\begin{equation}
\label{eq:Idelta}
I(\{s\},i \delta)=\sum_{i=0}^\infty \epsilon^i \int_0^\infty dx\, \phi(x,\{s\},i \delta)f^{(i+w_0)}(x,\{s\},i \delta)\,,\quad \phi\in\left\{\frac{1}{y(x,\{s\},i\delta)},\frac{1}{x}\right\},\quad x,\{s\},\delta >0\,,
\end{equation}
where $w_0$ is the transcendental weight at leading $\epsilon$ order. In the equation above we made explicit the dependence on the Feynman prescription $i \delta$ which removes the branch cut ambiguities of the integrand in the physical region. $y^2$ is a quartic polynomial of $x$ defining the relevant elliptic curve. The integrand of eq.~(\ref{eq:Idelta}) is symmetric under $y \rightarrow -y$. This is due to the fact that the square root $y$ appears when performing the integration with respect to the second-last Feynman parameter, by factorizing a certain second degree polynomial (this can be seen for example by analyzing the associated compatibility graph \cite{brownperiods}), and the two roots of the polynomial are indeed symmetric upon flipping the sign of $y$.    As we will show in the next sections the first master integral of the unequal masses sunrise topology (section \ref{sec:OffshellUnequalMassSunrise}) and the triangle with bubble integral (\ref{sec:TriangleWithBubble}) belong to this class. The analytic continuation of the second master integral of the sunrise topology can be done using the same techniques as described below. However, further analysis is required to obtain numerically stable representations due to the presence of simple poles in the leftmost integration kernels and we leave it for future work. 

Our task will be to identify a set of regions in the $x,\{s\}$ space and remove, in each region, the dependence on the Feynman regulator $\delta$ by explicitly performing the $\delta\rightarrow0$ limit:
\begin{equation}
I(\{s\})=\sum_{i=0}^\infty \epsilon^i \int_0^\infty dx\, \sum_{j=1}^{n_R}\theta_j(x,\{s\}) \phi_j(x,\{s\})f_j^{(i+w_0)}(x,\{s\})\,,
\end{equation}
where $n_R$ is the number of relevant regions $R_j$ with $j\in\{1,\dots, n_R\}$,  $\theta_j(x,\{s\})=1$ if $(x,\{s\})\in R_j$ and $\theta_j(x,\{s\})=0$ otherwise, and
\begin{align}
    f_j^{(i)}(x,\{s\})&=\lim_{\delta\rightarrow0}f^{(i)}(x,\{s\},i\delta)\,,\\
    \phi_j(x,\{s\})& =\lim_{\delta\rightarrow0}\phi(x,\{s\},i \delta)\,, \quad\quad  x,\{s\}\in R_j\,.
\end{align} 
In order to perform the limits above we first need to compute $\lim_{\delta\rightarrow0}y(x,\{s\},i\delta)$. The square root $y(x,\{s\},i\delta)$ for fixed $\{s\}$ can be seen as a multivalued complex function of $x$ and (vanishing) $\delta$, taking  two values differing by an overall sign. When defining the analytic continuation one  usually defines a single-valued continuous branch of the square root for every $x$ and $\delta$, minus branch cuts for $\delta=0$ and a set of intervals $x\subset \mathbb{R}$ (see e.g. the discussion of \cite{Bogner:2017vim}). There are multiple branches satisfying the constraints above, and one will have in general:
\begin{equation}
\label{eq:srprescriptiongen}
   \lim_{\delta\rightarrow0}y(x,\{s\},i\delta)=\left\{
    \begin{array}{cc}
       \pm  y(x,\{s\},0) \quad &\text{if}\quad  y^2(x,\{s\},0)>0 \,, \\
       \pm   i\sqrt{-y^2(x,\{s\},0)} \quad &\text{if}\quad y^2(x,\{s\},0)<0\,,
    \end{array}\right.
\end{equation}
where the actual signs depend on the definition of the branch under consideration. However, as discussed above, the integrand of eq.~(\ref{eq:Idelta}) is symmetric under $y\rightarrow -y$ and, in this case, the sign of the square root is immaterial and we set: 
\begin{equation}
\label{eq:srprescription}
   \lim_{\delta\rightarrow0}y(x,\{s\},i\delta)=\left\{
    \begin{array}{cc}
         y(x,\{s\},0) \quad & \text{if}\quad x,\{s\}\in R_j:\; y^2(x,\{s\},0)>0\,,  \\
          i\sqrt{-y^2(x,\{s\},0)} \quad & \text{if}\quad x,\{s\}\in R_j:\; y^2(x,\{s\},0)<0\,.
    \end{array}\right.
\end{equation}
The prescription above defines $\phi_j(x,\{s\})$ in every region $R_j$ and implies that in each region $y^2(x,\{s\},0)$ has definite sign.

In the examples discussed in the next sections we will consider polylogarithmic expressions up to weight three and, having a fast numerical evaluation in mind, we look for a representation of $f_j^{(i)}(x,\{s\})$ in terms of logarithms and classical polylogarithms of suitably chosen arguments. The functions $f_j^{(i)}(x,\{s\})$ can be found proceeding in the following algorithmic steps (see also~\cite{Goncharov:2010jf,Duhr:2011zq}). 

\begin{enumerate}

\item Function arguments are defined as monomials of the letters appearing in the symbol alphabet of $f_j^{(i)}(x,\{s\},0)$. In general also spurious letters might be needed when defining functions arguments (see for example~\cite{Duhr:2011zq}), and we have found this to be necessary for the upcoming examples. For the classical polylogarithms ${\rm Li}_{n}(a(x,\{s\}))$, one requires that $1-a(x,\{s\})$ factorizes over the alphabet.

If the alphabet contains algebraic functions the factorization can be checked as follows. We take the logarithm of the function argument under consideration, and we equate it to an ansatz in the form of a linear combination of the logarithms of the alphabet letters. In this way we obtain a system of linear equations for the free coefficients of the ansatz. We numerically sample the equations for many values of the kinematic variables. If the equations admit a solution the argument factorizes as desired over the alphabet and the solution defines the factorized form.

\item  For each weight, one considers a set of linearly independent functions from the set of functions defined in the previous step. We have the freedom to choose the set of linearly independent functions defining the functional basis at weight $i$. We require that our basis elements satisfy:
\begin{equation}
    \label{eq:fconstraints}
    \text{Li}_k(a(x,\{s\})):\;a(x,\{s\})\notin[1,\infty),\quad \log(a(x,\{s\})): a(x,\{s\})\notin(-\infty,0]\,.
\end{equation}
One then defines the most general ansatz for a $\mathbb{Q}$-linear combination of these functions and products thereof, of weight $i$. The coefficients of the ansatz are then fixed imposing that the symbol of the ansatz equals the symbol of $f^{(i)}(x,\{s\},0)$.

\item We determine the terms in the kernel of the symbol at weight $i$ by writing the most general ansatz in terms of the lower weight functions. We fix the free coefficients of the ansatz by  specializing it to several points in the region under consideration. We then equate the resulting expressions to $f^{(i)}(x,\{s\},i\delta)$ and obtain a system of linear equations for the free coefficients that can be solved numerically with arbitrary precision. This gives a numerical value for the free coefficients of the ansatz that can be subsequently fitted against a basis of transcendental constants of appropriate weight.

\end{enumerate}
The algorithm above does not rely on the rationality of the alphabet letters and generalizes the algorithm of~\cite{Duhr:2011zq} to algebraic cases. 
 
 \subsection{Identifying admissible regions}
\label{sec:regions}
In the previous section we have seen how to perform the $\delta\rightarrow0$ limit at the integrand level and express the result in terms of analytic functions. The limit can be safely taken if we are able to identify a set of regions in the  $x,\{s\}$ space that contain no branch points for the integrand $\phi(x,\{s\},0)f^{(i+w_0)}(x,\{s\},0)$. Let us show with an elementary example how a suitable set of regions can be identified. We consider the following elementary function:
\begin{equation}
    f(x,a,i\delta)=\frac{\log(x-a+i\delta)}{\sqrt{x-a+i\delta}}\,, \quad x,a,\delta>0\,.
\end{equation}
 In order to be able to perform the limit we decompose the phase space in regions where the square root and the logarithm have no branch points:
 \begin{equation}
     R_1:x-a>0\,,\quad R_2:x-a<0\,.
 \end{equation}
 We can then explicitly perform the limit in the form:
 \begin{equation}
    \lim_{\delta\rightarrow0} f(x,a,i\delta)=\theta(x-a)\frac{\log(x-a)}{\sqrt{x-a}}+\theta(a-x)\frac{1}{\sqrt{a-x}}\left(\pi-i \log(a-x)\right)\,.
 \end{equation}

As we have seen in the previous section we will be interested in functions whose algebraic dependence comes only from the elliptic curve, therefore two regions are identified by requiring that the elliptic curve $y^2$ has definite sign:
\begin{equation}
    A:y^2(x,\{s\},0)<0\,,\quad B:y^2(x,\{s\},0)>0\,.
\end{equation}
We then further partition these regions by requiring that the purely polylogarithmic expression, $f^{(i)}(x,\{s\},0)$, does not have branch points in the resulting subregions. The subregions are conveniently identified by studying the symbol alphabet letters. Specifically, the alphabet letters we will encounter for $f^{(i)}(x,\{s\},0)$ have the following general form:
\begin{equation}
    \alpha_i(x,\{s\})= g_i(x,\{s\})\,,\quad \beta_j (x,\{s\})= h_j(x,\{s\})+c_j(x,\{s\}) y(x,\{s\})\,,
\end{equation}
where $i\in \{1,\dots,n_\alpha\}$, $j\in \{1,\dots,n_\beta\}$, $n_\alpha$, $n_\beta$ is the number of letters of the form $\alpha_i(x,\{s\})$ and $\beta_j(x,\{s\})$ respectively, $g_i(x,\{s\})$,$\;h_i(x,\{s\})$,$\;c_i(x,\{s\})$  are polynomials and $c_i(x,\{s\})$ has definite sign. Since alphabet letters  $\alpha_i(x,\{s\})$ are real valued in region $A$ while letters $\beta_i(x,\{s\})$ have non-vanishing imaginary part with definite sign, subregions $A_i$ are identified by requiring that each of the $\alpha_i(x,\{s\})$ has definite sign:
\begin{align}
    A_1: \;& \alpha_1(x,\{s\})>0, \alpha_2(x,\{s\})>0,\dots, \alpha_{n_\alpha}(x,\{s\})>0\,,\nonumber\\
    A_2:\; & \alpha_1(x,\{s\})<0, \alpha_2(x,\{s\})>0,\dots, \alpha_{n_\alpha}(x,\{s\})>0\,,\nonumber\\
      &\quad\quad\quad\quad\quad\quad \vdots \nonumber\\
      A_{2^{n_\alpha-1}}:\; & \alpha_1(x,\{s\})<0, \alpha_2(x,\{s\})<0,\dots, \alpha_{n_\alpha}(x,\{s\})>0\,,\nonumber\\
     A_{2^{n_\alpha}}: \; & \alpha_1(x,\{s\})<0, \alpha_2(x,\{s\})<0,\dots, \alpha_{n_\alpha}(x,\{s\})<0\,.
\end{align}
In region $B$ both $\alpha_i(x,\{s\})$ and $\beta_j(x,\{s\})$ are real valued, and subregions $B_i$ are identified by requiring that $\alpha_i(x,\{s\})$ and $\beta_j(x,\{s\})$ have definite sign:
\begin{align}
    B_1: \;& \alpha_1(x,\{s\})>0, \dots, \alpha_{n_\alpha}(x,\{s\})>0,\beta_1(x,\{s\})>0,\dots, \beta_{n_\beta}(x,\{s\})>0\,,\nonumber\\
    B_2:\; & \alpha_1(x,\{s\})<0,\dots, \alpha_{n_\alpha}(x,\{s\})>0,\beta_1(x,\{s\})>0,\dots, \beta_{n_\beta}(x,\{s\})>0\,,\nonumber\\
      &\quad\quad\quad\quad\quad\quad \vdots \nonumber\\
      B_{2^{n_\alpha+n_\beta-1}}: \; & \alpha_1(x,\{s\})<0,\dots, \alpha_{n_\alpha}(x,\{s\})<0,\beta_1(x,\{s\})<0,\dots, \beta_{n_\beta}(x,\{s\})>0\,,\nonumber\\
     B_{2^{n_\alpha+n_\beta}}: \; & \alpha_1(x,\{s\})<0,\dots, \alpha_{n_\alpha}(x,\{s\})<0,\beta_1(x,\{s\})<0,\dots, \beta_{n_\beta}(x,\{s\})<0\,.
\end{align}
In general the partition above will overcount the number of regions that are actually needed. This is due to the fact that some of the zeros of the letters do not correspond to actual branch points of the polylogarithmic expression under consideration. While in principle one could perform a more refined analysis at this stage, for example by systematically studying the symbol map and the monodromy of the left-most symbol letters \cite{Duhr:2012fh}, in practice such overpartition is convenient when it comes to finding a basis of functions satisfying the constraints of eq.~(\ref{eq:fconstraints}) in a given region. Indeed, it is true in general that the 'larger' the region the fewer are the functions satisfying the desired properties in that region, and in complicated cases one might find that the set of admissible functions does not span the functional space under consideration.
    
\section{Examples}
\label{sec:examples}
In this section we provide a few examples that showcase the techniques discussed in the previous sections. We consider the unequal mass sunrise topology, and a triangle with bubble topology from both the direct integration and the differential equation point of view. By default we give our results in the Euclidean region. We provide the analytic continuation of the first master integral of the unequal mass sunrise, and of the triangle with bubble integral at order $\epsilon^0$ and $\epsilon^1$ in sections \ref{sec:SunriseAnalyticContinuation} and \ref{sec:TWBAnalyticContinuation}.

\subsection{The off-shell sunrise diagram with unequal masses}
\label{sec:OffshellUnequalMassSunrise}
We begin our discussion with the direct integration of the first master integral of the massive off-shell elliptic sunrise diagram with three different internal masses:
\begin{equation}
    S_{1,1,1}(s,m_1^2,m_2^2,m_3^2) \,\,\,\, = \,\,\,\      \vcenter{\hbox{\includegraphics[width=0.2\textwidth]{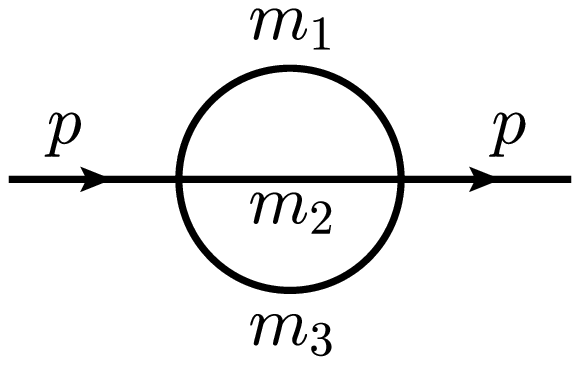}}}\, ,
\end{equation}
where the subscripts of $S_{1,1,1}$ denote the powers of the internal propagators, and where we let $s=p^2$. The same notation will be used in the remainder of the paper.
\subsubsection{Direct integration}
The Feynman parametrization of the sunrise has Symanzik polynomials:
\begin{align}
 \mathcal{U} &= \alpha_1\alpha_2+\alpha_1\alpha_3+\alpha_2\alpha_3, & \mathcal{F} &= \left(\alpha _1 \alpha _2+\alpha _3 \alpha _2+\alpha _1 \alpha _3\right) \left(\alpha _1 m_1^2+\alpha _2 m_2^2+\alpha _3 m_3^2\right)-\alpha _1 \alpha _2 \alpha _3 s\, ,
\end{align}
and is given by:
\begin{equation}
    \label{eq:SFeynPar}
    S_{\nu_1,\nu_2,\nu_3}(p^2,m_1^2,m_2^2,m_3^2) = \int_\Delta d^3\vec{\alpha}\, \alpha_1^{\nu_1-1}\alpha_2^{\nu_2-1}\alpha_3^{\nu_3-1}\mathcal{U}^{-\frac{3 d}{2}+\nu _1+\nu _2+\nu _3}\mathcal{F}^{d-\nu _1-\nu _2-\nu _3}\,.
\end{equation}
We consider the case $d=2-2\epsilon$, where $S_{1,1,1}$ is finite. (Also note that the dimensionally regulated divergent 4-dimensional integral can be obtained from direct integration by analytic regularization \cite{analyticregularization}, or from a dimensional recurrence relation.)  In the Euclidean region $p^2 < 0$ and the internal masses are positive real valued. Expanding the integrand around $d = 2-2\epsilon$ gives:
\begin{align}
    S_{1,1,1}(s,m_1^2,m_2^2,m_3^2) &= \sum_{k=0}^\infty \epsilon^k \int_\Delta d^3\vec{\alpha}\, \frac{1}{k!} \mathcal{F}^{-1}\log\left(\frac{\mathcal{U}^3}{\mathcal{F}^2}\right)^k\nonumber\\
    &\equiv S^{(0)}_{111} (s,m_1^2,m_2^2,m_3^2)+\epsilon\, S^{(1)}_{111} (s,m_1^2,m_2^2,m_3^2)+\nonumber\\&
    \qquad+\epsilon^2\, S^{(2)}_{111} (s,m_1^2,m_2^2,m_3^2)+
    \mathcal{O}(\epsilon^2)\,.
\end{align}
First we apply Cheng-Wu to put $\alpha_1$ to 1, and integrate with respect to $\alpha_2$. The $\mathcal{U}$ polynomial is linear in the integration variable, whereas the $\mathcal{F}$ polynomial is not. Therefore we factor:
\begin{equation}
    \mathcal{F} = m_2^2 \left(\alpha_3+1\right) (\alpha_2 - R_+)(\alpha_2 - R_-)\, ,
\end{equation}
where the roots are:
\begin{align}
    \label{eq:sunrisezeros}
    R_\pm(s,m_1^2,m_2^2,m_3^2) = \frac{-\alpha _3^2 m_3^2+\alpha _3 \left(-m_1^2-m_2^2-m_3^2+s\right)-m_1^2\pm\sqrt{P_S}}{2 \left(\alpha _3+1\right) m_2^2}\,,
\end{align}
and we have a fourth degree polynomial in the last integration parameter:
\begin{align}
    \label{eq:SunriseEllipticCurve0Inf}
    P_{S} = \left(\alpha _3^2 m_3^2+\alpha _3 m_1^2+\alpha _3 m_2^2+\alpha _3 m_3^2+m_1^2-\alpha _3 s\right)^2-4 \left(\alpha _3 m_2^2+m_2^2\right) \left(\alpha _3 m_1^2+\alpha _3^2 m_3^2\right)\,,
\end{align}
which defines an elliptic curve. Explicitly integrating with respect to $\alpha_2$ one finds:
\begin{align}
    S_{1,1,1}^{(0)}(s,m_1^2,m_2^2,m_3^2) &= \int_0^\infty d\alpha_3\, \frac{1}{\sqrt{P_S}}\log\left(\frac{R_-}{R_+}\right)\,.
\end{align}
At order $\epsilon^1$ we obtain:
\begin{align}
\label{eq:one-foldsunriseeps1}
S_{1,1,1}^{(1)}(s,&m_1^2,m_2^2,m_3^2) = \int _0^{\infty }d\alpha_3 \frac{1}{\sqrt{P_S}}\bigg(G_{-\frac{R_+}{R_--R_+}}G_{Q_S}-G_{0,\frac{1}{R_-+1}}+G_{0,\frac{1}{R_++1}}\nonumber\\&-2G_{\frac{1}{R_-+1},\frac{1}{R_++1}}+2G_{\frac{1}{R_++1},\frac{1}{R_-+1}}-3G_{\alpha_3+1,\frac{1}{R_-+1}}+3G_{\alpha_3+1,\frac{1}{R_++1}} \bigg)\, ,
\end{align}
where we introduced:
\begin{align}
Q_S=\frac{\left(\alpha _3+1\right) m_2^2 \left(\alpha _3 m_3^2+m_1^2\right)}{\left(\alpha _3+1\right) m_1^2 m_2^2+\alpha _3 \left(\left(\alpha _3+1\right) m_2^2 m_3^2-\alpha _3\right)}\,.
\end{align}
In deriving this result we combined some logarithmic terms encountered at an intermediate stage. Note that the inner polylogarithms are of uniform weight 2. Using HyperInt one may also obtain higher orders in $\epsilon$ with little difficulty. One can then verify by explicit computation that at order $k$ in $\epsilon$ the polylogarithmic part is of weight $k+1$.

\subsubsection{Differential equations for the inner polylogarithmic part}
Next we employ the ideas of section \ref{sec:IPPDifferentialEquations} to setup a system of differential equations for the inner polylogarithmic part. First we define the propagators explicitly:
\begin{align}
    D_1 = -k_1^2+m_1^2 && D_2=-k_2^2+m_2^2&&D_3 = -(k_1+k_2+p)^2+m_3^2\, .
\end{align}
We may then write using the Feynman trick that:
\begin{align}
    \frac{1}{D_1^{\nu_1}D_3^{\nu_3} } = \frac{\Gamma \left(\nu _1+\nu _3\right)}{\Gamma \left(\nu _1\right) \Gamma \left(\nu _3\right)}\int_0^\infty dx\, \frac{x^{\nu_3-1}}{(D_1+xD_3)^{\nu_1+\nu_3}} \, .
\end{align}
Letting $\tilde{D}_1\equiv D_1+xD_3$, we define a ''generalized'' topology where $x$ is interpreted as an external scale:
\begin{align}
    S^{\textrm{IPP}}_{\nu_1+\nu_3,\nu_2} &\equiv \frac{\Gamma(\nu_1+\nu_3)\Gamma(\nu_2)}{(i\pi^{d/2})^2 \Gamma(\nu_1+\nu_2+\nu_3-d)} \int d^dk_1 d^dk_2 \,\frac{1}{\tilde{D}_1^{\nu_1+\nu_3}D_2^{\nu_2}}\,,
\end{align}
which satisfies that:
\begin{align}
    \label{eq:sunriseIPPratio}
    S_{\nu_1,\nu_2,\nu_3} &= \int_0^\infty x^{-1+\nu_3} S^{\textrm{IPP}}_{\nu_1+\nu_3,\nu_2} \,dx\,.
\end{align}
To perform an IBP reduction of the integrals in the inner polylogarithmic part we extend the topology with additional propagators to $\{\tilde{D}_1,D_2,N_1,N_2,N_3\}$, where:
\begin{align}
 N_1 = -k_1^2\,, &&
 N_2 = -(k_1+k_2)^2\,, &&
 N_3 = -(k_1+p)^2\,,
\end{align}
and we obtain the IBP reduction using the C++ version of FIRE5. One may verify that in $d=2-2\epsilon$ the following master integrals form a canonical basis:
\begin{align}
B_1 = 2(m_3^2)^{2\epsilon} x \epsilon\tilde{S}^{\textrm{IPP}}_{2,0}\,, && 
B_2 = 2(m_3^2)^{2\epsilon} (1+x)\epsilon^2 \tilde{S}^{\textrm{IPP}}_{1,1}\,, &&
B_3 = \epsilon(m_3^2)^{2\epsilon + 1} y \tilde{S}^{\textrm{IPP}}_{2,1} \, ,
\end{align}
where $y^2=P_S^{(x)}/m_3^4$, and $P_S^{(x)}$ is the polynomial of eq. (\ref{eq:SunriseEllipticCurve0Inf}) with $\alpha_3$ replaced by $x$. We have introduced a constant normalization for the inner polylogarithmic part by defining:
\begin{align}
\tilde{S}^{\textrm{IPP}}_{\nu_1+\nu_3,\nu_2} \equiv \frac{1}{(i\pi^{d/2})^2 \Gamma(3-d)} \int d^dk_1 d^dk_2 \,\frac{1}{\tilde{D}_1^{\nu_1+\nu_3}D_2^{\nu_2}}\,.
\end{align}
We note that we included the prefactor $(m_3^2)^{2\epsilon}$ in the canonical basis integrals to make them dimensionless. We divided out the term $m_3^4$ in the elliptic curve to obtain the form:
\begin{align}
    y^2 = (x-a_1)(x-a_2)(x-a_3)(x-a_4)\,,
\end{align}
where the $a_i$ variables denote the roots of the elliptic curve. In principle these are defined up to permutations, and for our purposes the ordering will not play an important role. To fix some convention for the roots, we let:
\begin{align}
 a_1&=-\frac{m_1^2-\left(\sqrt{s}+m_2\right)^2+m_3^2+\sqrt{\left(\left(\sqrt{s}-m_1+m_2\right)^2-m_3^2\right) \left(\left(\sqrt{s}+m_1+m_2\right)^2-m_3^2\right)}}{2 m_3^2}\,, \nonumber\\
 a_2&=-\frac{m_1^2-\left(\sqrt{s}-m_2\right)^2+m_3^2+\sqrt{\left(\left(\sqrt{s}+m_1-m_2\right)^2-m_3^2\right) \left(\left(-\sqrt{s}+m_1+m_2\right)^2-m_3^2\right)}}{2 m_3^2}\,, \nonumber\\
 a_3&=\frac{-m_1^2+\left(\sqrt{s}-m_2\right)^2-m_3^2+\sqrt{\left(\left(\sqrt{s}+m_1-m_2\right)^2-m_3^2\right) \left(\left(-\sqrt{s}+m_1+m_2\right)^2-m_3^2\right)}}{2 m_3^2}\,, \nonumber\\
 a_4&=\frac{-m_1^2+\left(\sqrt{s}+m_2\right)^2-m_3^2+\sqrt{\left(\left(\sqrt{s}-m_1+m_2\right)^2-m_3^2\right) \left(\left(\sqrt{s}+m_1+m_2\right)^2-m_3^2\right)}}{2 m_3^2}
\,.
\end{align}
With $\vec{B}=(B_1,B_2,B_3)$, the canonical form differential equation is given by:
\begin{align}
	d\vec{B} = \epsilon\,d\mathbf{A}\,\vec{B}\,,
\end{align}
where the differential equation matrix is:
\begin{align}
\mathbf{A} = \left(
\begin{array}{ccc}
 l_8-2 l_4 & 0 & 0 \\
 \frac{1}{2} \left(l_6-l_5\right) & \frac{1}{2} \left(-3 l_5-l_6+4 l_7-2 l_8\right) & l_1-l_2 \\
 \frac{1}{4} \left(3 l_1+l_2\right) & \frac{3}{4} \left(l_2-l_1\right) & \frac{1}{2} \left(-4 l_3+l_5+3 l_6+4 l_7+6 l_8\right) \\
\end{array}
\right)\,,
\end{align}
and where the letters $l_i$ are given by:
\begin{align}
 &l_1 = \resizebox{.4\hsize}{!}{$\log\left(-\frac{x s+(x+1) m_1^2-x m_2^2+\left(x^2+x-y\right) m_3^2}{x s+(x+1) m_1^2-x m_2^2+\left(x^2+x+y\right) m_3^2}\right)$}\,, \span\omit\span\omit&&
 l_2 = \resizebox{.4\hsize}{!}{$\log\left(\frac{(x+1) m_1^2+x m_2^2+\left(x^2+x+y\right) m_3^2-s x}{(x+1) m_1^2+x m_2^2+\left(x^2+x-y\right) m_3^2-s x}\right)$}\,,\span\omit\span\omit\nonumber\\
 &l_3 =\log\left(y^2\right)\,, &&
 \quad l_4 =\log\left(\frac{m_1^2}{m_3^2}+x\right)\,, &&
 l_5 =\log\left(\frac{m_2^2}{m_3^2}\right)\,, &&
 \quad l_6 =\log\left(\frac{s}{m_3^2}\right)\,, \nonumber\\
 &l_7 =\log\left(x+1\right)\,, &&
 \quad l_8 =\log\left(x\right)\,.
\end{align}
We may obtain the symbol of the master integrals as long as we have their leading coefficients in the $\epsilon$-expansion. One may find by power counting that $B_3$ vanishes at finite order. The leading coefficients of $B_1$ and $B_2$ are exactly 1. Therefore the leading coefficient vector is given by $\vec{B}^{(0)}=(1,1,0)$. The symbol at all orders in $\epsilon$ can thus be written as:
\begin{align}
	\mathcal{S}\left(\vec{B}\right)=\sum_{k=0}^\infty \epsilon^k\left(R\left(\mathbf{A}^{\otimes^k}\cdot (1,1,0)^T\right)\right)\,.
\end{align}
One may explicitly verify that the resulting symbol matches the symbol obtained from applying the maximal iteration of the coproduct to the solutions from HyperInt.

\paragraph{Solution in terms of multiple elliptic polylogarithms}
We consider another method next, and solve the differential equation in terms of elliptic polylogarithms ($\text{E}_4$-functions.) To do so we first map the integration parameter $x$ to the domain $[0,1]$. Note that for the initial application of Feynman's trick we could have alternatively used the form:
\begin{align}
    \frac{1}{D_1^{\nu_1}D_3^{\nu_3} } = \frac{\Gamma \left(\nu _1+\nu _3\right)}{\Gamma \left(\nu _1\right) \Gamma \left(\nu _3\right)} \int_0^1 dx'\, \frac{(1-x')^{\nu_1}(x')^{\nu_3}}{((1-x')D_1+x'D_3)^{\nu_1+\nu_3}}\,.
\end{align}
We define $\hat{D}_1 \equiv (1-x')D_1+x'D_3$, and consider:
\begin{align}
\hat{S}^{\textrm{IPP}}_{\nu_1+\nu_3,\nu_2} \equiv \frac{1}{(i\pi^{d/2})^2 \Gamma(3-d)} \int d^dk_1 d^dk_2 \,\frac{1}{\hat{D}_1^{\nu_1+\nu_3}D_2^{\nu_2}}\,.
\end{align}
Under the identification $x=x'/(1-x')$ one has:
\begin{align}
\tilde{S}^{\textrm{IPP}}_{\nu_1+\nu_3,\nu_2} =(1-x')^{\nu_1+\nu_3} \hat{S}^{\textrm{IPP}}_{\nu_1+\nu_3,\nu_2} \,,
\end{align}
and we may rewrite the canonical basis in terms of $x'$ and $\hat{S}^{\text{IPP}}$ as:
\begin{align}
B_1 = 2(m_3^2)^{2\epsilon} (1-x')x' \epsilon\hat{S}^{\textrm{IPP}}_{2,0}\,, &&
B_2 = 2(m_3^2)^{2\epsilon} \epsilon^2 \hat{S}^{\textrm{IPP}}_{1,1}\,, &&
B_3 = \epsilon(m_3^2)^{2\epsilon}(m_2^2-s) y' \hat{S}^{\textrm{IPP}}_{2,1} \, ,
\end{align}
where we now have the elliptic curve:
\begin{align}
\label{eq:EllipticCurveSunrise01}
(y')^2 &= \frac{1}{\left(s-m_2^2\right)^2}\bigg(\left(x'\right)^2 \left(2 m_2^2 \left(x'-1\right) \left(m_3^2-s x'+s\right)+\left(m_3^2+s \left(x'-1\right)\right)^2+m_2^4 \left(x'-1\right)^2\right)+\nonumber\\&\quad\quad m_1^4 \left(x'-1\right)^2-2 m_1^2 \left(x'-1\right) x' \left(m_2^2 \left(x'-1\right)+m_3^2+s \left(x'-1\right)\right)\bigg) \nonumber\\
    &= (x'-a_1')(x'-a_2')(x'-a_3')(x'-a_4')\,.
\end{align}
The explicit expressions for the roots $a_i'$ are long and not particularly insightful expressions, and the reader may obtain them from the relation:
\begin{align}
    a_i' = \frac{a_i}{a_i+1}\,.
\end{align}
We point out that the upcoming expressions in terms of $\text{E}_4$-functions will be provided in the Euclidean region. This means we will use the following kinds of simplifications:
\begin{align}
    y(0)=\sqrt{\frac{m_1^4}{\left(m_2^2-s\right)^2}}=\frac{m_1^2}{m_2^2-s}\,.
\end{align}
We will solve the differential equation with respect to $x'$, which is given by:
\begin{align}
    \frac{\partial}{\partial x'}\vec{B} = \epsilon \frac{\partial \mathbf{A}}{\partial x'} \vec{B}\,.
\end{align}
The partial derivative of $\mathbf{A}$ with respect to $x'$ works out to:
\begin{align}
\resizebox{ \textwidth}{!}{$
\frac{\partial\mathbf{A}}{\partial x'} = \left(
\begin{array}{c|c|c}
 -\frac{2 \left(m_3^2-m_1^2\right)}{m_1^2+x' \left(m_3^2-m_1^2\right)}+\frac{1}{x'-1}+\frac{1}{x'} & 0 & 0 \\\hline
 \multirow{2}{*}{0} & -\frac{1}{x'}-\frac{1}{x'-1} & \multirow{2}{*}{$\frac{2 m_1^2}{x' y' \left(m_2^2-s\right)}+\frac{2 m_3^2}{(x'-1) y' \left(m_2^2-s\right)}$}\\&+\frac{2 \left(m_1^2-m_3^2\right)}{y' \left(s-m_2^2\right)} \\\hline 
 \frac{m_1^2}{2 x' y' \left(m_2^2-s\right)}+\frac{m_3^2 m_1^2}{y' \left(m_1^2-m_3^2\right) \left(x' m_1^2-m_1^2-x' m_3^2\right)}  & \frac{3 m_1^2}{2 x' y' \left( s- m_2^2\right)}+\frac{3 \left(m_1^2-m_3^2\right)}{2 y' \left(m_2^2-s\right)} & \frac{3}{x'}-\frac{2}{x'-a_1'}-\frac{2}{x'-a_2'}-\frac{2}{x'-a_3'}-\frac{2}{x'-a_4'}   \\ 
 +\frac{m_3^2}{2 (x'-1) y' \left(m_2^2-s\right)}+\frac{m_1^4+2 \left(s-m_2^2-m_3^2\right) m_1^2+m_3^4}{2 y' \left(s-m_2^2\right) \left(m_1^2-m_3^2\right)}-\frac{x'}{y'}
 &  
 -\frac{3 m_3^2}{2 (x'-1) y' \left(m_2^2-s\right)}
 &+\frac{3}{x'-1}
\end{array}
\right)
$}
\end{align}
All of the entries may be expressed in terms of integration kernels of the $\text{E}_4$-functions defined in \cite{Broedel:2017siw}, of which we wrote down the relevant ones down in eq. (\ref{eq:E4IntegrationKernels}). We may write down the formal solution of the differential equation as a path-ordered exponential:
\begin{align}
\label{eq:BSunrisePathOrderedExponential}
\vec{B}(x',s,m_1,m_2,m_3) = \mathbb{P} \exp\left(\epsilon \int_{x'_0}^{x'}\frac{\partial\mathbf{A}}{\partial x'}\,dx'  \right) \vec{B}(x'_0,p^2,m_1,m_2,m_3)\,,
\end{align}
and we find a particularly simple expression for the first master integral of the sunrise:
\begin{align}
\label{eq:S111iteratedintegral}
S_{1,1,1}(s,m_1,m_2,m_3) &= \frac{(m_3^2)^{-2\epsilon}}{(m_2^2-s)\epsilon}\int_0^1 dx' \frac{B_3}{y'} \nonumber\\
&=\frac{(m_3^2)^{-2\epsilon}}{(m_2^2-s)\epsilon}\int_0^1 dx' \frac{1}{y'} \sum_{k=1}^3\left(\mathbb{P} \exp\left(\epsilon \int_{x'_0}^{x'}\frac{\partial\mathbf{A}}{\partial x'}\,dx'  \right)_{3,k} B_k(x'_0,s,m_1,m_2,m_3)\right)\,,
\end{align}
which exposes the last integration kernel at all orders in $\epsilon$. In order to obtain a representation in terms of $\text{E}_4$-functions, we would like to pick the boundary condition $x'_0 = 0$, but we note that $\vec{B}(x'_0,s,m_1,m_2,m_3)$ is singular in this limit. Nonetheless, the limit as $x'_0\rightarrow 0$ of the right hand side of eq. (\ref{eq:BSunrisePathOrderedExponential}) should be finite, since the left hand side of the equation is finite. 

Since the iterated integrals arising from the path-ordered exponential are multiple elliptic polylogarithms, we know that we may regulate the base-point divergence, which is of a logarithmic kind, using the tangential basepoint prescription. To get a consistent finite result we should apply the exact same regularization to the boundary term $\vec{B}(x'_0,s,m_1,m_2,m_3)$, which will amount to taking the limit as $x'_0\rightarrow 0$ from the positive real axis, and throwing away divergences of the form $\log(x'_0)^k$, where $k$ is a positive integer.

Let's explicitly compute $\text{reglim}_{x'\rightarrow 0}\vec{B}(x',s,m_1,m_2,m_3)$. It is relatively easy to compute the corresponding expression for $B_1$, as the Feynman parametrization of $\hat{S}_{2,0}^{\text{IPP}}$ has no non-trivial integrations. Furthermore, note that $B_1$ does not depend on $s$. For the integrals $B_2$ and $B_3$ there is a non-trivial integration to be performed. To compute the regularized limits of $B_2$ and $B_3$ we first exploit a symmetry based argument to simplify this integration. 

If we put $x'=0$ in the momentum space representation we find that the topology becomes that of a squared tadpole, and the resulting integral is independent of $s$. However, we need to first compute the integral for nonzero $x'$, and then compute the regularized limit as $x'\rightarrow 0$ in order to get the correct boundary term. One may wonder if the dependence on $s$ also disappears in the regularized limit, so that:
\begin{align}
\label{eq:RegLimSymmetry}
\textrm{reglim}_{x'\rightarrow 0}\left(\vec{B}(x',s,m_1,m_2,m_3)\right) = \textrm{reglim}_{x'\rightarrow 0}\left(\vec{B}(x',0,m_1,m_2,m_3)\right)\,.
\end{align}
We may write a closed form expression for $\vec{B}(x',0,m_1,m_2,m_3)$ by integrating up the Feynman parametrization. This leads to the following expressions:
\begin{align}
\label{eq:Bwithpsq0}
B_1&(x',0,m_1,m_2,m_3) = C_1\,, \nonumber\\
B_2&(x',0,m_1,m_2,m_3) = C_1 \left(\frac{ 2^{1-2 \epsilon } \sqrt{\pi } \epsilon  \Gamma (\epsilon )}{\Gamma \left(\epsilon +\frac{1}{2}\right)}\left(\frac{A_1^2 A_2}{\left(1-x'\right) x'}\right)^{-\epsilon }-\, _2F_1\left(1,2 \epsilon ;\epsilon +1;A_1\right)\right)\,,\nonumber\\
B_3&(x',0,m_1,m_2,m_3) = C_1 \sqrt{A_2^2} \left(\frac{A_1 \, _2F_1\left(1,2 \epsilon +1;\epsilon +1;A_1\right)}{\left(x'-1\right) x'}-\frac{ 4^{-\epsilon }\sqrt{\pi } \epsilon  \left(1-A_1\right)^{-\epsilon -1} A_1^{1-\epsilon } \Gamma (\epsilon )}{\left(x'-1\right) x' \Gamma \left(\epsilon +\frac{1}{2}\right)}\right)\,,
\end{align}
where we labeled the following terms:
\begin{align}
    A_1 &= \frac{m_2^2 \left(x'-1\right) x'}{m_1^2 \left(x'-1\right)-m_3^2 x'}\,, &&
    A_2 = \frac{m_1^2 \left(-x'\right)+x' \left(m_2^2 \left(x'-1\right)+m_3^2\right)+m_1^2}{m_2^2}\,, \nonumber\\
    C_1 &= \left(m_3^2\right)^{2 \epsilon } \left(\frac{A_1^2}{m_2^4 \left(1-x'\right) x'}\right)^{\epsilon }\,.
\end{align}
Using these results we have tested numerically whether eq. (\ref{eq:RegLimSymmetry}) is justified for $B_2$ and $B_3$. We took a random point in the external scales, with positive masses and a negative value for $s$. Furthermore, we took a number of increasingly small samples for $x'$, up to $10^{-20}$. We then computed $B_k(x',s,m_1,m_2,m_3)$ for $k=2,3$ order by order in $\epsilon$ up to $\mathcal{O}(\epsilon^3)$ for each value of $x'$ by numerical integration. We note that in doing so it is important to first perform analytic regularization \cite{analyticregularization} of the Feynman parametrization so that the numerical integral converges order by order in $\epsilon$. We then computed numerical values for $B_k(x',0,m_1,m_2,m_3)$ from the expressions in eq. (\ref{eq:Bwithpsq0}). One finds that the difference $B_k(x',s,m_1,m_2,m_3) - B_k(x',0,m_1,m_2,m_3)$ becomes increasingly small for increasingly small $x'$ (while the individual terms actually blow up as $x'$ decreases.) By repeating this analysis for a few more points in the external scales we believe that eq. (\ref{eq:RegLimSymmetry}) is correct.

Next we take the regularized limit as $x'\rightarrow 0$. There are terms of the form $(x')^\epsilon$, which we first expand in $\epsilon$, so that $(x')^\epsilon = 1+\epsilon  \log (x')+\frac{1}{2} \epsilon ^2 \log ^2(x')+\mathcal{O}(\epsilon^3)$, and then we throw away the logarithmic divergences. In other words we let the terms $(x')^\epsilon\rightarrow 1$. The final expressions for the regularized limits are very simple pure functions of uniform transcendental weight:
{\def\arraystretch{2.2}
\begin{align}
\label{eq:reglimBsunrise}
\textrm{reglim}_{x'\rightarrow 0}\,\vec{B}(x',0,m_1,m_2,m_3) = \left(
\begin{array}{c}
 \left(\frac{m_3^2}{m_1^2}\right)^{2 \epsilon } \\
 -\frac{2^{1-2 \epsilon } \sqrt{\pi } \Gamma (1-\epsilon ) \Gamma (\epsilon ) \left(\frac{m_1^2}{m_2^2}\right)^{\epsilon } \left(\frac{m_1^2}{m_3^2}\right)^{-2 \epsilon }}{\Gamma (-\epsilon ) \Gamma \left(\epsilon +\frac{1}{2}\right)}-\left(\frac{m_1^2}{m_3^2}\right)^{-2 \epsilon } \\
 -\frac{4^{-\epsilon } \sqrt{\pi } \Gamma (1-\epsilon ) \Gamma (\epsilon ) \left(\frac{m_1^2}{m_2^2}\right)^{\epsilon } \left(\frac{m_1^2}{m_3^2}\right)^{-2 \epsilon }}{\Gamma (-\epsilon ) \Gamma \left(\epsilon +\frac{1}{2}\right)}-\left(\frac{m_1^2}{m_3^2}\right)^{-2 \epsilon } \\
\end{array}
\right)\,.
\end{align}
}

From eqs. (\ref{eq:BSunrisePathOrderedExponential}), (\ref{eq:S111iteratedintegral}) and (\ref{eq:reglimBsunrise}) we have all the elements to express $\vec{B}$ and in particular $S_{1,1,1}$ in terms of multiple elliptic polylogarithms. Note that in accordance with \cite{Broedel:2017siw} we shuffle-regulate the $\text{E}_4$-functions that end with a kernel of the type $\psi_1(0,x) = 1/x$. For example we may write:
\begin{align}
	\text{E}_4\left(\begin{xsmallmatrix}{.05em}1&1\\2&0\end{xsmallmatrix}; 1\right) &= \text{E}_4\left(\begin{xsmallmatrix}{.05em}1&1\\2&0\end{xsmallmatrix}; 1\right) + \text{E}_4\left(\begin{xsmallmatrix}{.05em}1\\2\end{xsmallmatrix}; 1\right)\text{E}_4\left(\begin{xsmallmatrix}{.05em}1\\0\end{xsmallmatrix}; 1\right) - \text{E}_4\left(\begin{xsmallmatrix}{.05em}1\\2\end{xsmallmatrix}; 1\right)\shuffle \text{E}_4\left(\begin{xsmallmatrix}{.05em}1\\0\end{xsmallmatrix}; 1\right) \nonumber\\
	&= -\text{E}_4\left(\begin{xsmallmatrix}{.05em}1&1\\0&2\end{xsmallmatrix}; 1\right)\,,
\end{align}
where we explicitly worked out the shuffle product in one of the terms and used that $\text{E}_4\left(\begin{xsmallmatrix}{.05em}1\\0\end{xsmallmatrix}; 1\right) = \log(1) = 0$. In terms of $\text{E}_4$-functions the solution of the unequal mass sunrise in the Euclidean region is given up to order $\mathcal{O}(\epsilon)^2$ by:
{\small
\begin{align}
&c_4 \left(m_2^2-s\right) (m_3^2)^{2 \epsilon } S_{1,1,1} =\nonumber\\& -\text{E}_4\left(\begin{xsmallmatrix}{.15em}0&-1\\0&\hfill 0\\\end{xsmallmatrix},1\right)-\text{E}_4\left(\begin{xsmallmatrix}{.15em}0&-1\\0&\hfill 1\\\end{xsmallmatrix},1\right)-\text{E}_4\left(\begin{xsmallmatrix}{.15em}0&-1\\0&\hfill \infty\\\end{xsmallmatrix},1\right)+\text{E}_4\left(\begin{xsmallmatrix}{.15em}0&-1\\0&\hfill \frac{m_1^2}{m_1^2-m_3^2}\\\end{xsmallmatrix},1\right)+\text{E}_4\left(\begin{xsmallmatrix}{.15em}1&0\\0&0\\\end{xsmallmatrix},1\right)+\log\left(\frac{m_1^2}{m_2^2}\right)\text{E}_4\left(\begin{xsmallmatrix}{.15em}0\\0\\\end{xsmallmatrix},1\right)\nonumber\\&-\frac{m_1^2\left(m_2^2-2m_3^2-s\right)+m_1^4+m_3^4}{c_4\left(m_1^2-m_3^2\right)\left(s-m_2^2\right)}\text{E}_4\left(\begin{xsmallmatrix}{.15em}0&0\\0&0\\\end{xsmallmatrix},1\right) + \epsilon\bigg(
-\frac{\pi^2}{6}\text{E}_4\left(\begin{xsmallmatrix}{.15em}0\\0\\\end{xsmallmatrix},1\right)+2\text{E}_4\left(\begin{xsmallmatrix}{.15em}0&-1&1\\0&\hfill 0&1\\\end{xsmallmatrix},1\right)-\text{E}_4\left(\begin{xsmallmatrix}{.15em}0&-1&1\\0&\hfill 0&-\frac{m_1^2}{m_3^2-m_1^2}\\\end{xsmallmatrix},1\right)\nonumber\\&+2\text{E}_4\left(\begin{xsmallmatrix}{.15em}0&-1&1\\0&\hfill 1&1\\\end{xsmallmatrix},1\right)-\text{E}_4\left(\begin{xsmallmatrix}{.15em}0&-1&1\\0&\hfill 1&-\frac{m_1^2}{m_3^2-m_1^2}\\\end{xsmallmatrix},1\right)-\text{E}_4\left(\begin{xsmallmatrix}{.15em}0&-1&1\\0&\hfill \infty&1\\\end{xsmallmatrix},1\right)+2\text{E}_4\left(\begin{xsmallmatrix}{.15em}0&-1&1\\0&\hfill \infty&-\frac{m_1^2}{m_3^2-m_1^2}\\\end{xsmallmatrix},1\right)+\text{E}_4\left(\begin{xsmallmatrix}{.15em}0&-1&1\\0&\hfill \frac{m_1^2}{m_1^2-m_3^2}&1\\\end{xsmallmatrix},1\right)\nonumber\\&-2\text{E}_4\left(\begin{xsmallmatrix}{.15em}0&-1&1\\0&\hfill \frac{m_1^2}{m_1^2-m_3^2}&-\frac{m_1^2}{m_3^2-m_1^2}\\\end{xsmallmatrix},1\right)-5\text{E}_4\left(\begin{xsmallmatrix}{.15em}0&1&-1\\0&0&\hfill 0\\\end{xsmallmatrix},1\right)-5\text{E}_4\left(\begin{xsmallmatrix}{.15em}0&1&-1\\0&0&\hfill 1\\\end{xsmallmatrix},1\right)-2\text{E}_4\left(\begin{xsmallmatrix}{.15em}0&1&-1\\0&0&\hfill \infty\\\end{xsmallmatrix},1\right)+2\text{E}_4\left(\begin{xsmallmatrix}{.15em}0&1&-1\\0&0&\hfill \frac{m_1^2}{m_1^2-m_3^2}\\\end{xsmallmatrix},1\right)\nonumber\\&-3\text{E}_4\left(\begin{xsmallmatrix}{.15em}0&1&-1\\0&1&\hfill 0\\\end{xsmallmatrix},1\right)-3\text{E}_4\left(\begin{xsmallmatrix}{.15em}0&1&-1\\0&1&\hfill 1\\\end{xsmallmatrix},1\right)-3\text{E}_4\left(\begin{xsmallmatrix}{.15em}0&1&-1\\0&1&\hfill \infty\\\end{xsmallmatrix},1\right)+3\text{E}_4\left(\begin{xsmallmatrix}{.15em}0&1&-1\\0&1&\hfill \frac{m_1^2}{m_1^2-m_3^2}\\\end{xsmallmatrix},1\right)+2\text{E}_4\left(\begin{xsmallmatrix}{.15em}0&1&-1\\0&a_1'&\hfill 0\\\end{xsmallmatrix},1\right)+2\text{E}_4\left(\begin{xsmallmatrix}{.15em}0&1&-1\\0&a_1'&\hfill 1\\\end{xsmallmatrix},1\right)\nonumber\\&+2\text{E}_4\left(\begin{xsmallmatrix}{.15em}0&1&-1\\0&a_1'&\hfill \infty\\\end{xsmallmatrix},1\right)-2\text{E}_4\left(\begin{xsmallmatrix}{.15em}0&1&-1\\0&a_1'&\hfill \frac{m_1^2}{m_1^2-m_3^2}\\\end{xsmallmatrix},1\right)+2\text{E}_4\left(\begin{xsmallmatrix}{.15em}0&1&-1\\0&a_2'&\hfill 0\\\end{xsmallmatrix},1\right)+2\text{E}_4\left(\begin{xsmallmatrix}{.15em}0&1&-1\\0&a_2'&\hfill 1\\\end{xsmallmatrix},1\right)+2\text{E}_4\left(\begin{xsmallmatrix}{.15em}0&1&-1\\0&a_2'&\hfill \infty\\\end{xsmallmatrix},1\right)\nonumber\\&-2\text{E}_4\left(\begin{xsmallmatrix}{.15em}0&1&-1\\0&a_2'&\hfill \frac{m_1^2}{m_1^2-m_3^2}\\\end{xsmallmatrix},1\right)+2\text{E}_4\left(\begin{xsmallmatrix}{.15em}0&1&-1\\0&a_3'&\hfill 0\\\end{xsmallmatrix},1\right)+2\text{E}_4\left(\begin{xsmallmatrix}{.15em}0&1&-1\\0&a_3'&\hfill 1\\\end{xsmallmatrix},1\right)+2\text{E}_4\left(\begin{xsmallmatrix}{.15em}0&1&-1\\0&a_3'&\hfill \infty\\\end{xsmallmatrix},1\right)-2\text{E}_4\left(\begin{xsmallmatrix}{.15em}0&1&-1\\0&a_3'&\hfill \frac{m_1^2}{m_1^2-m_3^2}\\\end{xsmallmatrix},1\right)\nonumber\\&+2\text{E}_4\left(\begin{xsmallmatrix}{.15em}0&1&-1\\0&a_4'&\hfill 0\\\end{xsmallmatrix},1\right)+2\text{E}_4\left(\begin{xsmallmatrix}{.15em}0&1&-1\\0&a_4'&\hfill 1\\\end{xsmallmatrix},1\right)+2\text{E}_4\left(\begin{xsmallmatrix}{.15em}0&1&-1\\0&a_4'&\hfill \infty\\\end{xsmallmatrix},1\right)-2\text{E}_4\left(\begin{xsmallmatrix}{.15em}0&1&-1\\0&a_4'&\hfill \frac{m_1^2}{m_1^2-m_3^2}\\\end{xsmallmatrix},1\right)+5\text{E}_4\left(\begin{xsmallmatrix}{.15em}0&1&1\\0&0&1\\\end{xsmallmatrix},1\right)-2\text{E}_4\left(\begin{xsmallmatrix}{.15em}0&1&1\\0&0&a_1'\\\end{xsmallmatrix},1\right)\nonumber\\&-2\text{E}_4\left(\begin{xsmallmatrix}{.15em}0&1&1\\0&0&a_2'\\\end{xsmallmatrix},1\right)-2\text{E}_4\left(\begin{xsmallmatrix}{.15em}0&1&1\\0&0&a_3'\\\end{xsmallmatrix},1\right)-2\text{E}_4\left(\begin{xsmallmatrix}{.15em}0&1&1\\0&0&a_4'\\\end{xsmallmatrix},1\right)-\text{E}_4\left(\begin{xsmallmatrix}{.15em}0&1&1\\0&0&-\frac{m_1^2}{m_3^2-m_1^2}\\\end{xsmallmatrix},1\right)-2\text{E}_4\left(\begin{xsmallmatrix}{.15em}1&0&-1\\0&0&\hfill 0\\\end{xsmallmatrix},1\right)-2\text{E}_4\left(\begin{xsmallmatrix}{.15em}1&0&-1\\0&0&\hfill 1\\\end{xsmallmatrix},1\right)\nonumber\\&+\text{E}_4\left(\begin{xsmallmatrix}{.15em}1&0&-1\\0&0&\hfill \infty\\\end{xsmallmatrix},1\right)-\text{E}_4\left(\begin{xsmallmatrix}{.15em}1&0&-1\\0&0&\hfill \frac{m_1^2}{m_1^2-m_3^2}\\\end{xsmallmatrix},1\right)+3\text{E}_4\left(\begin{xsmallmatrix}{.15em}1&0&1\\0&0&1\\\end{xsmallmatrix},1\right)-2\text{E}_4\left(\begin{xsmallmatrix}{.15em}1&0&1\\0&0&a_1'\\\end{xsmallmatrix},1\right)-2\text{E}_4\left(\begin{xsmallmatrix}{.15em}1&0&1\\0&0&a_2'\\\end{xsmallmatrix},1\right)-2\text{E}_4\left(\begin{xsmallmatrix}{.15em}1&0&1\\0&0&a_3'\\\end{xsmallmatrix},1\right)\nonumber\\&-2\text{E}_4\left(\begin{xsmallmatrix}{.15em}1&0&1\\0&0&a_4'\\\end{xsmallmatrix},1\right)-\text{E}_4\left(\begin{xsmallmatrix}{.15em}1&1&0\\0&0&0\\\end{xsmallmatrix},1\right)-3\log\left(\frac{m_1^2}{m_2^2}\right)\text{E}_4\left(\begin{xsmallmatrix}{.15em}0&-1\\0&\hfill 0\\\end{xsmallmatrix},1\right)-3\log\left(\frac{m_1^2}{m_2^2}\right)\text{E}_4\left(\begin{xsmallmatrix}{.15em}0&-1\\0&\hfill 1\\\end{xsmallmatrix},1\right)+3\log\left(\frac{m_1^2}{m_2^2}\right)\text{E}_4\left(\begin{xsmallmatrix}{.15em}0&1\\0&1\\\end{xsmallmatrix},1\right)\nonumber\\&-2\log\left(\frac{m_1^2}{m_2^2}\right)\text{E}_4\left(\begin{xsmallmatrix}{.15em}0&1\\0&a_1'\\\end{xsmallmatrix},1\right)-2\log\left(\frac{m_1^2}{m_2^2}\right)\text{E}_4\left(\begin{xsmallmatrix}{.15em}0&1\\0&a_2'\\\end{xsmallmatrix},1\right)-2\log\left(\frac{m_1^2}{m_2^2}\right)\text{E}_4\left(\begin{xsmallmatrix}{.15em}0&1\\0&a_3'\\\end{xsmallmatrix},1\right)-2\log\left(\frac{m_1^2}{m_2^2}\right)\text{E}_4\left(\begin{xsmallmatrix}{.15em}0&1\\0&a_4'\\\end{xsmallmatrix},1\right)\nonumber\\&+\frac{1}{2}\log^2\left(\frac{m_1^2}{m_2^2}\right)\text{E}_4\left(\begin{xsmallmatrix}{.15em}0\\0\\\end{xsmallmatrix},1\right)+3\log\left(\frac{m_1^2}{m_3^2}\right)\text{E}_4\left(\begin{xsmallmatrix}{.15em}0&-1\\0&\hfill 0\\\end{xsmallmatrix},1\right)+3\log\left(\frac{m_1^2}{m_3^2}\right)\text{E}_4\left(\begin{xsmallmatrix}{.15em}0&-1\\0&\hfill 1\\\end{xsmallmatrix},1\right)-3\log\left(\frac{m_1^2}{m_3^2}\right)\text{E}_4\left(\begin{xsmallmatrix}{.15em}1&0\\0&0\\\end{xsmallmatrix},1\right)\nonumber\\&+\log\left(\frac{m_3^2}{m_1^2}\right)\text{E}_4\left(\begin{xsmallmatrix}{.15em}0&-1\\0&\hfill 0\\\end{xsmallmatrix},1\right)+\log\left(\frac{m_3^2}{m_1^2}\right)\text{E}_4\left(\begin{xsmallmatrix}{.15em}0&-1\\0&\hfill 1\\\end{xsmallmatrix},1\right)-2\log\left(\frac{m_3^2}{m_1^2}\right)\text{E}_4\left(\begin{xsmallmatrix}{.15em}0&-1\\0&\hfill \infty\\\end{xsmallmatrix},1\right)+2\log\left(\frac{m_3^2}{m_1^2}\right)\text{E}_4\left(\begin{xsmallmatrix}{.15em}0&-1\\0&\hfill \frac{m_1^2}{m_1^2-m_3^2}\\\end{xsmallmatrix},1\right)\nonumber\\&-\log\left(\frac{m_3^2}{m_1^2}\right)\text{E}_4\left(\begin{xsmallmatrix}{.15em}1&0\\0&0\\\end{xsmallmatrix},1\right)-2\log\left(\frac{m_1^2}{m_3^2}\right)\log\left(\frac{m_1^2}{m_2^2}\right)\text{E}_4\left(\begin{xsmallmatrix}{.15em}0\\0\\\end{xsmallmatrix},1\right)-\frac{3\left(m_1^2-m_3^2\right)}{c_4\left(s-m_2^2\right)}\log\left(\frac{m_1^2}{m_2^2}\right)\text{E}_4\left(\begin{xsmallmatrix}{.15em}0&0\\0&0\\\end{xsmallmatrix},1\right)\nonumber\\&+\frac{3\left(m_1^2-m_3^2\right)}{c_4\left(s-m_2^2\right)}\log\left(\frac{m_1^2}{m_3^2}\right)\text{E}_4\left(\begin{xsmallmatrix}{.15em}0&0\\0&0\\\end{xsmallmatrix},1\right)+\frac{m_1^2\left(-m_2^2-4m_3^2+s\right)+2m_1^4+2m_3^4}{c_4\left(m_1^2-m_3^2\right)\left(s-m_2^2\right)}\text{E}_4\left(\begin{xsmallmatrix}{.15em}0&0&1\\0&0&1\\\end{xsmallmatrix},1\right)\nonumber\\&+\frac{2\left(m_1^2\left(m_2^2-2m_3^2-s\right)+m_1^4+m_3^4\right)}{c_4\left(m_1^2-m_3^2\right)\left(s-m_2^2\right)}\text{E}_4\left(\begin{xsmallmatrix}{.15em}0&1&0\\0&a_1'&0\\\end{xsmallmatrix},1\right)+\frac{2\left(m_1^2\left(m_2^2-2m_3^2-s\right)+m_1^4+m_3^4\right)}{c_4\left(m_1^2-m_3^2\right)\left(s-m_2^2\right)}\text{E}_4\left(\begin{xsmallmatrix}{.15em}0&1&0\\0&a_2'&0\\\end{xsmallmatrix},1\right)\nonumber\\&+\frac{2\left(m_1^2\left(m_2^2-2m_3^2-s\right)+m_1^4+m_3^4\right)}{c_4\left(m_1^2-m_3^2\right)\left(s-m_2^2\right)}\text{E}_4\left(\begin{xsmallmatrix}{.15em}0&1&0\\0&a_3'&0\\\end{xsmallmatrix},1\right)+\frac{2\left(m_1^2\left(m_2^2-2m_3^2-s\right)+m_1^4+m_3^4\right)}{c_4\left(m_1^2-m_3^2\right)\left(s-m_2^2\right)}\text{E}_4\left(\begin{xsmallmatrix}{.15em}0&1&0\\0&a_4'&0\\\end{xsmallmatrix},1\right)\nonumber\\&+\frac{2m_1^2\left(-m_2^2-m_3^2+s\right)+m_1^4+m_3^4}{c_4\left(m_1^2-m_3^2\right)\left(s-m_2^2\right)}\log\left(\frac{m_3^2}{m_1^2}\right)\text{E}_4\left(\begin{xsmallmatrix}{.15em}0&0\\0&0\\\end{xsmallmatrix},1\right)\nonumber\\&-\frac{-2m_1^2\left(m_2^2+m_3^2-s\right)+m_1^4+m_3^4}{c_4\left(m_1^2-m_3^2\right)\left(s-m_2^2\right)}\text{E}_4\left(\begin{xsmallmatrix}{.15em}0&0&1\\0&0&-\frac{m_1^2}{m_3^2-m_1^2}\\\end{xsmallmatrix},1\right)+\frac{3m_1^2\left(-m_2^2+2m_3^2+s\right)-3m_1^4-3m_3^4}{c_4\left(m_1^2-m_3^2\right)\left(s-m_2^2\right)}\text{E}_4\left(\begin{xsmallmatrix}{.15em}0&1&0\\0&1&0\\\end{xsmallmatrix},1\right)\nonumber\\&+\frac{m_1^2\left(m_2^2+4m_3^2-s\right)-2m_1^4-2m_3^4}{c_4\left(m_1^2-m_3^2\right)\left(s-m_2^2\right)}\text{E}_4\left(\begin{xsmallmatrix}{.15em}1&0&0\\0&0&0\\\end{xsmallmatrix},1\right)+\frac{2m_1^2\left(-m_2^2+5m_3^2+s\right)-5m_1^4-5m_3^4}{c_4\left(m_1^2-m_3^2\right)\left(s-m_2^2\right)}\text{E}_4\left(\begin{xsmallmatrix}{.15em}0&1&0\\0&0&0\\\end{xsmallmatrix},1\right)
\bigg) + \mathcal{O}(\epsilon^2)
\end{align}}
Lastly we remark on the three other master integrals in the top sector of the sunrise. One has for example:
\begin{align}
    S_{1,2,1} = \frac{1}{1+2\epsilon} \int_0^1 dx' \,\hat{S}_{2,2}^{\text{IPP}}.
\end{align}
We can furthermore express $\hat{S}_{2,2}^{\text{IPP}}$ in terms of the canonical basis integrals by IBP reduction, which leads to the following relation:
\begin{align}
\label{eq:IBPRelationShatIPP22}
(m_3^2)^{2\epsilon} \hat{S}^{\text{IPP}}_{2,2} &=  \left(\frac{c_{1,1}}{x-a_1'}+\frac{c_{1,2}}{x-a_2'}+\frac{c_{1,3}}{x-a_3'}+\frac{c_{1,4}}{x-a_4'}\right)B_1+ \left(\frac{c_{2,1}}{x-a_1'}+\frac{c_{2,2}}{x-a_2'}+\frac{c_{2,3}}{x-a_3'}+\frac{c_{2,4}}{x-a_4'}\right)B_2\nonumber\\&+ \left(\frac{c_{3,1}}{y \left(x-a_1'\right)}+\frac{c_{3,2}}{y \left(x-a_2'\right)}+\frac{c_{3,3}}{y \left(x-a_3'\right)}+\frac{c_{3,4}}{y \left(x-a_4'\right)}+\frac{c_{3,5}}{y}\right)B_3\,,
\end{align}
where we have the following coefficients:
\begin{align}
 c_{1,1}&=\frac{\left(a_1'-1\right) \left(a_1' \left(7 m_2^2+p^2\right)-m_1^2\right)+a_1' m_3^2}{4 a_{1,2}' a_{1,3}' a_{1,4}' m_2^2 \left(m_2^2-p^2\right)^2} \,,&     c_{2,1}&=\frac{3 \left(a_1'-1\right) \left(a_1' \left(p^2-m_2^2\right)-m_1^2\right)+3 a_1' m_3^2}{4 a_{1,2}' a_{1,3}' a_{1,4}' m_2^2 \left(m_2^2-p^2\right)^2}\,, \nonumber\\
 c_{3,1} &= \mathrlap{\frac{(4 \epsilon +1) \left(\left(a_1'-1\right) m_1^2-a_1' m_3^2\right) \left(\left(a_1'-1\right) \left(a_1' \left(m_2^2+3 p^2\right)-m_1^2\right)+a_1' m_3^2\right)}{\epsilon  a_{1,2}' a_{1,3}' a_{1,4}' \left(p^2-m_2^2\right)^4}\,,} \nonumber\\
 c_{3,5}&=\frac{m_2^2 (7 \epsilon +2)+p^2 \epsilon }{2 m_2^2 \epsilon  \left(m_2^2-p^2\right)^2}\,,
\end{align}
and where the other coefficients are given by cyclic permutations: $c_{i,j} = c_{i,j-1}|_{a_k'\rightarrow a_{k+1}'}$, for $i=1,2,3$ and $j = 2,3,4$, and where we let $a_5'$ refer to $a_1'$. It is clear from eqs. (\ref{eq:BSunrisePathOrderedExponential}), (\ref{eq:reglimBsunrise}) and (\ref{eq:IBPRelationShatIPP22}) that $S_{1,2,1}$ can be integrated in terms of $\text{E}_4$-functions. The first integrations are all expressible in terms of the kernels in eq. (\ref{eq:E4IntegrationKernels}), and while the last integration contains kernels of the type $dx'/(y'(x'-a_i'))$, it may be written in terms of kernels of $\text{E}_4$-functions by IBP relations \cite{Broedel:2017siw}.

\subsubsection{Analytic continuation}
\label{sec:SunriseAnalyticContinuation}
In this section we perform the analytic continuation to the physical region $s>0,m_1^2>m_2^2>m_3^2>0$ of the first sunrise master integral $S_{1,1,1}(s,m_1^2,m_2^2,m_3^2)$ using the methods introduced in section \ref{sec:continuation}. The analytic continuation of the $\epsilon^0$ order is elementary as it requires only elementary identities among logarithms and we do not discuss it here, while we provide the result in appendix \ref{app:contS}. We discuss the analytic continuation of the order $\epsilon^1$ coefficient, eq.~(\ref{eq:one-foldsunriseeps1}), which for the reader's convenience we write here in the following form:
\begin{equation}
    S_{1,1,1}^{(1)}(s,m_1^2,m_2^2,m_3^2,i\delta) = \frac{1}{m_3^2}\int_0^\infty \frac{1}{y} f_{S}^{(2)}(x,s,m_1^2,m_2^2,m_3^2,i\delta) dx\,,
\end{equation}
where the $i\delta$ is introduced by applying the Feynman prescription $s\rightarrow s+i\delta$. The symbol alphabet letters of $f^{(2)}(x,s,m_1^2,m_2^2,m_3^2,0)$ can be expressed in terms of the following linearly independent letters:
\begin{align}
    \alpha _1& =x,\;\alpha _2=x+1,\;\alpha _3=m_2^2,\;\alpha _4=m_3^2 x+m_1^2,\;\alpha _5=x \left(m_3^2-s\right)+m_1^2-m_2^2\,,\nonumber\\
    \alpha _6&=s,\;\alpha _7=m_3^2 y,\;\beta _1 =m_3^2 x^2+m_1^2 (x-1)-m_2^2 x-m_3^2 x+s x+m_3^2 y\,,\nonumber\\
    \beta _2&=-m_3^2 x^2+m_1^2 (-(x+1))+m_2^2 x-m_3^2 x+s x+m_3^2 y\,,\nonumber\\
    \beta _3&=-m_3^2 x^2+m_1^2 (-(x+1))+m_2^2 (x+2)-m_3^2 x+s x+m_3^2 y\,.
\end{align}
The alphabet of $f^{(2)}(x,s,m_1^2,m_2^2,m_3^2,0)$ contains only 8 linearly independent letters, however, as discussed in section \ref{sec:continuation}, spurious letters are in general needed when defining function arguments, and in the present case we needed the extended alphabet above to be able to find a representation for $\lim_{\delta\rightarrow 0} f^{(2)}(x,s,m_1^2,m_2^2,m_3^2,i\delta)$ in the different relevant regions in terms of classical polylogarithms and logarithms (see e.g. \cite{Frellesvig:2016ske}).

As explained in section \ref{sec:continuation} two regions are identified by requiring that the algebraic function, the elliptic curve $y(x)$ in our case, does not have branch points:
\begin{equation}
 A_S:y^2(x,s,m_1^2,m_2^2,m_3^2)<0\,,\quad B_S:y^2(x,s,m_1^2,m_2^2,m_3^2)>0\,. 
\end{equation}
We then notice that in the region $A_S$ neither of the symbol letters vanish, so no further partitioning of $A_S$ is required. On the other hand region $B_S$ is partitioned as follows: 
\begin{align}
    B_{S,1}& :  \alpha _5<0, \beta _1<0, \beta _2<0, \beta _3>0\,,\; \; B_{S,3}:\alpha _5<0, \beta _1>0, \beta _2>0, \beta _3>0\,,\nonumber\\
    B_{S,2}& :  \alpha _5<0\,, \beta _1>0, \beta _2<0, \beta _3<0\,,\;\;  B_{S,4}: \alpha _5>0, \beta _1>0, \beta _2<0, \beta _3>0\,.
\end{align}
Note that, as prescribed in section \ref{sec:continuation}, each subregion is defined by requiring that all the letters have define sign, however for some subregions the set of constraints have no intersection, and only the subregions above need to be considered in this case. For later convenience we rename the regions as:
\begin{equation}
    R_{S,1}=A_S,\;R_{S,2}=B_{S,1},\;R_{S,3}=B_{S,2},\;R_{S,4}=B_{S,3},\;R_{S,5}=B_{S,4}\,.\;
\end{equation}
We get:
\begin{align}
\label{eq:contS}
    S_{1,1,1}^{(1)}(s,m_1^2,m_2^2,m_3^2)=& \frac{1}{m_3^2}\int_0^\infty \frac{\theta_1(x,s, m_1^2,m_2^2,m_3^2)}{i\sqrt{-y^2(x,s, m_1^2,m_2^2,m_3^2)}} f_{S,1}^{(2)}(x,s, m_1^2,m_2^2,m_3^2)dx \nonumber\\+&\frac{1}{m_3^2}\int_0^\infty \sum_{j=2}^5 \frac{\theta_j(x,s, m_1^2,m_2^2,m_3^2)}{y(x,s, m_1^2,m_2^2,m_3^2)} f_{S,j}^{(2)}(x,s, m_1^2,m_2^2,m_3^2)dx\,,
\end{align}
where $\theta_i(x,s, m_1^2,m_2^2,m_3^2)=1$ if $x,s, m_1^2,m_2^2,m_3^2\in R_{S,i}$, $\theta_i(x,s, m_1^2,m_2^2,m_3^2)=0$ otherwise, and we have for example:
\begin{align}
    f_{S,2}^{(2)}&(x,s, m_1^2,m_2^2,m_3^2) = \text{Li}_2\left(\frac{2 \alpha _1 \alpha _3}{\beta _2}\right)-\text{Li}_2\left(\frac{\beta _2}{2 \alpha _1 \alpha _6}\right)-4 \text{Li}_2\left(-\frac{\alpha _7 \beta _2}{2 \alpha _1^2 \alpha _3
   \alpha _6}\right)+4 \text{Li}_2\left(-\frac{2 \alpha _5 \alpha _7 \beta _2}{\alpha _1 \alpha _6 \beta _1 \beta _3}\right)\nonumber\\
   & +2 \log \left(\alpha _1\right) \log \left(-\beta _1\right)+8 \log \left(\alpha _1\right) \log \left(-\beta _2\right)+2 \log \left(\alpha _1\right) \log \left(\beta _3\right)-4 \log \left(\alpha
   _2\right) \log \left(-\beta _1\right)\nonumber\\
   &-8 \log \left(-\alpha _5\right) \log \left(-\beta _1\right)+4 \log \left(\alpha _6\right) \log \left(-\beta _1\right)+7 \log \left(\alpha _3\right) \log
   \left(-\beta _2\right)+8 \log \left(-\alpha _5\right) \log \left(-\beta _2\right)\nonumber\\
   &+\log \left(\alpha _6\right) \log \left(-\beta _2\right)-4 \log \left(\alpha _2\right) \log \left(\beta _3\right)-8
   \log \left(-\alpha _5\right) \log \left(\beta _3\right)+4 \log \left(\alpha _6\right) \log \left(\beta _3\right)-6 \log ^2\left(\alpha _1\right)\nonumber\\
   &+\log ^2\left(\alpha _2\right)-\frac{5}{2} \log
   ^2\left(\alpha _3\right)-\log ^2\left(\alpha _4\right)+4 \log ^2\left(-\alpha _5\right)-\frac{1}{2} \log ^2\left(\alpha _6\right)+3 \log \left(\alpha _2\right) \log \left(\alpha _1\right)\nonumber\\
   &-8 \log
   \left(\alpha _3\right) \log \left(\alpha _1\right)+\log \left(\alpha _4\right) \log \left(\alpha _1\right)-2 \log \left(-\alpha _5\right) \log \left(\alpha _1\right)-5 \log \left(\alpha _6\right)
   \log \left(\alpha _1\right)\nonumber\\
   &-12 \log (2) \log \left(\alpha _1\right)+8 \log (2) \log \left(\alpha _2\right)+2 \log \left(\alpha _2\right) \log \left(\alpha _3\right)-7 \log (2) \log \left(\alpha
   _3\right)+4 \log \left(\alpha _2\right) \log \left(-\alpha _5\right)\nonumber\\
   &+8 \log (2) \log \left(-\alpha _5\right)-4 \log \left(\alpha _3\right) \log \left(\alpha _6\right)-4 \log \left(-\alpha
   _5\right) \log \left(\alpha _6\right)-9 \log (2) \log \left(\alpha _6\right)+4 \log ^2\left(-\beta _1\right)\nonumber\\
   &+4 \log ^2\left(\beta _3\right)-8 \log (2) \log \left(-\beta _1\right)-8 \log
   \left(-\beta _1\right) \log \left(-\beta _2\right)+16 \log (2) \log \left(-\beta _2\right)+8 \log \left(-\beta _1\right) \log \left(\beta _3\right)\nonumber\\
   &-8 \log \left(-\beta _2\right) \log \left(\beta
   _3\right)-8 \log (2) \log \left(\beta _3\right)\,.
\end{align}
The expressions for $f_{S,i}^{(2)}(x,s, m_1^2,m_2^2,m_3^2)$, $i\in\{1,3,4,5\}$ are provided in appendix \ref{app:contS}. Note that in regions $B_{S,j}$ all the expressions have explicit imaginary parts and all the logarithms and dilogarithms are real valued, while this is not the case for region $A_S$ where the functions are complex valued in general due to the presence of $i\sqrt{-y^2}$ in the arguments.  

As already pointed out, the prescription of section \ref{sec:regions} usually  leads to an overcounting of the regions. This redundancy can sometimes be avoided by using the same set of functions (logarithms and dilogarithms in this case) for multiple subregions, and verifying that the resulting expressions are the same in different subregions. However these functions must satisfy the constraints of eq.~(\ref{eq:fconstraints}), and in complicated cases, as the one under consideration, it is difficult to find a set of functions that satisfy these constraints on multiple subregions. Nevertheless this was possible for the triangle with bubble integral discussed in the next section.

\begin{figure}[t]
\centering
\includegraphics[width=0.75\textwidth]{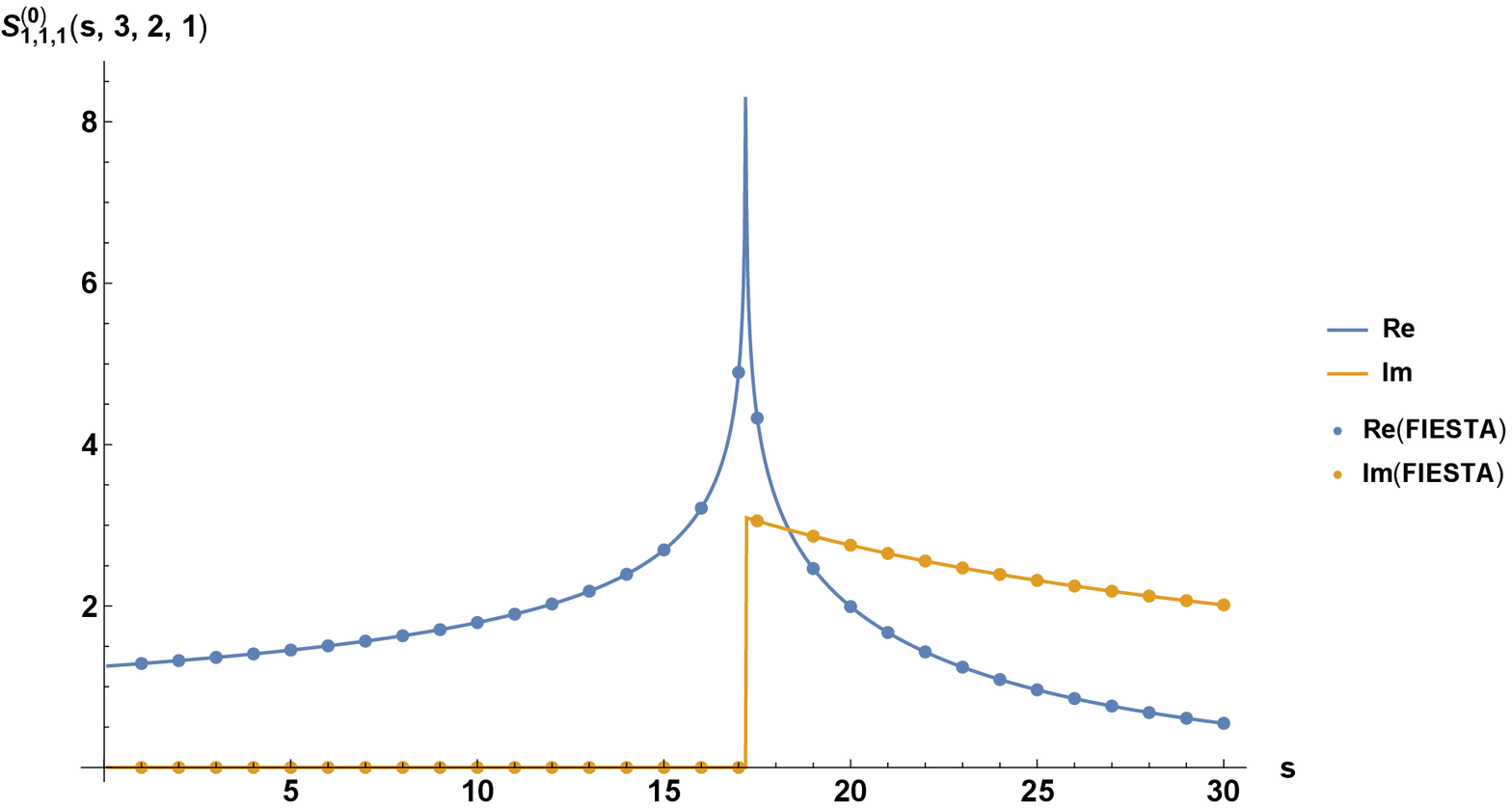}\\
\vspace{2em}
\includegraphics[width=0.75\textwidth]{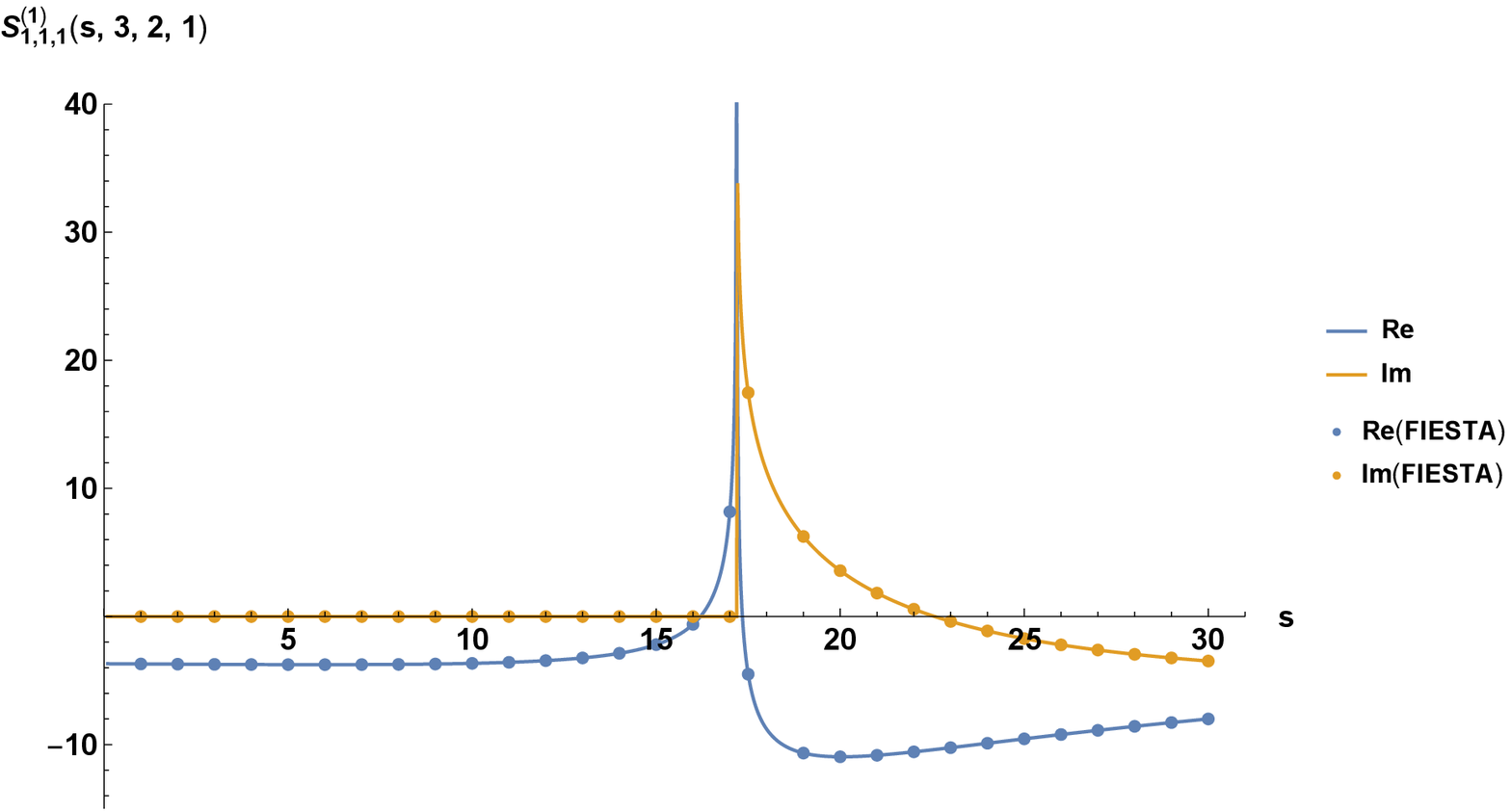}
\caption{Plots of the first two epsilon orders of $S_{1,1,1}(s,m_1^2,m_2^2,m_3^2)$}
\label{fig:sun0sun1}
\end{figure}

Let us stress that the analytic continuation eq.~(\ref{eq:contS}) is suitable for fast and precise numerical evaluations, for example we have:
\begin{equation}
    S^{(1)}_{1,1,1}(20,3,2,1)= -10.9508889661198906+i 3.5786350181321100
\end{equation}
In order to validate our results we performed extensive numerical checks against the computer program FIESTA. The results are summarized in Fig. \ref{fig:sun0sun1}.

\subsection{Triangle with bubble}
\label{sec:TriangleWithBubble}
We consider the below triangle diagram, with a massive bubble insertion, relevant for the two-loop QCD corrections to heavy quark pair production:
\begin{equation}
    \label{eq:TWBDiagrammatic}
    T_{1,2,1,1}(s,m^2) \,\,\,\, = \,\,\,\      (m^2-s)\quad\vcenter{\hbox{\includegraphics[width=0.22\textwidth]{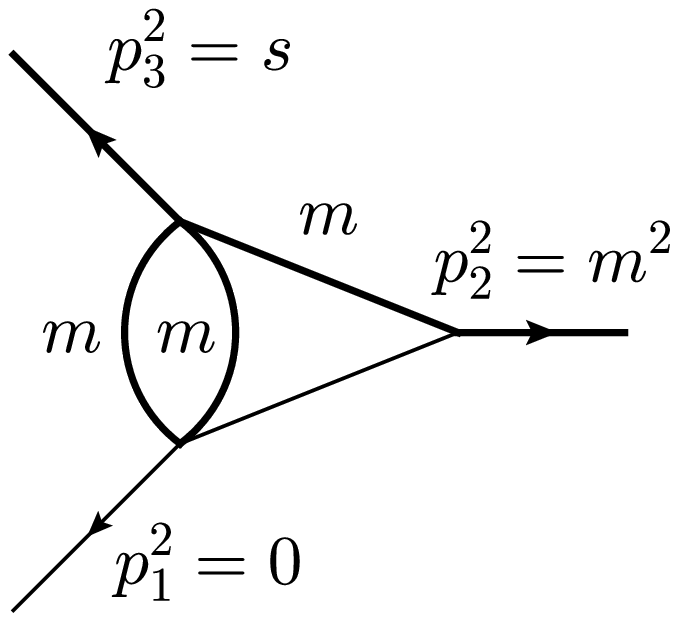}}}\, .
\end{equation}
This diagram has the massive sunrise as a subtopology (as seen from contracting the massless internal propagator.) Hence we expect that the diagram cannot be expressed using multiple polylogarithms. Indeed an explicit calculation confirms this. In order to make the diagram finite in 4 dimensions we have put a dot on one of the massive propagators of the bubble. We note that there is only 1 master integral in the top sector of the topology to which this diagram belongs.
\subsubsection{Direct integration}
We take the following convention for the propagators:
\begin{align}
    D_1 = -(k_1+p_2)^2 + m^2\,, && D_2 = -(k_2-p_3)^2+m^2\,, && D_3 = -(k_1 + k_2 + p_2)^2 + m^2\,, && D_4 = -k_1^2\,.
\end{align}
Then we have in particular:
\begin{align}
    T_{1,2,1,1} = \frac{(m^2-s)}{\left(i\pi^{\frac{d}{2}}\right)^2\Gamma(5-d)} \int \frac{d^dk_1 d^dk_2}{D_1 D_2^2 D_3 D_4}\,.
\end{align}
The Feynman parametrization has the Symanzik polynomials:
\begin{align}
\mathcal{U} &= \alpha _1 \alpha _2+\alpha _3 \alpha _2+\alpha _4 \alpha _2+\alpha _1 \alpha _3+\alpha _3 \alpha _4\,, & \mathcal{F} = \big(&\alpha _2 \alpha _1^2+\alpha _3 \alpha _1^2+\alpha _2^2 \alpha _1+\alpha _3^2 \alpha _1+3 \alpha _2 \alpha _3 \alpha _1+\alpha _2 \alpha _3^2+\nonumber\\&&&\alpha _2^2 \alpha _3+\alpha _2^2 \alpha _4+\alpha _3^2 \alpha _4+2 \alpha _2 \alpha _3 \alpha _4\big) m^2-\alpha _1 \alpha _2 \alpha _3 s\,,
\end{align}
and is given by:
\begin{align}
    T_{\nu_1,\nu_2,\nu_3,\nu_4} = (m^2-s) \int_\Delta d^n\vec{\alpha}\, \left(\prod_{i=1}^n \alpha_i^{\nu_i-1}\right) \mathcal{U}^{\nu_1+\nu_2+\nu_3+\nu_4-\frac{3d}{2}}\mathcal{F}^{-\nu_1-\nu_2-\nu_3-\nu_4+d}\,.
\end{align}
We'll work in the Euclidean region where $s < 0$ and $m^2 > 0$. Expanding the integrand of $T_{1,2,1,1}(s,m^2)$ around $d = 4-2\epsilon$ gives:
\begin{align}
    T_{1,2,1,1}(s,m^2) &= (m^2-s)\sum_{k=0}^\infty \epsilon^k \int_\Delta d^4\vec{\alpha}\, \frac{\alpha_2}{k!} (\mathcal{U}\mathcal{F})^{-1}\log\left(\frac{\mathcal{U}^3}{\mathcal{F}^2}\right)^k\nonumber\\
    &\equiv T_{1,2,1,1}^{(0)}(s,m^2)+\epsilon\, T_{1,2,1,1}^{(1)}(s,m^2)+\mathcal{O}(\epsilon^2)\,.
\end{align}
At order $\epsilon^0$ one obtains:
\begin{align}
    T_{1,2,1,1}^{(0)}(s,m^2) &= (m^2-s) \int_\Delta d^4\vec{\alpha}\, \frac{\alpha_2}{UF}= (m^2-s)\int_0^\infty \frac{d\alpha_1d\alpha_3d\alpha_4}{\left.\left[UF\right]\right|_{\alpha_2=1}}\,,
\end{align}
where we found it convenient to apply Cheng-Wu to set $\alpha_2 = 1$. We can integrate with respect to the massless propagator to obtain:
\begin{align}
    T_{1,2,1,1}^{(0)}(s,m^2) &= (m^2-s)\int_0^\infty d\alpha_1d\alpha_3\, \frac{\log \left(\frac{m^2 (\alpha_1+\alpha_3+1) (\alpha_1 \alpha_3+\alpha_1+\alpha_3)-\alpha_1 \alpha_3 s}{(\alpha_3+1) m^2 (\alpha_1 \alpha_3+\alpha_1+\alpha_3)}\right)}{\alpha_1 (\alpha_3+1) \left(m^2 (\alpha_1 \alpha_3+\alpha_1+\alpha_3)-\alpha_3 s\right)}\,.
\end{align}
The polynomial $\left(m^2 (\alpha_1+\alpha_3+1) (\alpha_1 \alpha_3+\alpha_1+\alpha_3)-\alpha_1 \alpha_3 s\right)$ does not factor linearly in either of the remaining integration parameters without introducing a square root containing the other integration variable. Its roots are special cases of those encountered for the sunrise, namely $R^{(\alpha_1)}_{\pm}(s,m^2,m^2,m^2)$, where $R^{(\alpha_1)}_{\pm}(s,m_1^2,m_2^2,m_3^2)$ corresponds to eq. (\ref{eq:sunrisezeros}) with $\alpha_3$ replaced by $\alpha_1$. Performing the integration on $\alpha_3$ yields us with:
\begin{align}
    \label{eq:twbint1}
    T_{1,2,1,1}^{(0)}(s,m^2) =& \int _0^{\infty }\frac{1}{\alpha _1} \left(-G_{\alpha _1+1,b(1)}+G_{\frac{1}{R_-+1},b(1)}+G_{\frac{1}{R_++1},b(1)}-G_{b(1)} G_{-\frac{1}{\alpha _1}}\right)d\alpha _1
\end{align}
where we used the shorthand notation $R_{\pm} = R^{(\alpha_1)}_{\pm}(s,m^2,m^2,m^2)$, and have introduced the term:
\begin{align}
b(1) = \frac{(\alpha_1+1)m^2-s}{m^2-s}\, .
\end{align}
Lastly, at order $\epsilon^1$ we can integrate along the same path and we obtain the following result:
\begin{align}
&T_{1,2,1,1}^{(1)}(s,m^2) =\int_0^\infty d\alpha_1\,\frac{1}{\alpha_1}\bigg(2G_{b(1)}G_{b(2)}G_{b(3)}-3G_{b(1)}G_{b(2)}G_{\frac{1}{1-\alpha_1}}-2G_{b(1)}G_{b(3)}G_{\frac{1}{1-\alpha_1}}+2G_{b(2)}G_{0,b(1)}\nonumber\\&-2G_{\frac{1}{1-\alpha_1}}G_{0,b(1)}-G_{b(1)}G_{0,b(2)}+G_{b(1)}G_{0,\frac{1}{1-\alpha_1}}+G_{b(2)}G_{b(1),b(1)}-G_{\frac{1}{1-\alpha_1}}G_{b(1),b(1)}\nonumber\\&-2G_{b(1)}G_{b(2),b(2)}+2G_{b(1)}G_{b(2),\frac{1}{1-\alpha_1}}-2G_{b(3)}G_{\frac{1}{R_-+1},b(1)}+3G_{\frac{1}{1-\alpha_1}}G_{\frac{1}{R_-+1},b(1)}-2G_{b(3)}G_{\frac{1}{R_++1},b(1)}\nonumber\\&+3G_{\frac{1}{1-\alpha_1}}G_{\frac{1}{R_++1},b(1)}+3G_{b(1)}G_{\frac{1}{1-\alpha_1},b(2)}+3G_{b(1)}G_{\frac{1}{1-\alpha_1},\frac{1}{1-\alpha_1}}+2G_{b(3)}G_{\alpha_1+1,b(1)}-3G_{\frac{1}{1-\alpha_1}}G_{\alpha_1+1,b(1)}\nonumber\\&-G_{0,\frac{1}{R_-+1},b(1)}-G_{0,\frac{1}{R_++1},b(1)}+G_{0,\alpha_1+1,b(1)}-2G_{\frac{1}{R_-+1},0,b(1)}-G_{\frac{1}{R_-+1},b(1),b(1)}+2G_{\frac{1}{R_-+1},\frac{1}{R_-+1},b(1)}\nonumber\\&+2G_{\frac{1}{R_-+1},\frac{1}{R_++1},b(1)}-2G_{\frac{1}{R_-+1},\alpha_1+1,b(1)}-2G_{\frac{1}{R_++1},0,b(1)}-G_{\frac{1}{R_++1},b(1),b(1)}+2G_{\frac{1}{R_++1},\frac{1}{R_-+1},b(1)}\nonumber\\&+2G_{\frac{1}{R_++1},\frac{1}{R_++1},b(1)}-2G_{\frac{1}{R_++1},\alpha_1+1,b(1)}+2G_{\alpha_1+1,0,b(1)}+G_{\alpha_1+1,b(1),b(1)}-3G_{\alpha_1+1,\frac{1}{R_-+1},b(1)}\nonumber\\&-3G_{\alpha_1+1,\frac{1}{R_++1},b(1)}+3G_{\alpha_1+1,\alpha_1+1,b(1)}\bigg)\,,
\end{align}
where we introduced the additional terms:
\begin{equation}
    b(2) = \frac{1}{1 - \alpha_1(1 + \alpha_1)}\,,\quad b(3) = \frac{1}{1 - m^2 \alpha_1(1 + \alpha_1 )}\,.
\end{equation}
The higher orders in $\epsilon$ may be obtained from the same integration sequence.

\subsubsection{Differential equations for the inner polylogarithmic part}
We combine 2 massive propagators and define:
\begin{align}
    T^{\text{IPP}}_{a_1+a_2,a_3,a_4} \equiv \frac{m^2(1+t) \Gamma(a_1+a_2)\Gamma(a_3)\Gamma(a_4)}{\left(i \pi^{\frac{d}{2}}\right)^2 \Gamma(a-d)}\int \frac{d^dk_1 d^dk_2}{(x D_1+D_2)^{a_1+a_2} D_3^{a_3} D_4^{a_4}}\,,
\end{align}
where we let $t = -s/m^2$ be a scale with zero mass dimension. That way:
\begin{align}
    T_{a_1,a_2,a_3,a_4} = \int_0^\infty dx\, x^{a_1-1} T^{\text{IPP}}_{a_1+a_2,a_3,a_4}\,.
\end{align}
We then have in particular:
\begin{align}
    T_{1,2,1,1} = \int_0^\infty T^{\text{IPP}}_{3,1,1}\,dx\,, && T_{1,1,2,1} = \int_0^\infty  T^{\text{IPP}}_{2,2,1}\,dx\,,
\end{align}
and we note that $T_{1,2,1,1} = T_{1,1,2,1}$ by the symmetry of the diagram. Nonetheless, $T^{\text{IPP}}_{3,1,1}$ = $T^{\text{IPP}}_{2,2,1}$ are different polylogarithmic expressions, as they represent different integration sequences of the same integral. We adopt the notation:
\begin{align}
    y^2 = 1+x \left(2+2 t+3 x+t (6+t) x+2 (1+t) x^2+x^3\right) = m^{-4}P_S^{(x)}(-m^2 t,m^2,m^2,m^2)\,,
\end{align}
where $P_S^{(x)}$ corresponds to eq. (\ref{eq:SunriseEllipticCurve0Inf}) with $\alpha_3$ replaced by $x$. A canonical basis for the IPP is given by:
\begin{align}
    \vec{B} = \left(\begin{array}{ccccccccccc}
         \text{c}_ {2,2,1}^1 T_ {2,2,1}^{\text{IPP}} &+& \text{c}_ {3,1,1}^1 T_{3,1,1}^{\text{IPP}} &&  &&  &&  &&  \\
  && \text{c}_ {3,1,1}^2 T_ {3,1,1}^{\text{IPP}} &&  &&  &&  &&  \\
  &&  && \text{c}_ {4,0,1}^3 T_ {4,0,1}^{\text{IPP}} &&  &&  &&  \\
  &&  &&  && \text{c}_ {2,1,0}^4 T_ {2,1,0}^{\text{IPP}} &+& \text{c}_{3,1,0}^4 T_ {3,1,0}^{\text{IPP}} &+& \text{c}_ {4,0,0}^4 T_ {4,0,0}^{\text{IPP}} \\
  &&  &&  && \text{c}_ {2,1,0}^5 T_ {2,1,0}^{\text{IPP}} &+& \text{c}_{3,1,0}^5 T_ {3,1,0}^{\text{IPP}} &+& \text{c}_ {4,0,0}^5 T_ {4,0,0}^{\text{IPP}} \\
  &&  &&  &&  &&  && \text{c}_ {4,0,0}^6 T_ {4,0,0}^{\text{IPP}}
    \end{array}\right)\,,
\end{align}
where the coefficients are:
\begin{align}
\def\arraystretch{1.5}
\begin{array}{ll}
 \text{c}_{2,2,1}^1=(m^2)^{1+2 \epsilon } x (1+t+x) \epsilon ^2 \,,& \text{c}_{3,1,1}^1=2 (m^2)^{1+2 \epsilon
   } x^2 \epsilon ^2 \,,\\
 \text{c}_{3,1,1}^2=(m^2)^{1+2 \epsilon } (1+t) x \epsilon ^2 \,,& \text{c}_{4,0,1}^3=(m^2)^{1+2 \epsilon }
   x^2 \epsilon  \,,\\
 \text{c}_{2,1,0}^4=\frac{(m^2)^{-1+2 \epsilon } (1+x)^2 (1+x (1+t+x)) \epsilon ^2 (-2+3 \epsilon )}{2 y} \,,&
   \text{c}_{3,1,0}^4=-\frac{(m^2)^{2 \epsilon } (1+x) \epsilon}{y}  (t^2 x^2 \epsilon +(1+x+x^2)^2
   \epsilon \\\text{c}_{4,0,0}^4=\frac{3 (m^2)^{2 \epsilon } x \epsilon}{y (-1+2 \epsilon )}   (2 x (1+x-t x+x^2)+ \epsilon&\quad+2 t x (\epsilon +x (2+(-5+x) \epsilon ))) \,,\\
    \quad+x
   (-1+t+6 t x+(-1+t) x^2+x^3) \epsilon ) \,,& \text{c}_{2,1,0}^5=\frac{1}{2}
   (m^2)^{-1+2 \epsilon } (1+x)^2 \epsilon ^2 (-2+3 \epsilon ) \,,\\
 \text{c}_{3,1,0}^5=-(m^2)^{2 \epsilon } (1+x) (1+x (1+t+x)) \epsilon ^2 \,,& \text{c}_{4,0,0}^5=\frac{3
   (m^2)^{2 \epsilon } x (1+x)^2 \epsilon ^2}{-1+2 \epsilon } \,.\\
\end{array}
\end{align}
The resulting canonical form differential equation is then given by:
\begin{align}
    d \vec{B} = \epsilon\,d\mathbf{A} \vec{B}\,,
\end{align}
where
\begin{align}
 \mathbf{A} = \left(
\begin{array}{cccccc}
 -l_2-2 l_4+l_6 & 2 l_2+2 l_6 & 6 l_2+6 l_4 & \frac{l_9}{4} & -\frac{l_2}{2}-\frac{l_4}{2}-\frac{l_5}{2} &
   \frac{l_2}{2}+\frac{l_4}{2}+\frac{l_5}{2} \\
 l_2+l_6 & -2 l_4+2 l_6 & -6 l_2 & -\frac{l_7}{4} & \frac{l_1}{4}+\frac{l_2}{2} & -\frac{l_1}{4}-\frac{l_2}{2} \\
 0 & 0 & 2 l_2 & 0 & 0 & \frac{l_5}{3} \\
 0 & 0 & 0 & \frac{3 l_1}{2}+3 l_2-4 l_3+2 l_5 & -\frac{3 l_7}{2} & \frac{3 l_7}{2}+2 l_8 \\
 0 & 0 & 0 & \frac{l_7}{2} & -\frac{l_1}{2}-l_2+2 l_5 & \frac{l_1}{2} \\
 0 & 0 & 0 & 0 & 0 & l_2-2 l_5 \\
\end{array}
\right)\,,
\end{align}
and where the letters are:
\begin{align}
 &l_1=\log\left(t\right) \,,&&
 l_2=\log\left(x\right) \,,&&
 l_3=\log\left(y\right) \,,&&
 l_4=\log\left(t+1\right) \,,\nonumber\\
 &l_5=\log\left(x+1\right) \,,&&
 l_6=\log\left(t+x+1\right)\,, &&
 l_7=\log\left(\frac{x^2+t x+x-y+1}{x^2+t x+x+y+1}\right)\,,\nonumber\\
 &l_8=\log\left(\frac{(t+x+2) x+x+y+1}{(t+x+2) x+x-y+1}\right)\,,\span\omit\span\omit\span\omit\span\omit\nonumber\\
 &l_9=\log\left(\frac{x^4+2 t x^3+2 x^3+t^2 x^2+4 t x^2+x^2+\left(x^2+t x+x-1\right) y+1}{x^4+2 t x^3+2 x^3+t^2 x^2+4 t x^2+x^2-\left(x^2+t x+x-1\right) y+1}\right)\,. \span\omit\span\omit\span\omit\span\omit
\end{align}
We show next how to solve the resulting differential equation in terms of $\text{E}_4$-functions. We perform the change of variables $x = x'/(x'-1)$, and let:
\begin{align}
    y' &= \frac{1+2 (-1+t) x'+\left(3+t^2\right) \left(x'\right)^2-2 (1+t)^2 \left(x'\right)^3+(1+t)^2
   \left(x'\right)^4}{(1+t)^2} \nonumber\\
   &= (x'-a_1')(x'-a_2')(x'-a_3')(x'-a_4')\,,
\end{align}
where we picked the following convention for the roots:
\begin{align}
\begin{array}{ll}
 a_1'=\frac{1}{2} \left(1-\sqrt{\frac{4 \left(t+2 \sqrt{-t}-1\right)}{t^2+2 t+1}+1}\right)\,, & a_2'=\frac{1}{2} \left(1-\sqrt{\frac{4 \left(t-2 \sqrt{-t}-1\right)}{t^2+2 t+1}+1}\right)\,, \\
 a_3'=\frac{1}{2} \left(1+\sqrt{\frac{4 \left(t-2 \sqrt{-t}-1\right)}{t^2+2 t+1}+1}\right)\,, & a_4'=\frac{1}{2} \left(1+\sqrt{\frac{4 \left(t+2 \sqrt{-t}-1\right)}{t^2+2 t+1}+1}\right)\,, \\
\end{array}
\end{align}
which satisfies $0<a_1'<a_2'<a_3'<a_4'<1$, in the physical region $t < -9$. The differential equation matrix is given by:
\begin{align}
    \frac{\partial\mathbf{A}}{\partial x'} = \left(
\begin{array}{cccccc}
 \psi _{1,1} & \psi _{1,2} & \psi _{1,3} & \psi _{1,4} & \psi _{1,5} & \psi _{1,6} \\
 \psi _{2,1} & \psi _{2,2} & \psi _{2,3} & \psi _{2,4} & \psi _{2,5} & \psi _{2,6} \\
 0 & 0 & \psi _{3,3} & 0 & 0 & \psi _{3,6} \\
 0 & 0 & 0 & \psi _{4,4} & \psi _{4,5} & \psi _{4,6} \\
 0 & 0 & 0 & \psi _{5,4} & \psi _{5,5} & 0 \\
 0 & 0 & 0 & 0 & 0 & \psi _{6,6} \\
\end{array}
\right)\,,
\end{align}
where the non-zero entries are:
{\small
\def\arraystretch{1.5}
\begin{longtable}{LL}
 \psi_{1,1}=\psi_1\left(1+\frac{1}{t},x'\right)-\psi_1\left(0,x'\right) \,,& \psi_{2,6}=\frac{1}{2} \psi_1\left(1,x'\right)-\frac{1}{2} \psi_1\left(0,x'\right) \,,\\
 \psi_{1,2}=2 \psi_1\left(0,x'\right)-4 \psi_1\left(1,x'\right)+2 \psi_1\left(1+\frac{1}{t},x'\right) \,,& \psi_{3,3}=2 \psi_1\left(0,x'\right)-2 \psi_1\left(1,x'\right) \,,\\
 \psi_{1,3}=6 \psi_1\left(0,x'\right)-6 \psi_1\left(1,x'\right) \,,& \psi_{3,6}=-\frac{1}{3} \psi_1\left(1,x'\right) \,,\\
 \psi_{1,4}=-\psi_{-1}\left(1,x'\right)-\frac{1}{2} \psi_{-1}\left(\infty ,x'\right)+\frac{(t-1) \psi_0\left(0,x'\right)}{2 c_4 (t+1)} & \psi_{4,4}=3 \psi_1\left(0,x'\right)+3 \psi_1\left(1,x'\right)-2 \psi_1\left(a_1',x'\right) \\
 \quad+\frac{1}{2} \left(\psi_{-1}\left(0,x'\right)+\psi_1\left(0,x'\right)\right) \,,& \quad-2 \psi_1\left(a_2',x'\right)-2 \psi_1\left(a_3',x'\right)-2 \psi_1\left(a_4',x'\right) \,,\\
 \psi_{1,5}=\psi_1\left(1,x'\right)-\frac{1}{2} \psi_1\left(0,x'\right) \,,& \psi_{4,5}=-3 \psi_{-1}\left(1,x'\right)-3 \left(\psi_{-1}\left(0,x'\right)+\psi_1\left(0,x'\right)\right) \,,\\
 \psi_{1,6}=\frac{1}{2} \psi_1\left(0,x'\right)-\psi_1\left(1,x'\right) \,,& \psi_{4,6}=\psi_{-1}\left(0,x'\right)+\psi_{-1}\left(1,x'\right)-4 \psi_{-1}\left(\infty ,x'\right) \\
 \psi_{2,1}=\psi_1\left(0,x'\right)-2 \psi_1\left(1,x'\right)+\psi_1\left(1+\frac{1}{t},x'\right) \,,& \quad+\frac{2 \psi_0\left(0,x'\right)}{c_4}+\psi_1\left(0,x'\right) \,,\\
 \psi_{2,2}=2 \psi_1\left(1+\frac{1}{t},x'\right)-2 \psi_1\left(1,x'\right) \,,& \psi_{5,4}=\psi_{-1}\left(0,x'\right)+\psi_{-1}\left(1,x'\right)+\psi_1\left(0,x'\right) \,,\\
 \psi_{2,3}=6 \psi_1\left(1,x'\right)-6 \psi_1\left(0,x'\right) \,,& \psi_{5,5}=-\psi_1\left(0,x'\right)-\psi_1\left(1,x'\right) \,,\\
 \psi_{2,4}=\frac{1}{2} \left(-\psi_{-1}\left(0,x'\right)-\psi_1\left(0,x'\right)\right)-\frac{1}{2} \psi_{-1}\left(1,x'\right) \,,& \psi_{6,6}=\psi_1\left(0,x'\right)+\psi_1\left(1,x'\right) \,,\\
 \psi_{2,5}=\frac{1}{2} \psi_1\left(0,x'\right)-\frac{1}{2} \psi_1\left(1,x'\right) \,.& \\
\end{longtable}}
\noindent We let:
\begin{align}
    \hat{T}^{\text{IPP}}_{a_1+a_2,a_3,a_4} \equiv \frac{m^2(1+t)}{\left(i \pi^{\frac{d}{2}}\right)^2\Gamma(5-d)} \int \frac{d^dk_1 d^dk_2}{(x' D_1+(1-x')D_2)^{a_1+a_2} D_3^{a_3} D_4^{a_4}}\,.
\end{align}
So that:
\begin{align}
    \label{eq:TWBInTermsOfB2}
    T_{1,2,1,1} = 2 \int_0^1 dx'\,(1-x') \hat{T}_{3,1,1}^{\text{IPP}} = -\frac{2 \left(m^2\right)^{-2 \epsilon }}{\epsilon ^2} \int_0^1 \frac{B_2}{\left(-1+x'\right) x'} \, dx'\,.
\end{align}
Furthermore, we may write the full solution of $\vec{B}$ as a path-ordered exponential:
\begin{align}
\label{eq:BTWBPathOrderedExponential}
\vec{B}(x',t,m^2) = \mathbb{P} \exp\left(\epsilon \int_{x'_0}^{x'}\frac{\partial\mathbf{A}}{\partial x'}\,dx'  \right) \vec{B}(x'_0,t,m^2)\,,
\end{align}
and combining this with eq. (\ref{eq:TWBInTermsOfB2}) yields:
\begin{align}
    \label{eq:TWBPathOrderedExponential}
    T_{1,1,2,1} = -\frac{2 \left(m^2\right)^{-2 \epsilon }}{\epsilon ^2} \int _0^1 dx'\,\left(\psi _1\left(1,x'\right)-\psi _1\left(0,x'\right)\right) \mathbb{P} \exp\left(\epsilon \int_{x'_0}^{x'}\frac{\partial\mathbf{A}}{\partial x'}\,dx'  \right) \vec{B}(x'_0,t,m^2)\,.
\end{align}
We are interested in finding the boundary term:
\begin{align}
    \text{reglim}_{x'_0\rightarrow 0}\,\vec{B}(x'_0,t,m^2)\,,
\end{align}
so that we may express eq. (\ref{eq:TWBPathOrderedExponential}) in terms of $\text{E}_4$-functions. One may verify that the top sector integrals $B_1$ and $B_2$ contain the terms $\hat{T}_{3,1,1}^{\text{IPP}}$ and $\hat{T}_{2,2,1}^{\text{IPP}}$ with prefactors that are proportional to an overall factor $x'$. Furthermore, we have the relations:
\begin{align}
    \label{eq:T1211InTermsOfIPP01}
    T_{1,2,1,1} = 2 \int_0^1 (1-x') \hat{T}^{\text{IPP}}_{3,1,1}\,dx'\,, && T_{1,1,2,1} = \int_0^1 \hat{T}^{\text{IPP}}_{2,2,1}\,dx'\,,
\end{align}
and since $T_{1,2,1,1} = T_{1,1,2,1}$ is a finite integral, the integrands in eq. (\ref{eq:T1211InTermsOfIPP01}) should have integrable singularities at the point 0. Therefore we find that:
\begin{align}
    \lim_{x' \rightarrow 0} B_1 = \lim_{x' \rightarrow 0} B_2 = 0\,,
\end{align}
which we also verified numerically. One may compute $B_3$ for arbitrary $x'$ by integrating the Feynman parametrization and one may observe that the limit $x'\rightarrow 0$ vanishes as well. The canonical basis integrals $B_4, B_5$ and $B_6$ belong to the sunrise subsector and their regularized limit may be obtained in the same manner as was done for the unequal mass sunrise topology. The results are:
\begin{align}
    \left(\begin{array}{c}B_4 \\ B_5 \\ B_6 \end{array}\right) = \left(
\begin{array}{c}
 \frac{\Gamma (1-\epsilon )^2 \Gamma (\epsilon )^2}{\Gamma (-\epsilon )^2 \Gamma (2 \epsilon +1)}-1 \\
 \frac{\Gamma (1-\epsilon )^2 \Gamma (\epsilon )^2}{\Gamma (-\epsilon )^2 \Gamma (2 \epsilon +1)}-\frac{1}{2} \\
 \frac{1}{2} \\
\end{array}
\right)\,.
\end{align}
We now have almost all the ingredients to write $T_{1,1,2,1}$ in terms of $\text{E}_4$-functions but there is a complication. Looking at eq. (\ref{eq:TWBPathOrderedExponential}), one notices the appearance of the kernel $\psi_1(1,x')$ in the last entry. This kernel yields $\text{E}_4$-functions that are individually divergent. First we consider the ``naive'' solution at finite order, which still contains divergent pieces:
\begin{align}
    &T_{1,1,2,1} = \frac{\left(m^2\right)^{-2 \epsilon }}{c_4}\bigg(
    c_4\text{E}_4\left(\begin{xsmallmatrix}{.15em}1&-1&-1\\0&\hfill 0&\hfill 0\\\end{xsmallmatrix},1\right)+c_4\text{E}_4\left(\begin{xsmallmatrix}{.15em}1&-1&-1\\0&\hfill 0&\hfill 1\\\end{xsmallmatrix},1\right)+2c_4\text{E}_4\left(\begin{xsmallmatrix}{.15em}1&-1&-1\\0&\hfill 0&\hfill \infty\\\end{xsmallmatrix},1\right)+c_4\text{E}_4\left(\begin{xsmallmatrix}{.15em}1&-1&-1\\0&\hfill 1&\hfill 0\\\end{xsmallmatrix},1\right)\nonumber\\&+c_4\text{E}_4\left(\begin{xsmallmatrix}{.15em}1&-1&-1\\0&\hfill 1&\hfill 1\\\end{xsmallmatrix},1\right)+2c_4\text{E}_4\left(\begin{xsmallmatrix}{.15em}1&-1&-1\\0&\hfill 1&\hfill \infty\\\end{xsmallmatrix},1\right)-c_4\text{E}_4\left(\begin{xsmallmatrix}{.15em}1&-1&-1\\1&\hfill 0&\hfill 0\\\end{xsmallmatrix},1\right)-c_4\text{E}_4\left(\begin{xsmallmatrix}{.15em}1&-1&-1\\1&\hfill 0&\hfill 1\\\end{xsmallmatrix},1\right)-2c_4\text{E}_4\left(\begin{xsmallmatrix}{.15em}1&-1&-1\\1&\hfill 0&\hfill \infty\\\end{xsmallmatrix},1\right)\nonumber\\&-c_4\text{E}_4\left(\begin{xsmallmatrix}{.15em}1&-1&-1\\1&\hfill 1&\hfill 0\\\end{xsmallmatrix},1\right)-c_4\text{E}_4\left(\begin{xsmallmatrix}{.15em}1&-1&-1\\1&\hfill 1&\hfill 1\\\end{xsmallmatrix},1\right)-2c_4\text{E}_4\left(\begin{xsmallmatrix}{.15em}1&-1&-1\\1&\hfill 1&\hfill \infty\\\end{xsmallmatrix},1\right)-\text{E}_4\left(\begin{xsmallmatrix}{.15em}1&-1&0\\0&\hfill 0&0\\\end{xsmallmatrix},1\right)-\text{E}_4\left(\begin{xsmallmatrix}{.15em}1&-1&0\\0&\hfill 1&0\\\end{xsmallmatrix},1\right)\nonumber\\&+\text{E}_4\left(\begin{xsmallmatrix}{.15em}1&-1&0\\1&\hfill 0&0\\\end{xsmallmatrix},1\right)+\text{E}_4\left(\begin{xsmallmatrix}{.15em}1&-1&0\\1&\hfill 1&0\\\end{xsmallmatrix},1\right)-c_4\text{E}_4\left(\begin{xsmallmatrix}{.15em}1&1&-1\\0&0&\hfill 0\\\end{xsmallmatrix},1\right)-c_4\text{E}_4\left(\begin{xsmallmatrix}{.15em}1&1&-1\\0&0&\hfill 1\\\end{xsmallmatrix},1\right)+2c_4\text{E}_4\left(\begin{xsmallmatrix}{.15em}1&1&-1\\0&0&\hfill \infty\\\end{xsmallmatrix},1\right)\nonumber\\&+c_4\text{E}_4\left(\begin{xsmallmatrix}{.15em}1&1&-1\\0&1&\hfill 0\\\end{xsmallmatrix},1\right)+c_4\text{E}_4\left(\begin{xsmallmatrix}{.15em}1&1&-1\\0&1&\hfill 1\\\end{xsmallmatrix},1\right)-2c_4\text{E}_4\left(\begin{xsmallmatrix}{.15em}1&1&-1\\1&0&\hfill \infty\\\end{xsmallmatrix},1\right)-\text{E}_4\left(\begin{xsmallmatrix}{.15em}1&1&0\\0&0&0\\\end{xsmallmatrix},1\right)+\text{E}_4\left(\begin{xsmallmatrix}{.15em}1&1&0\\1&0&0\\\end{xsmallmatrix},1\right)-c_4\text{E}_4\left(\begin{xsmallmatrix}{.15em}1&1&1\\0&0&1\\\end{xsmallmatrix},1\right)\nonumber\\&+c_4\text{E}_4\left(\begin{xsmallmatrix}{.15em}1&1&1\\1&1&1\\\end{xsmallmatrix},1\right)
    \bigg)+\mathcal{O}(\epsilon)\,.
\end{align}

We would like to deal with the divergent terms by using the shuffle product to remove every occurrence of the kernel $\psi_1(1,x')$ in the first entry. A complication is that the kernel $\psi_{-1}(1,x')$ may then appear in the first entry, which also diverges at 1. We deal with problem in a similar manner to \cite{Broedel:2017siw}, where such issues arise in the analysis of the second master integral of the equal mass sunrise. First, we define a new kernel:
\begin{align}
    \psi_{-\tilde{1}}(1,x) = \frac{y(1)}{(x-1)y} - \frac{1}{(x-1)}\,,
\end{align}
which is a regulated version of $\psi_{-1}(1,x')$. We then express our $\text{E}_4$-functions in terms of this new kernel. After doing so one may extract out the divergent pieces from each $\text{E}_4$-function by shuffle regularization. The only remaining divergent terms are:
\begin{align}
    \text{E}_4\left(\begin{xsmallmatrix}{.05em}1\\1\\\end{xsmallmatrix},1\right) \,,&& \text{E}_4\left(\begin{xsmallmatrix}{.05em}1&1\\1&1\\\end{xsmallmatrix},1\right)\,,
\end{align}
and their prefactors should vanish as we know $T_{1,1,2,1}$ is finite. One finds for example the contribution:
\begin{dmath}
    \text{E}_4\left(\begin{xsmallmatrix}{.05em}1&1\\1&1\\\end{xsmallmatrix},1\right)\bigg(-c_4\text{E}_4\left(\begin{xsmallmatrix}{.15em}-1\\\hfill 0\\\end{xsmallmatrix},1\right)-2c_4\text{E}_4\left(\begin{xsmallmatrix}{.15em}-1\\\hfill \infty\\\end{xsmallmatrix},1\right)+\text{E}_4\left(\begin{xsmallmatrix}{.15em}0\\0\\\end{xsmallmatrix},1\right)-c_4\text{E}_4\left(\begin{xsmallmatrix}{.15em}-\tilde{1}\\\hfill 1\\\end{xsmallmatrix},1\right)\bigg)\,,
\end{dmath}
and it may be numerically verified up to high precision that the combination of $\text{E}_4$-functions multiplying $\text{E}_4\left(\begin{xsmallmatrix}{.05em}1&1\\1&1\\\end{xsmallmatrix},1\right)$ evaluates to zero. We decide to restore the kernel $\psi_{-1}(1,x)$ in all entries but the first, and we obtain the following representation in terms of $\text{E}_4$-functions that are individually finite:
\begin{align}
&T_{1,1,2,1} = \frac{\left(m^2\right)^{-2 \epsilon }}{c_4}\bigg(
c_4\text{E}_4\left(\begin{xsmallmatrix}{.15em}-1&-1&1\\\hfill 0&\hfill 0&1\\\end{xsmallmatrix},1\right)+c_4\text{E}_4\left(\begin{xsmallmatrix}{.15em}-1&-1&1\\\hfill 0&\hfill 1&1\\\end{xsmallmatrix},1\right)+2c_4\text{E}_4\left(\begin{xsmallmatrix}{.15em}-1&-1&1\\\hfill 0&\hfill \infty&1\\\end{xsmallmatrix},1\right)-\text{E}_4\left(\begin{xsmallmatrix}{.15em}-1&0&1\\\hfill 0&0&1\\\end{xsmallmatrix},1\right)\nonumber\\&+c_4\text{E}_4\left(\begin{xsmallmatrix}{.15em}-1&1&-1\\\hfill 0&1&\hfill 0\\\end{xsmallmatrix},1\right)+c_4\text{E}_4\left(\begin{xsmallmatrix}{.15em}-1&1&-1\\\hfill 0&1&\hfill 1\\\end{xsmallmatrix},1\right)+2c_4\text{E}_4\left(\begin{xsmallmatrix}{.15em}-1&1&-1\\\hfill 0&1&\hfill \infty\\\end{xsmallmatrix},1\right)-\text{E}_4\left(\begin{xsmallmatrix}{.15em}-1&1&0\\\hfill 0&1&0\\\end{xsmallmatrix},1\right)-c_4\text{E}_4\left(\begin{xsmallmatrix}{.15em}-1&1&1\\\hfill 0&1&1\\\end{xsmallmatrix},1\right)\nonumber\\&-2c_4\text{E}_4\left(\begin{xsmallmatrix}{.15em}-1&1&1\\\hfill \infty&1&1\\\end{xsmallmatrix},1\right)+\text{E}_4\left(\begin{xsmallmatrix}{.15em}0&1&1\\0&1&1\\\end{xsmallmatrix},1\right)+c_4\text{E}_4\left(\begin{xsmallmatrix}{.15em}1&-1&-1\\0&\hfill 0&\hfill 0\\\end{xsmallmatrix},1\right)+c_4\text{E}_4\left(\begin{xsmallmatrix}{.15em}1&-1&-1\\0&\hfill 0&\hfill 1\\\end{xsmallmatrix},1\right)+2c_4\text{E}_4\left(\begin{xsmallmatrix}{.15em}1&-1&-1\\0&\hfill 0&\hfill \infty\\\end{xsmallmatrix},1\right)\nonumber\\&+c_4\text{E}_4\left(\begin{xsmallmatrix}{.15em}1&-1&-1\\0&\hfill 1&\hfill 0\\\end{xsmallmatrix},1\right)+c_4\text{E}_4\left(\begin{xsmallmatrix}{.15em}1&-1&-1\\0&\hfill 1&\hfill 1\\\end{xsmallmatrix},1\right)+2c_4\text{E}_4\left(\begin{xsmallmatrix}{.15em}1&-1&-1\\0&\hfill 1&\hfill \infty\\\end{xsmallmatrix},1\right)-\text{E}_4\left(\begin{xsmallmatrix}{.15em}1&-1&0\\0&\hfill 0&0\\\end{xsmallmatrix},1\right)-\text{E}_4\left(\begin{xsmallmatrix}{.15em}1&-1&0\\0&\hfill 1&0\\\end{xsmallmatrix},1\right)\nonumber\\&+2c_4\text{E}_4\left(\begin{xsmallmatrix}{.15em}1&-1&1\\0&\hfill \infty&1\\\end{xsmallmatrix},1\right)-\text{E}_4\left(\begin{xsmallmatrix}{.15em}1&0&1\\0&0&1\\\end{xsmallmatrix},1\right)-c_4\text{E}_4\left(\begin{xsmallmatrix}{.15em}1&1&-1\\0&0&\hfill 0\\\end{xsmallmatrix},1\right)-c_4\text{E}_4\left(\begin{xsmallmatrix}{.15em}1&1&-1\\0&0&\hfill 1\\\end{xsmallmatrix},1\right)+2c_4\text{E}_4\left(\begin{xsmallmatrix}{.15em}1&1&-1\\0&0&\hfill \infty\\\end{xsmallmatrix},1\right)\nonumber\\&+c_4\text{E}_4\left(\begin{xsmallmatrix}{.15em}1&1&-1\\0&1&\hfill 0\\\end{xsmallmatrix},1\right)+c_4\text{E}_4\left(\begin{xsmallmatrix}{.15em}1&1&-1\\0&1&\hfill 1\\\end{xsmallmatrix},1\right)+2c_4\text{E}_4\left(\begin{xsmallmatrix}{.15em}1&1&-1\\0&1&\hfill \infty\\\end{xsmallmatrix},1\right)-\text{E}_4\left(\begin{xsmallmatrix}{.15em}1&1&0\\0&0&0\\\end{xsmallmatrix},1\right)-\text{E}_4\left(\begin{xsmallmatrix}{.15em}1&1&0\\0&1&0\\\end{xsmallmatrix},1\right)-c_4\text{E}_4\left(\begin{xsmallmatrix}{.15em}1&1&1\\0&0&1\\\end{xsmallmatrix},1\right)\nonumber\\&+c_4\text{E}_4\left(\begin{xsmallmatrix}{.15em}-\tilde{1}&-1&1\\\hfill 1&\hfill 0&1\\\end{xsmallmatrix},1\right)+c_4\text{E}_4\left(\begin{xsmallmatrix}{.15em}-\tilde{1}&-1&1\\\hfill 1&\hfill 1&1\\\end{xsmallmatrix},1\right)+2c_4\text{E}_4\left(\begin{xsmallmatrix}{.15em}-\tilde{1}&-1&1\\\hfill 1&\hfill \infty&1\\\end{xsmallmatrix},1\right)-\text{E}_4\left(\begin{xsmallmatrix}{.15em}-\tilde{1}&0&1\\\hfill 1&0&1\\\end{xsmallmatrix},1\right)+c_4\text{E}_4\left(\begin{xsmallmatrix}{.15em}-\tilde{1}&1&-1\\\hfill 1&1&\hfill 0\\\end{xsmallmatrix},1\right)\nonumber\\&+c_4\text{E}_4\left(\begin{xsmallmatrix}{.15em}-\tilde{1}&1&-1\\\hfill 1&1&\hfill 1\\\end{xsmallmatrix},1\right)+2c_4\text{E}_4\left(\begin{xsmallmatrix}{.15em}-\tilde{1}&1&-1\\\hfill 1&1&\hfill \infty\\\end{xsmallmatrix},1\right)-\text{E}_4\left(\begin{xsmallmatrix}{.15em}-\tilde{1}&1&0\\\hfill 1&1&0\\\end{xsmallmatrix},1\right)-c_4\text{E}_4\left(\begin{xsmallmatrix}{.15em}-\tilde{1}&1&1\\\hfill 1&1&1\\\end{xsmallmatrix},1\right)
\bigg)+\mathcal{O}(\epsilon)\,.
\end{align}
The higher orders in $\epsilon$ may be obtained in the same manner.

\subsubsection{Analytic continuation}
\label{sec:TWBAnalyticContinuation}
In this section we apply the methods of section \ref{sec:continuation} to perform the analytic continuation of the triangle with bubble integral $T_{1,2,1,1}(s,m^2)$ to the physical region $s>0,m^2>0$. We will explicitly discuss the analysis for the order $\epsilon^0$ contribution $T^{(0)}_{1211}$. We also performed the analytic continuation of the order $\epsilon^1$, however the final result involves rather complicated expressions and we don't present its derivation here. Our starting point is the representation of Eq. (\ref{eq:twbint1}) that we rewrite here for the reader's convenience in the following form:
\begin{equation}
    T^{(0)}_{1,2,1,1}(s,m^2,i\delta)=\int_0^\infty \frac{1}{x} f_T^{(2)}(x,s,m^2,i\delta) dx\,,
\end{equation}
where we renamed the integration variable to $x$, and where $f^{(2)}(x,s,m^2,i\delta)$ is obtained from eq. (\ref{eq:twbint1}) by applying the Feynman prescription $s\rightarrow s+i\delta$. Referring to section \ref{sec:regions}, the symbol alphabet letters of $f^{(2)}(x,s,m^2,0)$ can be expressed in terms of the following linearly independent letters (see section \ref{sec:SunriseAnalyticContinuation} for further discussion):
\begin{align}
    \rho _1& =x,\rho _2=x+1,\rho _3=m^2,\rho _4=m^2-s,\rho _5=s\,,\nonumber\\
    \sigma _1& =m^2 \left(x^2+x+y-1\right)-s x,\sigma _2=s x-m^2 \left(x^2+x-y+1\right)\,,\nonumber\\\sigma _3&=s x-m^2 \left(x^2+3
   x-y+1\right)\,.
\end{align}

Following the prescription of section \ref{sec:regions}, we identify the following subregions of region $A_T:  y^2(x,s,m^2)<0$:
\begin{equation}
    A_{T,1}:\; \rho_{4}>0\,,\quad A_{T,2}:\;\rho_{4}<0\,,
\end{equation}
while for $B_T:  y^2(x,s,m^2)>0$, we have the following subregions:
\begin{align}
    B_{T,1}:& \;\rho_{4}<0,\sigma_{1}<0,\sigma_{2}<0,\sigma_{3}<0\,,\quad     B_{T,3}:\;\rho _4<0, \sigma _2>0, \sigma _3>0, \sigma _1<0\,,\nonumber\\
    B_{T,2}:& \;\rho _4<0, \sigma _2<0, \sigma _3<0, \sigma _1>0\,,\quad B_{T,4}:\; \rho _4>0, \sigma _2<0, \sigma _3<0, \sigma _1>0\,.
\end{align}
By applying the algorithm of section \ref{sec:continuation} we are able to perform the $\delta\rightarrow 0$ limit of $f^{(2)}(x,s,m^2,i\delta)$ in terms of classical polylogarithms and logarithms in each of these regions. We obtain:
\begin{equation}
\label{eq:contT}
   T^{(0)}_{1,2,1,1}(s,m^2)=\int_0^\infty \frac{1}{x}\sum_{i=1}^3 \theta_{i}(x,s,m^2) f_{T,i}^{(2)}(x,s,m^2)\,,
\end{equation}
where
\begin{equation}
\label{eq:regionsTriangle}
    R_{T,1}=A_{T,1}\cup A_{T,2}\,,\;\;R_{T,2}=B_{T,1}\cup B_{T,2}\cup B_{T,4}\,,\;\;R_{T,3}=B_{T,3}\,,
\end{equation}
\begin{figure}[t]
\centering
\includegraphics[width=0.75\textwidth]{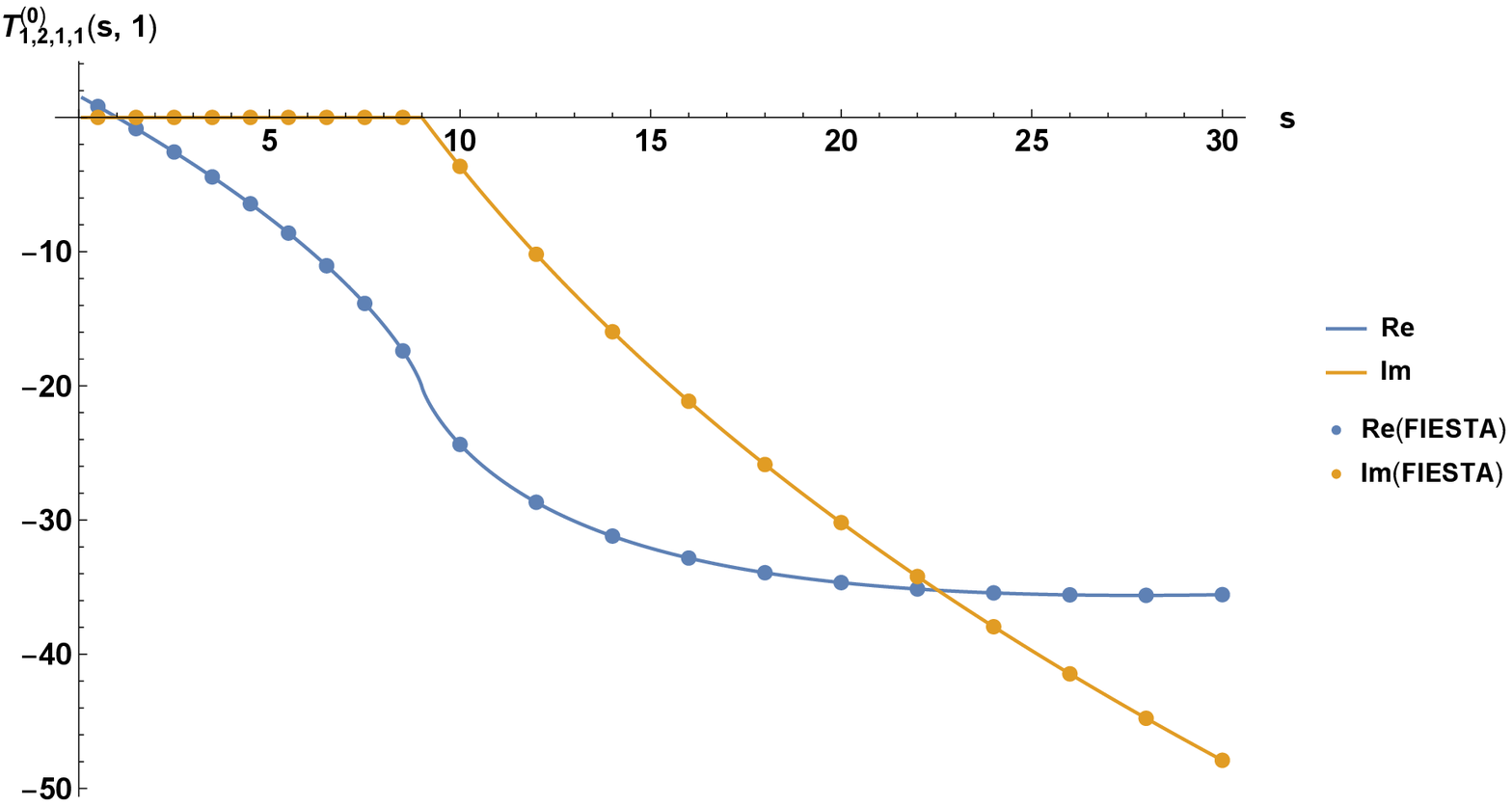}\\
\vspace{2em}
\includegraphics[width=0.75\textwidth]{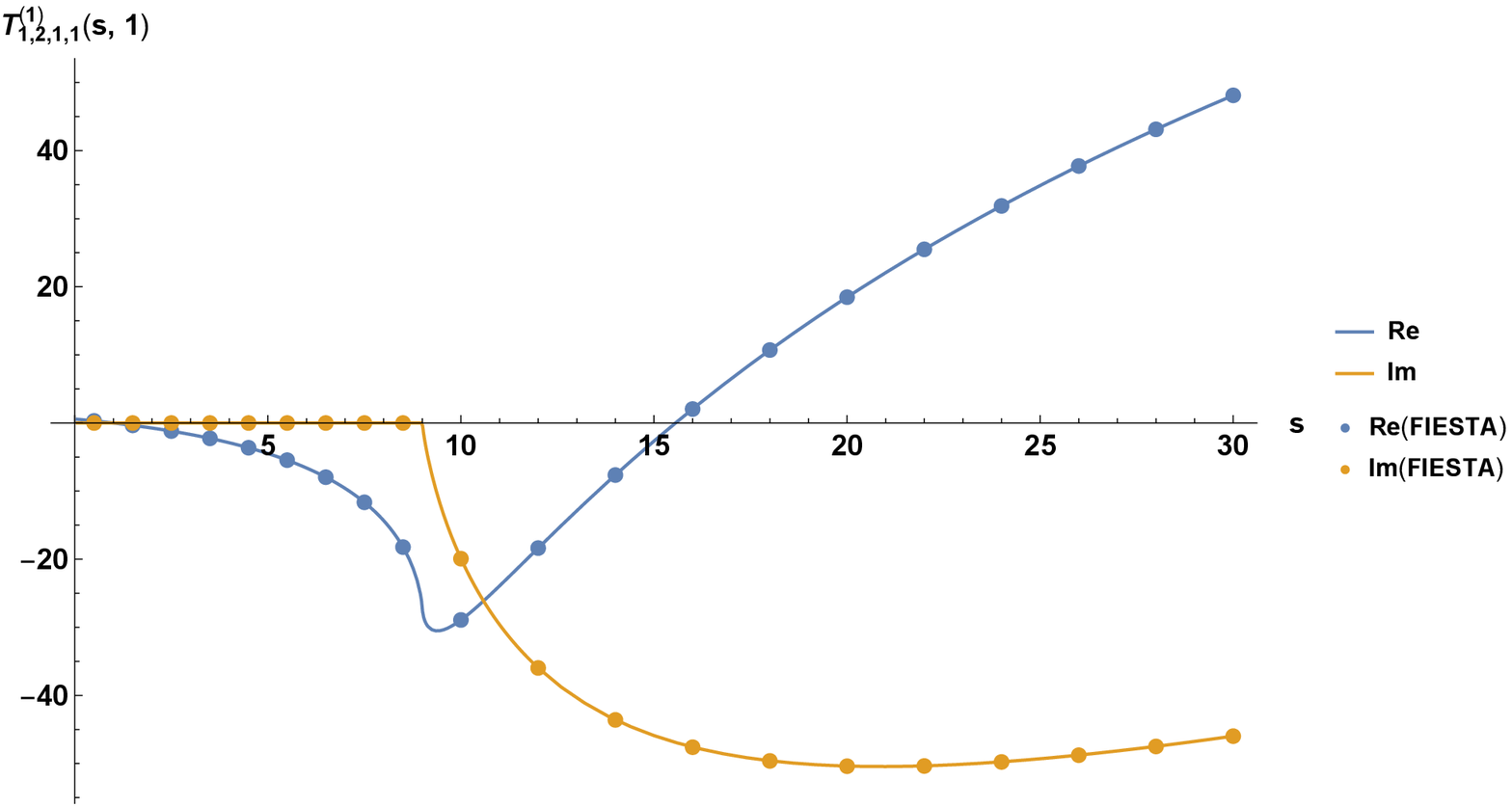}
\caption{Plots of the first two epsilon orders of $T_{1,2,1,1}(s,m^2)$}
\label{fig:tri0tri1}
\end{figure}
and $\theta_{i}(x,s,m^2)=1$ if $x,s,m^2\in R_i$ and $\theta_{i}(x,s,m^2)=0$ otherwise. The expression for $f_{1}^{(2)}(x,s,m^2)$ is, by using the prescription of eq. (\ref{eq:srprescription}), related to the one of $f_{3}^{(2)}(x,s,m^2)$ by:
\begin{equation}
\label{eq:f_T1}
f_{T,1}^{(2)}(x,s,m^2)=f_{T,3}^{(2)}(x,s,m^2)|_{y\rightarrow i\sqrt{-y^2}} \,,   
\end{equation}
while:
\begin{align}
\label{eq:f_T2}
  f_{T,2}^{(2)}&(x,s,m^2) =-\text{Li}_2\left(-\frac{4 \rho _1^2 \rho _2 \rho _3 \rho _4}{\sigma _1 \sigma _3}\right)+\text{Li}_2\left(\frac{2 \rho _2 \rho _3 \rho _4 \sigma _2}{\rho _5 \sigma _1 \sigma
   _3}\right)-\text{Li}_2\left(\frac{\sigma _1 \sigma _3}{4 \rho _2^2 \rho _3^2}\right)+\text{Li}_2\left(-\frac{\sigma _1 \sigma _3}{2 \rho _2^2 \rho _3 \sigma
   _2}\right)\nonumber\\
   &+\text{Li}_2\left(\frac{\rho _4}{\rho _3}\right)-\log \left(\rho _3\right) \log \left(-\sigma _2\right)-2 \log \left(\rho _1\right) \log \left(-\sigma _2\right)-\log \left(\rho _5\right) \log \left(-\sigma _2\right)+\frac{1}{2} \log
   ^2\left(\rho _3\right)\nonumber\\
   &+\frac{1}{2} \log ^2\left(\rho _5\right)+2 \log \left(\rho _1\right) \log \left(\rho _3\right)-\log \left(\rho _2\right) \log \left(\rho _3\right)+\log (2) \log
   \left(\rho _3\right)+2 \log (2) \log \left(\rho _1\right)\nonumber\\
   &+\log \left(\rho _2\right) \log \left(\rho _5\right)+\log (2) \log \left(\rho _5\right)+\log ^2\left(-\sigma _2\right)-2 \log (2)
   \log \left(-\sigma _2\right)+\log ^2(2)\,,
\end{align}
and:
\begin{align}
\label{eq:f_T3}
    f_{T,3}^{(2)}&(x,s,m^2)= -\text{Li}_2\left(-\frac{4 \rho _1^2 \rho _2 \rho _3 \rho _4}{\sigma _1 \sigma _3}\right)+\text{Li}_2\left(\frac{2 \rho _2 \rho _3 \rho _4 \sigma _2}{\rho _5 \sigma _1 \sigma
   _3}\right)-\text{Li}_2\left(\frac{\sigma _1 \sigma _3}{4 \rho _2^2 \rho _3^2}\right)+\text{Li}_2\left(-\frac{\sigma _1 \sigma _3}{2 \rho _2^2 \rho _3 \sigma
   _2}\right)\nonumber\\
   &+\text{Li}_2\left(\frac{\rho _4}{\rho _3}\right)-\log \left(\rho _3\right) \log \left(\sigma _2\right)-2 \log \left(\rho _1\right) \log \left(\sigma _2\right)-\log \left(\rho _5\right) \log \left(\sigma _2\right)+\frac{1}{2} \log ^2\left(\rho
   _3\right)+\frac{1}{2} \log ^2\left(\rho _5\right)\nonumber\\
   &+2 \log \left(\rho _1\right) \log \left(\rho _3\right)-\log \left(\rho _2\right) \log \left(\rho _3\right)+\log (2) \log \left(\rho
   _3\right)+i \pi  \log \left(\rho _3\right)+2 \log (2) \log \left(\rho _1\right)\nonumber\\
   &+2 i \pi  \log \left(\rho _1\right)+\log \left(\rho _2\right) \log \left(\rho _5\right)+\log (2) \log
   \left(\rho _5\right)+i \pi  \log \left(\rho _5\right)+\log ^2\left(\sigma _2\right)-2 \log (2) \log \left(\sigma _2\right)\nonumber\\
   &-2 i \pi  \log \left(\sigma _2\right)-\pi ^2+\log ^2(2)+2 i \pi  \log (2)\,.
\end{align}
Let us comment on the origin of the three regions $R_{T,i}$ of eq.~(\ref{eq:regionsTriangle}). We have seen that the prescription of section \ref{sec:regions} usually leads to an upper bound for the set of relevant regions. In the case under consideration that prescription identifies 6 regions, $A_{T,i}$, $B_{T,j}$ with $i\in\{1,2\}$ and $j\in\{1,2,3,4\}$. By using the algorithm of section \ref{sec:continuation} we were able to identify a basis of functions satisfying eq.~(\ref{eq:fconstraints}) for $A_{T,1}\cup A_{T,2}$, $B_{T,1}\cup B_{T,2}\cup B_{T,4}$ and $B_{T,3}$ respectively. The computation of $\lim_{\delta\rightarrow 0}f_{T}^{(2)}(x,s,m^2,i\delta)$ in each of these "enlarged" regions leads to the same expression for all the corresponding subregions, therefore only the 3 regions $R_{T,i}$ are needed.

Since $x,s,y$ are real valued in $R_{T,2}$ and $R_{T,3}$, and since by construction the logarithms and dilogarithms of eqs.~(\ref{eq:f_T2}), (\ref{eq:f_T3}) satisfy eq.~(\ref{eq:fconstraints}), the terms $f_{T,2}^{(2)}(x,s,m^2)$ and $f_{T,3}^{(2)}(x,s,m^2)$ have explicit imaginary parts and all the logarithms and dilogarithms are real valued. This is not the case for $f_{T,1}^{(2)}(x,s,m^2)$ where the dependence on $i\sqrt{-y^2}$ implies that individual functions are complex valued in general.

Let us mention again that the analytic continuation eq.~(\ref{eq:contT}) is suitable for fast and precise numerical evaluations, for example we have:
\begin{equation}
    T^{(0)}_{1,2,1,1}(15,1)= -32.095121541517732621840491 - i 18.624629780558552746660283
\end{equation}
In order to validate our results we performed extensive numerical checks against the computer program FIESTA. The results are summarized in Fig \ref{fig:tri0tri1}.

\section{A non-planar triangle from Higgs+Jet}
\label{sec:non-planar triangle}

In this section we show that linear reducibility does not directly imply that a representation in terms of eMPL exists, and further exploration of the methods discussed in this paper is required. Nevertheless we provide evidence that a simple all orders structure, analogous to the one discussed in the previous section, holds. We consider a non-planar triangle in $d=4-2\epsilon$ with two off-shell legs, relevant for the two-loop QCD corrections to Higgs plus jet production:
\begin{equation}
    N_{1,1,1,1,1,1}(s,p_2^2,m^2) \,\,\,\, = \,\,\,\      (1+2 \epsilon)(p_2^2-s)\quad \vcenter{\hbox{\includegraphics[width=0.2\textwidth]{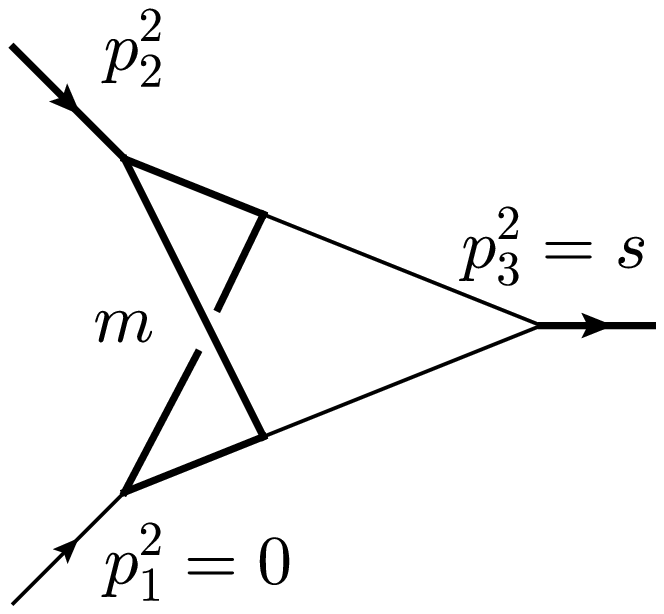}}}\, ,
\end{equation}
whose homogenous solutions were found in \cite{Primo:2016ebd}. This integral has an elliptic maximal cut, but no elliptic subtopologies are present. Note the $(1+2 \epsilon)(p_2^2-s)$ prefactor, which is needed to obtain a uniformly transcendental inner polylogarithmic expression. We consider the Euclidean region $s<0, p_2^2 < 0$ and $m^2>0$. The parametric representation in $d=4-2 \epsilon$ dimensions reads:
\begin{align}
    N_{1,1,1,1,1,1} (s,p_2^2,m^2) &=(1+2 \epsilon)(p_2^2-s)\sum_{k=0}^\infty \epsilon^k \int_\Delta d^6 \vec{\alpha}\, \frac{1}{k!}\, \alpha_1\, \mathcal{F}^{-2} \,\log\left( \frac{\mathcal{U}^3}{\mathcal{F}^2}\right)^k\nonumber\\
    &\equiv N^{(0)}_{1,1,1,1,1,1} (s,p_2^2,m^2)+\epsilon\, N^{(1)}_{1,1,1,1,1,1} (s,p_2^2,m^2)+\mathcal{O}(\epsilon^2)\,,
\end{align}
where the Symanzik polynomials are:
\begin{align}
  \mathcal{U}  =& \,\alpha _3 \alpha _4+\alpha _6 \alpha _4+\alpha _3 \alpha _5+\alpha _2 \left(\alpha _3+\alpha _4+\alpha _5\right)+\alpha _3 \alpha _6+\alpha _5 \alpha _6+\alpha _1 \left(\alpha _2+\alpha _4+\alpha
   _5+\alpha _6\right)\,,\nonumber\\
  \mathcal{F}  =&\, \big(\alpha _3+\alpha _4+\alpha _5\big) \alpha _2^2 m^2+\big(\alpha _3+\alpha _4+\alpha _5\big) \alpha _6^2 m^2+\alpha _3 \big(\alpha _4^2 m^2+\alpha _5^2 m^2+\alpha _5 \alpha _4 \big(2 m^2-p_2^2\big)\big)\nonumber\\&+\alpha _6 \big(\alpha _5^2 m^2+\alpha _4 \big(2 \alpha _3+\alpha _4\big) m^2+\big(\alpha _3+\alpha _4\big) \alpha _5 \big(2 m^2-p_2^2\big)\big)+\alpha _2 \big(\alpha _4^2 m^2+\alpha _5 \big(\alpha _5+2 \alpha _6\big) m^2\nonumber\\&+\alpha _4 \big(2 \alpha _6 m^2+\alpha _5 \big(2 m^2-p_2^2\big)\big)+\alpha _3 \big(2 \alpha _4 m^2+2 \alpha _6 m^2+\alpha _5 \big(2 m^2-s\big)\big)\big)+\alpha _1 \big(\alpha _2^2 m^2+\alpha _4^2 m^2\nonumber\\&+\alpha _5^2 m^2+\alpha _6^2 m^2+2 \alpha _4 \alpha _5 m^2+\alpha _2 \big(2 \big(\alpha _5+\alpha _6\big) m^2+\alpha _4 \big(2 m^2-p_2^2\big)-\alpha _3 s\big)+\alpha _6 \big(2 \alpha _5 m^2\nonumber\\&+\alpha _4 \big(2 m^2-s\big)\big)-\alpha _4 \alpha _5 p_2^2-\alpha _3 \big(\alpha _4+\alpha _5+\alpha _6\big) s\big)\,.
\end{align}
In order to achieve linear reducibility up to the last integration we apply Cheng-Wu by setting $\alpha_2\rightarrow1-\alpha_4-\alpha_5-\alpha_6$, and we integrate along the sequence $\alpha_3$, $\alpha_1$, $\alpha_6$, $\alpha_4$, which defines the integration domain to be:
\begin{equation}
    \int_\Delta d^6 \vec{\alpha}\rightarrow \int_0^1 d \alpha_5\int_0^{1-\alpha_5}d\alpha_4\int_0^{1-\alpha_4-\alpha_5}d\alpha_6\int_0^\infty d\alpha_1\int_0^\infty d\alpha_3\,.
\end{equation}
The first two integrations yield:
\begin{align}
  N^{(0)}_{1,1,1,1,1,1} &  (s,p_2^2,m^2) = \int_0^1 d \alpha_5\int_0^{1-\alpha_5}d\alpha_4\int_0^{1-\alpha_4-\alpha_5}d\alpha_6\nonumber\\
  &\frac{\left(p_2^2-s\right) \log \left(\frac{\left(\alpha _4+\alpha _5-1\right) s \left(\alpha _5 m^2+\alpha _4 \left(m^2-\alpha _5 p_2^2\right)\right)}{\left(m^2+\alpha _4 \left(\left(\alpha _4+\alpha
   _6-1\right) p_2^2-\alpha _6 s\right)\right) \left(m^2+\alpha _5 \left(\left(\alpha _4+\alpha _6\right) \left(s-p_2^2\right)-s\right)+\alpha _5^2 s\right)}\right)}{\left(m^2-\alpha _5 \left(\alpha
   _4+\alpha _6\right) p_2^2+\alpha _5 \alpha _6 s\right) \left(m^2+\alpha _4 \left(\alpha _4+\alpha _6-1\right) p_2^2-\alpha _4 \left(\alpha _4+\alpha _5+\alpha _6-1\right) s\right)}\,.
\end{align}
We proceed with the $\alpha_6$ integration:
\begin{align}
   &N^{(0)}_{1,1,1,1,1,1}   (s,p_2^2,m^2) = \int_0^1 d \alpha_5\int_0^{1-\alpha_5}d\alpha_4 \frac{1}{\alpha _5
   m^2-s \alpha _4^2 \alpha _5+\alpha _4 \left(m^2-s \alpha _5^2+\left(s-p_2^2\right) \alpha _5\right)}\Bigg[\nonumber\\
   &-G_{\frac{s}{m^2+s \alpha _5^2+s+\left(\left(s-p_2^2\right) \alpha _4-s\right) \alpha _5}} G_{\frac{-m^2+\left(s-p_2^2\right) \alpha _5^2+\left(p_2^2+s \left(\alpha _4-1\right)\right) \alpha
   _5}{\left(s-p_2^2\right) \alpha _5 \left(\alpha _4+\alpha _5-1\right)}}+G_{\frac{m^2-p_2^2 \alpha _4 \alpha _5}{\left(s-p_2^2\right) \alpha _4 \left(\alpha
   _4+\alpha _5-1\right)}} G_{\frac{s}{m^2+s \alpha _5^2+s+\left(\left(s-p_2^2\right) \alpha _4-s\right) \alpha _5}}\nonumber\\
   &+G_{\frac{m^2+p_2^2 \left(\alpha _4-1\right) \alpha _4}{\left(\alpha _5^2-\alpha _5+1\right) m^2+\alpha _4^2 \left(m^2-p_2^2
   \left(\alpha _5-1\right)\right)+\alpha _4 \left(\left(2 \alpha _5-1\right) m^2+p_2^2 \left(-\alpha _5^2+\alpha _5-1\right)\right)}} G_{\frac{-m^2+\left(s-p_2^2\right) \alpha _5^2+\left(p_2^2+s
   \left(\alpha _4-1\right)\right) \alpha _5}{\left(s-p_2^2\right) \alpha _5 \left(\alpha _4+\alpha _5-1\right)}}\nonumber\\
   &-G_{\frac{m^2-p_2^2 \alpha _4 \alpha _5}{\left(s-p_2^2\right) \alpha _4 \left(\alpha
   _4+\alpha _5-1\right)}}G_{\frac{m^2+p_2^2 \left(\alpha _4-1\right) \alpha _4}{\left(\alpha
   _5^2-\alpha _5+1\right) m^2+\alpha _4^2 \left(m^2-p_2^2 \left(\alpha _5-1\right)\right)+\alpha _4 \left(\left(2 \alpha _5-1\right) m^2+p_2^2 \left(-\alpha _5^2+\alpha
   _5-1\right)\right)}}\nonumber\\
   &-G_{\frac{m^2+p_2^2 \left(\alpha _5-1\right) \alpha _5}{\left(p_2^2-s\right) \alpha _5 \left(\alpha _4+\alpha _5-1\right)},\frac{m^2-p_2^2 \alpha _4 \alpha
   _5}{\left(s-p_2^2\right) \alpha _4 \left(\alpha _4+\alpha _5-1\right)}}+G_{\frac{m^2+p_2^2 \left(\alpha _5-1\right) \alpha _5}{\left(p_2^2-s\right) \alpha _5 \left(\alpha _4+\alpha
   _5-1\right)},\frac{m^2+\alpha _5 \left(p_2^2 \left(\alpha _5-1\right)-s \left(\alpha _4+\alpha _5-1\right)\right)}{\left(p_2^2-s\right) \alpha _5 \left(\alpha _4+\alpha _5-1\right)}}\nonumber\\&
   -G_{\frac{m^2+s
   \alpha _4^2+\alpha _4 \left(\left(s-p_2^2\right) \alpha _5-s\right)}{\left(s-p_2^2\right) \alpha _4 \left(\alpha _4+\alpha _5-1\right)},\frac{m^2-p_2^2 \alpha _4 \alpha _5}{\left(s-p_2^2\right)
   \alpha _4 \left(\alpha _4+\alpha _5-1\right)}}+G_{\frac{m^2+s \alpha _4^2+\alpha _4 \left(\left(s-p_2^2\right) \alpha _5-s\right)}{\left(s-p_2^2\right) \alpha _4 \left(\alpha _4+\alpha
   _5-1\right)},\frac{m^2+\alpha _5 \left(p_2^2 \left(\alpha _5-1\right)-s \left(\alpha _4+\alpha _5-1\right)\right)}{\left(p_2^2-s\right) \alpha _5 \left(\alpha _4+\alpha _5-1\right)}}\Bigg].
\end{align}
At this point it is straightforward to perform the last integration with respect to $\alpha_4$. In this way the final result is an integral with respect to $\alpha_5$ between $0$ and $1$. In order to keep the notation consistent with the previous examples we perform the following variable change,
\begin{equation}
    \alpha_5(\beta_5)=\frac{\beta_5}{1+\beta_5}\,,
\end{equation}
which maps the upper integration bound to $\infty$. The final expression for $N^{(0)}_{1,1,1,1,1,1}(s,p_2^2,m^2)$ is quite lengthy and we provide it in appendix \ref{app:full results}. Its expression is, as expected, of the form:
\begin{equation}
    N^{(0)}_{1,1,1,1,1,1}(s,p_2^2,m^2)=\int_0^\infty d\beta_5 \frac{1}{\sqrt{P_N(\beta_5,s,p_2^2,m^2)}} N^{(3)}(\beta_5,s,p_2^2,m^2)\,,
\end{equation}
where the elliptic curve is
\begin{equation}
    P_N(\beta_5,s,p_2^2,m^2)=\left(\beta _5+1\right)^4 m^4-2 \beta _5 \left(\beta _5+1\right)^2 m^2 \left(\beta _5 p_2^2+p_2^2-2 \beta _5 s-s\right)+\beta _5^2 \left(\beta _5 p_2^2+p_2^2-s\right)^2\,,
\end{equation}
and $\text{N}^{(3)}(\beta_5,s,p_2^2,m^2)$ is a pure polylogarithmic function of uniform weight 3. The very same integration procedure can be applied to compute $N^{(1)}_{1,1,1,1,1,1}(s,p_2^2,m^2)$ in the form:
\begin{equation}
    N^{(1)}_{1,1,1,1,1,1}(s,p_2^2,m^2)=\int_0^\infty d\beta_5 \frac{1}{\sqrt{P_N(\beta_5,s,p_2^2,m^2)}}N^{(4)}(\beta_5,s,p_2^2,m^2)\,.
\end{equation}
The explicit result is lengthy and can be obtained upon request from the authors.

As opposed to the previous examples, the IPP dependes on multiple algebraic functions, namely the square root of the following polynomials:
\begin{align}
    P_N(\beta_5,s,p_2^2,m^2) &=\left(\beta _5+1\right)^4 m^4-2 \beta _5 \left(\beta _5+1\right)^2 m^2 \left(\beta _5 p_2^2+p_2^2-2 \beta _5 s-s\right)+\beta _5^2 \left(\beta _5 p_2^2+p_2^2-s\right)^2\,,\nonumber\\
    Q_N(\beta_5,s,p_2^2,m^2) &=\left(\beta _5 p_2^2+p_2^2-s\right)^2-4 \left(\beta _5+1\right)^2 m^2 \left(p_2^2-s\right)\,,\nonumber\\
    R_N(\beta_5,s,p_2^2,m^2) &=\beta _5^2 \left(p_2^4-4 m^2 s\right)+2 \beta _5 s \left(p_2^2-4 m^2\right)+s \left(s-4 m^2\right)\,.
\end{align}
It is clear that a naive attempt to translate this result to eMPLs will introduce integration kernels that are not rational functions on an elliptic curve, but more complicated algebraic functions depending on the set of square roots above. While we cannot exclude that an eMPL representation exists, we believe that a systematic study of these more complicated cases is still missing and we leave it for future work.

\section{Conclusions and outlook}
\label{sec:conclusionsandoutlook}
In this paper we have investigated a class of elliptic Feynman integrals, dubbed linearly reducible elliptic Feynman integrals. By direct integration of the Feynman parametrization one may express such integrals order by order in the dimensional regulator as a 1-dimensional integral over a polylogarithmic integrand, which we call the inner polylogarithmic part (IPP). The resulting 1-dimensional integral representation can be analytically continued to the physical region in a form suitable for fast and precise numerical evaluations. When the IPP depends on one elliptic curve and no other algebraic functions, linearly reducible elliptic Feynman integrals can also be expressed in terms of multiple elliptic polylogarithms.

We have also shown that the IPP can be mapped to a (generalized) polylogarithmic Feynman integral that can be subsequently solved using the differential equations method. In particular we studied the IPP of the unequal mass sunrise topology, and a triangle with bubble topology, and provided a canonical basis of master integrals where the system of differential equations is in $d\log$ $\epsilon$-factorized form. For this basis, the differential equations with respect to the last integration parameter were found to be in an $\epsilon$-form where the integration kernels coincide with the integration kernels of the class of eMPLs of \cite{Broedel:2017kkb}. This allows one to systematically solve the IPP in terms of eMPLs, directly from the system of differential equations. Once such a representation is achieved the remaining last integration can be performed in terms of eMPLs as well. 

We expect that the methods discussed in this paper may provide new insights for problems where iterated integrals over multiple (and more complicated) algebraic functions need to be considered. Furthermore we aim to apply our methods to more complicated Feynman integrals in the future.

\section*{Acknowledgements}
The authors would like to thank Ruth Britto for many interesting discussions, support, and comments on the manuscript. We thank Roberto Bonciani for useful discussions about the $t\bar{t}$ elliptic integral sectors. We thank Robert Schabinger for fruitful discussions about the properties of the parametric representation of Feynman integrals. We thank Vladimir Smirnov for useful discussions about the integration algorithm of the computer program FIESTA. We thank Armin Schweitzer for useful discussions about the analytic continuation of the sunrise integral. FM would like to thank Trinity College Dublin for hospitality during the preparation of this work. MH would like to thank ETH Zurich for hospitality during the preparation of this work. This work was supported by a STSM Grant from the COST Action CA16201 PARTICLEFACE. This project has received funding from the European Research Council (ERC) under grant agreements No 647356 (CutLoops) and No 694712 (pertQCD).

\appendix

\section{Analytic continuation of the sunrise integral}
\label{app:contS}
In this appendix we provide the explicit expressions for the analytic continuation of the sunrise integral discussed in \ref{sec:SunriseAnalyticContinuation} up and including the order $\epsilon^1$:
\begin{align}
    S_{1,1,1}(s,m_1^2,m_2^2,m_3^2)= & \sum_{k=0}^1 \left[\frac{\epsilon^k}{m_3^2}\int_0^\infty \frac{\theta_1(x,s, m_1^2,m_2^2,m_3^2)}{i\sqrt{-y^2(x,s, m_1^2,m_2^2,m_3^2)}} f_{S,1}^{(k+1)}(x,s, m_1^2,m_2^2,m_3^2)dx\right. \nonumber\\+&\left. \frac{\epsilon^k}{m_3^2}\int_0^\infty \sum_{j=2}^5 \frac{\theta_j(x,s, m_1^2,m_2^2,m_3^2)}{y(x,s, m_1^2,m_2^2,m_3^2)} f_{S,j}^{(k+1)}(x,s, m_1^2,m_2^2,m_3^2)dx\right]+\mathcal{O}(\epsilon^2)\,.
\end{align}
At weight one we have:
\begin{align}
    f_{S,1}^{(1)}&=\log \left(\alpha _1\right)+\log \left(\alpha _2\right)+\log \left(\alpha _3\right)+\log \left(\alpha _4\right)+2 \log \left(\alpha _5\right)-2 \log \left(\beta _1\right)-2 \log \left(\beta
   _3\right)+4 \log (2)|_{y\rightarrow i\sqrt{-y^2}},\nonumber\\
   f_{S,2}^{(1)}&=\log \left(\alpha _1\right)+\log \left(\alpha _2\right)+\log \left(\alpha _3\right)+\log \left(\alpha _4\right)+2 \log \left(-\alpha _5\right)-2 \log \left(-\beta _1\right)-2 \log \left(\beta
   _3\right)+4 \log (2),\nonumber\\
   f_{S,3}^{(1)}&=\log \left(\alpha _1\right)+\log \left(\alpha _2\right)+\log \left(\alpha _3\right)+\log \left(\alpha _4\right)+2 \log \left(-\alpha _5\right)-2 \log \left(\beta _1\right)-2 \log \left(-\beta
   _3\right)+4 \log (2),\nonumber\\
   f_{S,4}^{(1)}&=\log \left(\alpha _1\right)+\log \left(\alpha _2\right)+\log \left(\alpha _3\right)+\log \left(\alpha _4\right)+2 \log \left(-\alpha _5\right)-2 \log \left(\beta _1\right)-2 \log \left(\beta
   _3\right)+2 i \pi +4 \log (2),\nonumber\\
   f_{S,5}^{(1)}&=\log \left(\alpha _1\right)+\log \left(\alpha _2\right)+\log \left(\alpha _3\right)+\log \left(\alpha _4\right)+2 \log \left(\alpha _5\right)-2 \log \left(\beta _1\right)-2 \log \left(\beta
   _3\right)+4 \log (2)\,.
\end{align}
At weight two we have:
\begin{align}
    f_{S,1}^{(2)}&(x,s, m_1^2,m_2^2,m_3^2)=3 \text{Li}_2\left(-\frac{2 \alpha _1^2 \alpha _3 \alpha _6}{\alpha _7 \beta _2}\right)+\text{Li}_2\left(\frac{4 \alpha _1 \alpha _2 \alpha _3 \alpha _4 \alpha _5}{\alpha _7 \beta _1 \beta
   _3}\right)-3 \text{Li}_2\left(-\frac{\alpha _1 \alpha _6 \beta _1 \beta _3}{2 \alpha _5 \alpha _7 \beta _2}\right)\nonumber\\
  & -2 \log \left(\alpha _1\right) \log \left(\beta _1\right)+3 \log \left(\alpha _1\right) \log \left(\beta _2\right)-2 \log \left(\alpha _1\right) \log \left(\beta _3\right)-3 \log \left(\alpha
   _2\right) \log \left(\beta _1\right)\nonumber\\
   &+\log \left(\alpha _4\right) \log \left(\beta _1\right)-2 \log \left(\alpha _5\right) \log \left(\beta _1\right)+4 \log \left(\alpha _7\right) \log \left(\beta
   _1\right)+3 \log \left(\alpha _3\right) \log \left(\beta _2\right)\nonumber\\
   &+3 \log \left(\alpha _5\right) \log \left(\beta _2\right)-3 \log \left(\alpha _2\right) \log \left(\beta _3\right)+\log
   \left(\alpha _4\right) \log \left(\beta _3\right)-2 \log \left(\alpha _5\right) \log \left(\beta _3\right)\nonumber\\
   &+4 \log \left(\alpha _7\right) \log \left(\beta _3\right)+\frac{1}{2} \log ^2\left(\alpha
   _1\right)+\log ^2\left(\alpha _2\right)-\frac{1}{2} \log ^2\left(\alpha _3\right)-\log ^2\left(\alpha _4\right)+\log ^2\left(\alpha _5\right)+\frac{1}{2} \log ^2\left(\alpha _7\right)\nonumber\\
   &+3 \log
   \left(\alpha _2\right) \log \left(\alpha _1\right)+\log \left(\alpha _4\right) \log \left(\alpha _1\right)+2 \log \left(\alpha _5\right) \log \left(\alpha _1\right)-4 \log \left(\alpha _7\right)
   \log \left(\alpha _1\right)+\log (2) \log \left(\alpha _1\right)\nonumber\\
   &+i \pi  \log \left(\alpha _1\right)+6 \log (2) \log \left(\alpha _2\right)+i \pi  \log \left(\alpha _2\right)+2 \log \left(\alpha
   _2\right) \log \left(\alpha _3\right)-3 \log (2) \log \left(\alpha _3\right)\nonumber\\
   &+i \pi  \log \left(\alpha _3\right)-2 \log (2) \log \left(\alpha _4\right)+i \pi  \log \left(\alpha _4\right)+3 \log
   \left(\alpha _2\right) \log \left(\alpha _5\right)-\log \left(\alpha _4\right) \log \left(\alpha _5\right)\nonumber\\
   &+\log (2) \log \left(\alpha _5\right)+i \pi  \log \left(\alpha _5\right)-\log \left(\alpha
   _2\right) \log \left(\alpha _7\right)-4 \log \left(\alpha _3\right) \log \left(\alpha _7\right)-\log \left(\alpha _4\right) \log \left(\alpha _7\right)\nonumber\\
   &-4 \log \left(\alpha _5\right) \log
   \left(\alpha _7\right)-8 \log (2) \log \left(\alpha _7\right)-i \pi  \log \left(\alpha _7\right)+\log ^2\left(\beta _1\right)+\log ^2\left(\beta _3\right)-\log (2) \log \left(\beta _1\right)\nonumber\\
   &-i \pi
    \log \left(\beta _1\right)-3 \log \left(\beta _1\right) \log \left(\beta _2\right)+6 \log (2) \log \left(\beta _2\right)+2 \log \left(\beta _1\right) \log \left(\beta _3\right)-3 \log \left(\beta
   _2\right) \log \left(\beta _3\right)\nonumber\\
   &-\log (2) \log \left(\beta _3\right)-i \pi  \log \left(\beta _3\right)-\frac{\pi ^2}{3}-2 \log ^2(2)+2 i \pi  \log (2)|_{y\rightarrow i\sqrt{-y^2}}\,,
\end{align}
\begin{align}
    f_{S,2}^{(2)}&(x,s, m_1^2,m_2^2,m_3^2) = \text{Li}_2\left(\frac{2 \alpha _1 \alpha _3}{\beta _2}\right)-\text{Li}_2\left(\frac{\beta _2}{2 \alpha _1 \alpha _6}\right)-4 \text{Li}_2\left(-\frac{\alpha _7 \beta _2}{2 \alpha _1^2 \alpha _3
   \alpha _6}\right)+4 \text{Li}_2\left(-\frac{2 \alpha _5 \alpha _7 \beta _2}{\alpha _1 \alpha _6 \beta _1 \beta _3}\right)\nonumber\\
   & +2 \log \left(\alpha _1\right) \log \left(-\beta _1\right)+8 \log \left(\alpha _1\right) \log \left(-\beta _2\right)+2 \log \left(\alpha _1\right) \log \left(\beta _3\right)-4 \log \left(\alpha
   _2\right) \log \left(-\beta _1\right)\nonumber\\
   &-8 \log \left(-\alpha _5\right) \log \left(-\beta _1\right)+4 \log \left(\alpha _6\right) \log \left(-\beta _1\right)+7 \log \left(\alpha _3\right) \log
   \left(-\beta _2\right)+8 \log \left(-\alpha _5\right) \log \left(-\beta _2\right)\nonumber\\
   &+\log \left(\alpha _6\right) \log \left(-\beta _2\right)-4 \log \left(\alpha _2\right) \log \left(\beta _3\right)-8
   \log \left(-\alpha _5\right) \log \left(\beta _3\right)+4 \log \left(\alpha _6\right) \log \left(\beta _3\right)-6 \log ^2\left(\alpha _1\right)\nonumber\\
   &+\log ^2\left(\alpha _2\right)-\frac{5}{2} \log
   ^2\left(\alpha _3\right)-\log ^2\left(\alpha _4\right)+4 \log ^2\left(-\alpha _5\right)-\frac{1}{2} \log ^2\left(\alpha _6\right)+3 \log \left(\alpha _2\right) \log \left(\alpha _1\right)\nonumber\\
   &-8 \log
   \left(\alpha _3\right) \log \left(\alpha _1\right)+\log \left(\alpha _4\right) \log \left(\alpha _1\right)-2 \log \left(-\alpha _5\right) \log \left(\alpha _1\right)-5 \log \left(\alpha _6\right)
   \log \left(\alpha _1\right)\nonumber\\
   &-12 \log (2) \log \left(\alpha _1\right)+8 \log (2) \log \left(\alpha _2\right)+2 \log \left(\alpha _2\right) \log \left(\alpha _3\right)-7 \log (2) \log \left(\alpha
   _3\right)+4 \log \left(\alpha _2\right) \log \left(-\alpha _5\right)\nonumber\\
   &+8 \log (2) \log \left(-\alpha _5\right)-4 \log \left(\alpha _3\right) \log \left(\alpha _6\right)-4 \log \left(-\alpha
   _5\right) \log \left(\alpha _6\right)-9 \log (2) \log \left(\alpha _6\right)+4 \log ^2\left(-\beta _1\right)\nonumber\\
   &+4 \log ^2\left(\beta _3\right)-8 \log (2) \log \left(-\beta _1\right)-8 \log
   \left(-\beta _1\right) \log \left(-\beta _2\right)+16 \log (2) \log \left(-\beta _2\right)+8 \log \left(-\beta _1\right) \log \left(\beta _3\right)\nonumber\\
   &-8 \log \left(-\beta _2\right) \log \left(\beta
   _3\right)-8 \log (2) \log \left(\beta _3\right)\,.
\end{align}
\begin{align}
 f_{S,3}^{(2)}&(x,s,m_1^2,m_2^2,m_3^2)=\text{Li}_2\left(\frac{2 \alpha _1 \alpha _3}{\beta _2}\right)-\text{Li}_2\left(\frac{\beta _2}{2 \alpha _1 \alpha _6}\right)-4 \text{Li}_2\left(-\frac{\alpha _7 \beta _2}{2 \alpha _1^2 \alpha _3
   \alpha _6}\right)+4 \text{Li}_2\left(-\frac{2 \alpha _5 \alpha _7 \beta _2}{\alpha _1 \alpha _6 \beta _1 \beta _3}\right)\nonumber\\
   & +2 \log \left(\alpha _1\right) \log \left(\beta _1\right)+8 \log \left(\alpha _1\right) \log \left(-\beta _2\right)+2 \log \left(\alpha _1\right) \log \left(-\beta _3\right)-4 \log \left(\alpha
   _2\right) \log \left(\beta _1\right)\nonumber\\
   &-8 \log \left(-\alpha _5\right) \log \left(\beta _1\right)+4 \log \left(\alpha _6\right) \log \left(\beta _1\right)+7 \log \left(\alpha _3\right) \log
   \left(-\beta _2\right)+8 \log \left(-\alpha _5\right) \log \left(-\beta _2\right)\nonumber\\
   &+\log \left(\alpha _6\right) \log \left(-\beta _2\right)-4 \log \left(\alpha _2\right) \log \left(-\beta
   _3\right)-8 \log \left(-\alpha _5\right) \log \left(-\beta _3\right)+4 \log \left(\alpha _6\right) \log \left(-\beta _3\right)\nonumber\\
   &-6 \log ^2\left(\alpha _1\right)+\log ^2\left(\alpha
   _2\right)-\frac{5}{2} \log ^2\left(\alpha _3\right)-\log ^2\left(\alpha _4\right)+4 \log ^2\left(-\alpha _5\right)-\frac{1}{2} \log ^2\left(\alpha _6\right)+3 \log \left(\alpha _2\right) \log
   \left(\alpha _1\right)\nonumber\\
   &-8 \log \left(\alpha _3\right) \log \left(\alpha _1\right)+\log \left(\alpha _4\right) \log \left(\alpha _1\right)-2 \log \left(-\alpha _5\right) \log \left(\alpha
   _1\right)-5 \log \left(\alpha _6\right) \log \left(\alpha _1\right)\nonumber\\
   &-12 \log (2) \log \left(\alpha _1\right)+8 \log (2) \log \left(\alpha _2\right)+2 \log \left(\alpha _2\right) \log \left(\alpha
   _3\right)-7 \log (2) \log \left(\alpha _3\right)+4 \log \left(\alpha _2\right) \log \left(-\alpha _5\right)\nonumber\\
   &+8 \log (2) \log \left(-\alpha _5\right)-4 \log \left(\alpha _3\right) \log \left(\alpha
   _6\right)-4 \log \left(-\alpha _5\right) \log \left(\alpha _6\right)-9 \log (2) \log \left(\alpha _6\right)+4 \log ^2\left(\beta _1\right)\nonumber\\
   &+4 \log ^2\left(-\beta _3\right)-8 \log (2) \log
   \left(\beta _1\right)-8 \log \left(\beta _1\right) \log \left(-\beta _2\right)+16 \log (2) \log \left(-\beta _2\right)+8 \log \left(\beta _1\right) \log \left(-\beta _3\right)\nonumber\\
   &-8 \log \left(-\beta
   _2\right) \log \left(-\beta _3\right)-8 \log (2) \log \left(-\beta _3\right)\,,
\end{align}
\begin{align}
    f_{S,4}^{(2)}&(x,s,m_1^2,m_2^2,m_3^2)=\text{Li}_2\left(\frac{2 \alpha _1 \alpha _3}{\beta _2}\right)-\text{Li}_2\left(\frac{\beta _2}{2 \alpha _1 \alpha _6}\right)-4 \text{Li}_2\left(-\frac{\alpha _7 \beta _2}{2 \alpha _1^2 \alpha _3
   \alpha _6}\right)+4 \text{Li}_2\left(-\frac{2 \alpha _5 \alpha _7 \beta _2}{\alpha _1 \alpha _6 \beta _1 \beta _3}\right)\nonumber\\
   &+2 \log \left(\alpha _1\right) \log \left(\beta _1\right)+8 \log \left(\alpha _1\right) \log \left(\beta _2\right)+2 \log \left(\alpha _1\right) \log \left(\beta _3\right)-4 \log \left(\alpha
   _2\right) \log \left(\beta _1\right)\nonumber\\
   &-8 \log \left(-\alpha _5\right) \log \left(\beta _1\right)+4 \log \left(\alpha _6\right) \log \left(\beta _1\right)+7 \log \left(\alpha _3\right) \log
   \left(\beta _2\right)+8 \log \left(-\alpha _5\right) \log \left(\beta _2\right)\nonumber\\
   &+\log \left(\alpha _6\right) \log \left(\beta _2\right)-4 \log \left(\alpha _2\right) \log \left(\beta _3\right)-8
   \log \left(-\alpha _5\right) \log \left(\beta _3\right)+4 \log \left(\alpha _6\right) \log \left(\beta _3\right)-6 \log ^2\left(\alpha _1\right)\nonumber\\
   &+\log ^2\left(\alpha _2\right)-\frac{5}{2} \log
   ^2\left(\alpha _3\right)-\log ^2\left(\alpha _4\right)+4 \log ^2\left(-\alpha _5\right)-\frac{1}{2} \log ^2\left(\alpha _6\right)+3 \log \left(\alpha _2\right) \log \left(\alpha _1\right)\nonumber\\
   &-8 \log
   \left(\alpha _3\right) \log \left(\alpha _1\right)+\log \left(\alpha _4\right) \log \left(\alpha _1\right)-2 \log \left(-\alpha _5\right) \log \left(\alpha _1\right)-5 \log \left(\alpha _6\right)
   \log \left(\alpha _1\right)\nonumber\\
   &-12 \log (2) \log \left(\alpha _1\right)+6 i \pi  \log \left(\alpha _1\right)+8 \log (2) \log \left(\alpha _2\right)+4 i \pi  \log \left(\alpha _2\right)+2 \log
   \left(\alpha _2\right) \log \left(\alpha _3\right)\nonumber\\
   &-7 \log (2) \log \left(\alpha _3\right)+i \pi  \log \left(\alpha _3\right)+4 \log \left(\alpha _2\right) \log \left(-\alpha _5\right)+8 \log (2)
   \log \left(-\alpha _5\right)-4 \log \left(\alpha _3\right) \log \left(\alpha _6\right)\nonumber\\
   &-4 \log \left(-\alpha _5\right) \log \left(\alpha _6\right)-9 \log (2) \log \left(\alpha _6\right)+3 i \pi 
   \log \left(\alpha _6\right)-8 i \pi  \log \left(\alpha _7\right)+4 \log ^2\left(\beta _1\right)+4 \log ^2\left(\beta _3\right)\nonumber\\
   &-8 \log (2) \log \left(\beta _1\right)-8 \log \left(\beta _1\right)
   \log \left(\beta _2\right)+16 \log (2) \log \left(\beta _2\right)+8 \log \left(\beta _1\right) \log \left(\beta _3\right)-8 \log \left(\beta _2\right) \log \left(\beta _3\right)\nonumber\\
   &-8 \log (2) \log
   \left(\beta _3\right)-4 \pi ^2\,,
\end{align}
\begin{align}
    f_{S,5}^{(2)}&(x,s,m_1^2,m_2^2,m_3^2)=\text{Li}_2\left(\frac{2 \alpha _1 \alpha _3}{\beta _2}\right)-\text{Li}_2\left(\frac{\beta _2}{2 \alpha _1 \alpha _6}\right)-4 \text{Li}_2\left(-\frac{\alpha _7 \beta _2}{2 \alpha _1^2 \alpha _3
   \alpha _6}\right)+4 \text{Li}_2\left(-\frac{2 \alpha _5 \alpha _7 \beta _2}{\alpha _1 \alpha _6 \beta _1 \beta _3}\right)\nonumber\\
   &+2 \log \left(\alpha _1\right) \log \left(\beta _1\right)+8 \log \left(\alpha _1\right) \log \left(-\beta _2\right)+2 \log \left(\alpha _1\right) \log \left(\beta _3\right)-4 \log \left(\alpha
   _2\right) \log \left(\beta _1\right)\nonumber\\
   &-8 \log \left(\alpha _5\right) \log \left(\beta _1\right)+4 \log \left(\alpha _6\right) \log \left(\beta _1\right)+7 \log \left(\alpha _3\right) \log
   \left(-\beta _2\right)+8 \log \left(\alpha _5\right) \log \left(-\beta _2\right)\nonumber\\
   &+\log \left(\alpha _6\right) \log \left(-\beta _2\right)-4 \log \left(\alpha _2\right) \log \left(\beta _3\right)-8
   \log \left(\alpha _5\right) \log \left(\beta _3\right)+4 \log \left(\alpha _6\right) \log \left(\beta _3\right)-6 \log ^2\left(\alpha _1\right)\nonumber\\
   &+\log ^2\left(\alpha _2\right)-\frac{5}{2} \log
   ^2\left(\alpha _3\right)-\log ^2\left(\alpha _4\right)+4 \log ^2\left(\alpha _5\right)-\frac{1}{2} \log ^2\left(\alpha _6\right)+3 \log \left(\alpha _2\right) \log \left(\alpha _1\right)\nonumber\\
   &-8 \log
   \left(\alpha _3\right) \log \left(\alpha _1\right)+\log \left(\alpha _4\right) \log \left(\alpha _1\right)-2 \log \left(\alpha _5\right) \log \left(\alpha _1\right)-5 \log \left(\alpha _6\right)
   \log \left(\alpha _1\right)-12 \log (2) \log \left(\alpha _1\right)\nonumber\\
   &+8 \log (2) \log \left(\alpha _2\right)+2 \log \left(\alpha _2\right) \log \left(\alpha _3\right)-7 \log (2) \log \left(\alpha
   _3\right)+4 \log \left(\alpha _2\right) \log \left(\alpha _5\right)+8 \log (2) \log \left(\alpha _5\right)\nonumber\\
   &-4 \log \left(\alpha _3\right) \log \left(\alpha _6\right)-4 \log \left(\alpha _5\right)
   \log \left(\alpha _6\right)-9 \log (2) \log \left(\alpha _6\right)+4 \log ^2\left(\beta _1\right)+4 \log ^2\left(\beta _3\right)\nonumber\\
   &-8 \log (2) \log \left(\beta _1\right)-8 \log \left(\beta _1\right)
   \log \left(-\beta _2\right)+16 \log (2) \log \left(-\beta _2\right)+8 \log \left(\beta _1\right) \log \left(\beta _3\right)\nonumber\\
   &-8 \log \left(-\beta _2\right) \log \left(\beta _3\right)-8 \log (2) \log
   \left(\beta _3\right)\,.
\end{align}

\section{Next-to linearly reducible example}
\label{app:next-to-lin}

In this appendix we study the following box diagram with a bubble insertion, relevant for the two-loop QCD corrections to heavy quark pair production:
\begin{equation}
    B_{2,1,1,1,1} (s,t,m^2)=\;\;(1+2 \epsilon) \vcenter{\hbox{\includegraphics[width=0.27\textwidth]{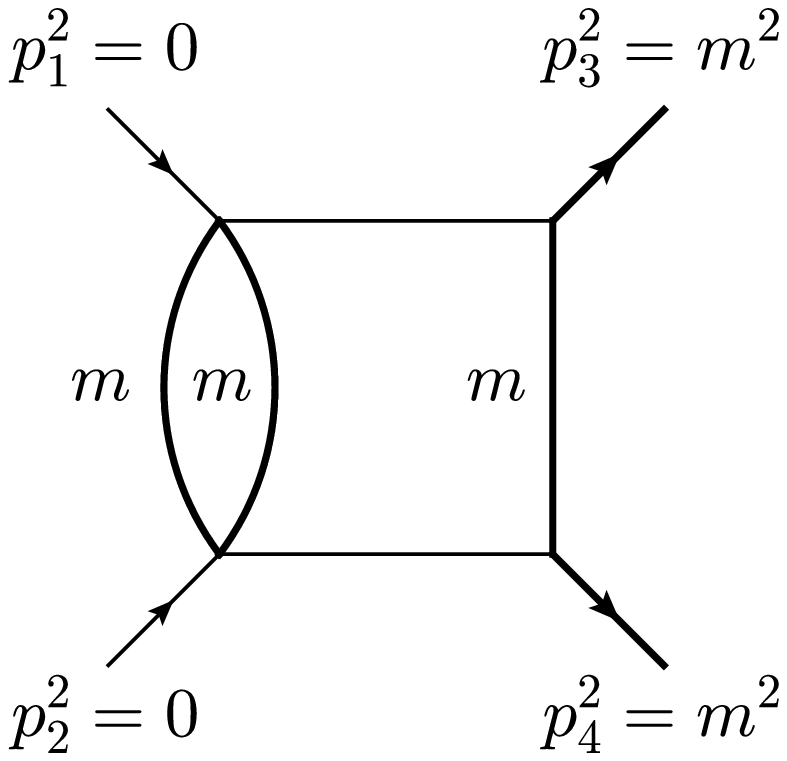}}},
\end{equation}
where the external invariants are defined as:
\begin{equation}
p_1\cdot p_2=\frac{s}{2}\,,\quad p_1\cdot p_3 = \frac{m^2-t}{2}\,,\quad p_2\cdot p_3 =\frac{s+t-m^2}{2}\,.
\end{equation}
Note the $(1+2 \epsilon)$ prefactor and the dotted bubble subintegral, which are needed to obtain a uniformly transcendental inner polylogarithmic expression. This integral depends on two elliptic curves, the one of the sunrise subtopology and the one of the integral itself (found by computing, e.g. the maximally cut integral). For this example linear reducibility seems to be possible only up to the second last integration variable. Nevertheless, we show that the solution can be expressed to all orders in terms two dimensional integrals. The integral is defined in $d=4-2 \epsilon$ dimensions as:
\begin{align}
    B_{2,1,1,1,1} (s,t,m^2)&=(1+2 \epsilon)\sum_{k=0}^\infty \epsilon^k \int_\Delta d^5 \vec{\alpha}\, \frac{1}{k!}\, \alpha_1\, \mathcal{F}^{-2} \,\log\left( \frac{\mathcal{U}^3}{\mathcal{F}^2}\right)^k\nonumber\\
    &\equiv T^{(0)}_{21111}(s,t,m^2)+\epsilon\, T^{(1)}_{21111}(s,t,m^2)+\mathcal{O}(\epsilon^2)\,,
\end{align}
where the Symanzik polynomials are:
\begin{align}
  \mathcal{U}  & = \, \,\alpha _2 \left(\alpha _3+\alpha _4+\alpha _5\right)+\alpha _1 \left(\alpha _2+\alpha _3+\alpha _4+\alpha _5\right)\,,\nonumber\\
  \mathcal{F} &  =  \, \left(\alpha _2+\alpha _3+\alpha _4+\alpha _5\right) \alpha _1^2 m^2+\alpha _2 \left(\alpha _4^2 m^2+\alpha _2 \left(\alpha _3+\alpha _4+\alpha _5\right) m^2-\alpha _3 \alpha _5 s\right)\nonumber\\
  &\;\quad\quad+\alpha _1
   \left(\alpha _2^2 m^2+\alpha _4^2 m^2+\alpha _2 \left(2 \alpha _3 m^2+2 \alpha _5 m^2+\alpha _4 \left(3 m^2-t\right)\right)-\alpha _3 \alpha _5 s\right)\,.
\end{align}
In order to achieve linear reducibility up to the second last integration we apply the Cheng-Wu theorem setting $\alpha_1=1$, and then we integrate along the sequence $\alpha_5$, $\alpha_3$, $\alpha_4$, where the first two parameters correspond to the massless propagators. At order $\epsilon^0$, the first integration with respect to $\alpha_5$ yields:
\begin{align}
     B^{(0)}_{2,1,1,1,1} & (s,t,m^2)  =  -\int_0^\infty  d \alpha_2 \,  d\alpha_3\, d \alpha_4  \frac{1 }{\left(\alpha _2+1\right) \left(\alpha _2 m^2+m^2-\alpha _3 s\right)}\nonumber\\
      &\times \frac{1}{\left(\alpha _2^2 (\alpha _3+\alpha _4+1\right) m^2 +\left(\alpha _4^2+\alpha _4+\alpha _3\right) m^2+\alpha _2
   \left(\left(\alpha _4^2+3 \alpha _4+2 \alpha _3+1\right) m^2-\alpha_2\alpha _4 t\right)}\,.
\end{align}
Performing the integration on $\alpha_3$ next, we get:
\begin{align}
   B^{(0)}_{2,1,1,1,1}(s, t, m^2)=\int_0^\infty  \frac{ d \alpha_2\, d \alpha_4\log \left(\frac{\alpha _2 \alpha _4 s t-\left(\alpha _2+\alpha _4+1\right) \left(\alpha _4 \alpha _2+\alpha _2+\alpha _4\right) m^2 s}{\left(\alpha _2+1\right)^3 m^4}\right)}{\left(\alpha_2+1\right)\left(\left(\alpha _2+1\right)^3 m^4+\left(\alpha _2+\alpha _4+1\right) \left(\alpha _4 \alpha _2+\alpha _2+\alpha _4\right) m^2 s-\alpha _2 \alpha _4 s\,t\right)}\,.
\end{align}
Note that the following two polynomials do not factor linearly in either integration parameter without the introduction of algebraic terms:
\begin{gather}
    \alpha _2 \alpha _4 s t-\left(\alpha _2+\alpha _4+1\right) \left(\alpha _4 \alpha _2+\alpha _2+\alpha _4\right) m^2 s\,,\nonumber\\
    \left(\alpha _2+1\right)^3 m^4+\left(\alpha _2+\alpha _4+1\right) \left(\alpha _4 \alpha _2+\alpha _2+\alpha _4\right) m^2 s-\alpha _2 \alpha _4 s\,t\,,
\end{gather}
and their zeros with respect to $\alpha_4$ are respectively:
\begin{align}
    Y^{(0)}_{\pm}(\alpha_2,s,t,m^2) & = \frac{\alpha _2 t-\left(\alpha _2^2+3 \alpha _2+1\right) m^2\pm \sqrt{W^{(0)}(\alpha_2,s,t,m^2)}}{2
   \left(\alpha _2+1\right) m^2},\nonumber\\
   Z^{(0)}_{\pm}(\alpha_2,s,t,m^2) & = \frac{\alpha _2 s\, t-\left(\alpha _2^2+3 \alpha _2+1\right) m^2 s\pm \sqrt{X^{(0)}(\alpha_2,s,t,m^2)}}{2 \left(\alpha _2+1\right) m^2 s}\,,
\end{align}
where we have the following fourth degree polynomials:
\begin{align}
    W^{(0)}(\alpha_2,s,t,m^2)  & = \left(\alpha _2^2+\alpha _2+1\right)^2 m^4-2 \alpha _2 \left(\alpha _2^2+3\alpha _2+1\right) m^2 t+\alpha _2^2 t^2,\nonumber \\
    X^{(0)} (\alpha_2,s,t,m^2) & = s^2 \left(\left(\alpha _2^2 +3\alpha _2 +1\right) m^2-\alpha _2 t\right)^2-4 s \left(\alpha _2+1\right)^2 m^4
   \left(\left(\alpha _2+1\right)^2 m^2+\alpha _2 s\right)\,.
\end{align}
Integrating with respect to $\alpha_4$ and using the notation above we finally get:
\begin{equation}
    B^{(0)}_{2,1,1,1,1} (s,t,  m^2) =\int_0^\infty d\alpha_2 \frac{1}{\sqrt{X^{(0)}(\alpha_2,s,t,m^2)}(\alpha_2+1)} B^{(2)}(\alpha_2,s,t, m^2)\,,
\end{equation}
where:
\begin{align}
  B^{(2)}(\alpha_2,s,t, m^2)& = G_{\frac{1}{Z^{(0)}_-+1},\frac{s}{\alpha _2 m^2+m^2+s}}- G_{\frac{1}{Z^{(0)}_++1},\frac{s}{\alpha _2 m^2+m^2+s}}+ G_{\frac{s}{\alpha _2 m^2+m^2+s},\frac{1}{Z^{(0)}_-+1}}\nonumber\\
  &- G_{\frac{s}{\alpha _2 m^2+m^2+s},\frac{1}{Z^{(0)}_++1}}+ G_{\frac{1}{Y^{(0)}_-+1},\frac{1}{Z^{(0)}_-+1}}- G_{\frac{1}{Y^{(0)}_-+1},\frac{1}{Z^{(0)}_++1}}+ G_{\frac{1}{Y^{(0)}_++1},\frac{1}{Z^{(0)}_-+1}}\nonumber\\
  &- G_{\frac{1}{Y^{(0)}_++1},\frac{1}{Z^{(0)}_++1}}- 2 G_{0,\frac{1}{Z^{(0)}_-+1}}+ 2 G_{0,\frac{1}{Z^{(0)}_++1}}- G_{\frac{1}{Z^{(0)}_-+1},\alpha _2+1}\nonumber\\
  &+ G_{\frac{1}{Z^{(0)}_++1},\alpha _2+1}- G_{\alpha _2+1,\frac{1}{Z^{(0)}_-+1}}+ G_{\alpha _2+1,\frac{1}{Z^{(0)}_++1}}\,.
\end{align}
We proceed with the order $\epsilon^1$. The integration with respect to $\alpha_5$ gives:
\begin{align}
    & B^{(1)}_{2,1,1,1,1}  (s,t,m^2)= \int_0^\infty   d \alpha_2 \,  d\alpha_3\, d \alpha_4\frac{1}{\left(\alpha _2+1\right)\left(\left(\alpha _2+1\right)
   m^2-\alpha _3 s\right)}\times\nonumber \\
    &\times \Bigg[\frac{\log \left(\frac{\left(\left(\alpha _2+1\right) \alpha _4^2 m^2+\left(\alpha _2+1\right) \left(\alpha _3+\alpha _2 \left(\alpha _3+1\right)\right) m^2+\alpha _4 \left(\alpha _2 \left(\alpha _2
   m^2+3 m^2-t\right)+m^2\right)\right)^2}{\left(\alpha _3+\alpha _4+\alpha _2 \left(\alpha _3+\alpha _4+1\right)\right)^3}\right)}{  \left(\alpha _2+1\right) \alpha _4^2 m^2+\left(\alpha _2+1\right) \left(\alpha _3+\alpha _2 \left(\alpha _3+1\right)\right) m^2+\alpha _4 \left(\alpha _2 \left(\alpha
   _2 m^2+3 m^2-t\right)+m^2\right)}\nonumber\\
   &+ \frac{3 \log \left(\frac{\left(\alpha _3+\alpha _4+\alpha _2 \left(\alpha _3+\alpha _4+1\right)\right) \left(\left(\alpha _2+1\right) m^2-\alpha _3 s\right)}{\left(\alpha _2+1\right) \alpha _4^2
   m^2+\left(\alpha _2+1\right) \left(\alpha _3+\alpha _2 \left(\alpha _3+1\right)\right) m^2+\alpha _4 \left(\alpha _2 \left(\alpha _2 m^2+3 m^2-t\right)+m^2\right)}\right)}{\left(\alpha _2+1\right) \alpha _4^2 m^2+\alpha _4 \left(\alpha _2 \left(m^2+\alpha _3 s-t\right)+\alpha _3 s\right)+\alpha _3
   \left(\alpha _3+\alpha _2 \left(\alpha _3+1\right)\right) s}\Bigg]\,.
\end{align}
The expression above contains two polynomials:
\begin{gather}
 \left(\alpha _2+1\right) \alpha _4^2 m^2+\left(\alpha _2+1\right) \left(\alpha _3+\alpha _2 \left(\alpha _3+1\right)\right) m^2+\alpha _4 \left(\alpha _2 \left(\alpha _2 m^2+3 m^2-t\right)+m^2\right)\,,\nonumber\\
 \left(\alpha _2+1\right) \alpha _4^2 m^2+\alpha _4 \left(\alpha _2 \left(m^2+\alpha _3 s-t\right)+\alpha _3 s\right)+\alpha _3
   \left(\alpha _3+\alpha _2 \left(\alpha _3+1\right)\right) s\,,
\end{gather}
whose zeros with respect to $\alpha_4$ contain square roots, and they are respectively:
\begin{align}
    Y^{(1)}_\pm (\alpha_2,\alpha_3,s,t,m^2)& =\frac{\alpha _2 t-\left(\alpha _2^2+3 \alpha _2 +1\right) m^2\pm\sqrt{W^{(1)}(\alpha_2,\alpha_3,s,t,m^2)}}{2 \left(\alpha _2+1\right) m^2}\,,\nonumber\\
    Z^{(1)}_\pm (\alpha_2,\alpha_3,s,t,m^2)&  =\frac{-\alpha _3 s-\alpha _2 \left(m^2+\alpha _3 s-t\right)\pm \sqrt{X^{(1)}(\alpha_2,\alpha_3,s,t,m^2)}}{2 \left(\alpha _2+1\right) m^2} \,,
\end{align}
where we have defined the following polynomials:
\begin{align}
    W^{(1)}(\alpha_2,\alpha_3,s,t,m^2)& =\left(\alpha _2 \left(\alpha _2 m^2+3 m^2-t\right)+m^2\right)^2-4 \left(\alpha _2+1\right)^2 \left(\alpha _3+\alpha _2
   \left(\alpha _3+1\right)\right) m^4\,,\nonumber\\
    X^{(1)}(\alpha_2,\alpha_3,s,t,m^2)& =\left(\alpha _2+1\right)^2 \alpha _3^2 s \left(s-4 m^2\right)-2 \left(\alpha _2+1\right) \alpha _3 \alpha _2 s \left(m^2+t\right)+\alpha _2^2
   \left(m^2-t\right)^2\,.
\end{align}
Using the notation above we arrive at the final result by integrating with respect to $\alpha_4$:
\begin{align}
      B^{(1)}_{2,1,1,1,1}  (s,t,m^2)&=\frac{1}{s}\int_0^\infty d\alpha_2 d\alpha_3 \left[\frac{B_X^{(1)}(\alpha_2,\alpha_3,s,t,m^2)}{\sqrt{X^{(1)}(\alpha_2,\alpha_3,s,t,m^2)}(\alpha_2+1)(\alpha_3-\frac{\alpha _2 m^2+m^2}{s})}\right.\nonumber\\
      &+\left. \frac{B_W^{(1)}(\alpha_2,\alpha_3,s,t,m^2)}{\sqrt{W^{(1)}(\alpha_2,\alpha_3,s,t,m^2)}(\alpha_2+1)(\alpha_3-\frac{\alpha _2 m^2+m^2}{s})}\right]\,,
\end{align}
where:
\begin{align}
 B_X^{(1)}(\alpha_2,\alpha_3&,s,t,m^2)=- 3 G_{0,\frac{1}{Z^{(1)}_-+1}}+ 3 G_{0,\frac{1}{Z^{(1)}_++1}}+ 3 G_{\frac{1}{Y^{(1)}_-+1},\frac{1}{Z^{(1)}_-+1}} + 3 G_{\frac{1}{Y^{(1)}_-+1},\frac{1}{Z^{(1)}_++1}}\nonumber\\
 &+ 3 G_{\frac{1}{Z^{(1)}_-+1},-\frac{\alpha _2+1}{\left(\alpha _2+1\right) \alpha _3-1}}- 3 G_{\frac{1}{Z^{(1)}_-+1},-\frac{\alpha _2 m^2+m^2-s \alpha _3}{\left(\alpha _2 m^2+m^2+s\right) \alpha _3-m^2}}+ 3 G_{\frac{1}{Y^{(1)}_++1},\frac{1}{Z^{(1)}_-+1}}\nonumber\\
 &- 3 G_{\frac{1}{Y^{(1)}_++1},\frac{1}{Z^{(1)}_++1}}- 3 G_{\frac{1}{Z^{(1)}_++1},-\frac{\alpha _2+1}{\left(\alpha _2+1\right) \alpha _3-1}}+ 3 G_{\frac{1}{Z^{(1)}_++1},-\frac{\alpha _2 m^2+m^2-s \alpha _3}{\left(\alpha _2 m^2+m^2+s\right) \alpha _3-m^2}}\nonumber\\
 &- 3 G_{-\frac{\alpha _2 m^2+m^2-s \alpha _3}{\left(\alpha _2 m^2+m^2+s\right) \alpha _3-m^2},\frac{1}{Z^{(1)}_-+1}}+ 3 G_{-\frac{\alpha _2 m^2+m^2-s \alpha _3}{\left(\alpha _2 m^2+m^2+s\right) \alpha _3-m^2},\frac{1}{Z^{(1)}_++1}}\,,
\end{align}
and:
\begin{align}
 B_W^{(1)}(\alpha_2,\alpha_3&,s,t,m^2)=G_{0,\frac{1}{Y^{(1)}_-+1}}-G_{0,\frac{1}{Y^{(1)}_++1}}- 2 G_{\frac{1}{Y^{(1)}_-+1},\frac{1}{Y^{(1)}_-+1}}+ 2 G_{\frac{1}{Y^{(1)}_-+1},\frac{1}{Y^{(1)}_++1}}\nonumber\\
 &- 3 G_{\frac{1}{Y^{(1)}_-+1},-\frac{1}{\alpha _3+\alpha _2 \left(\alpha _3+1\right)-1}}+ 2 G_{\frac{1}{Y^{(1)}_-+1},-\frac{1}{\alpha _3 m^2+\alpha _2^2 \left(\alpha _3+1\right) m^2+\alpha _2 \left(2 \alpha _3+1\right) m^2-1}}- 2 G_{\frac{1}{Y^{(1)}_++1},\frac{1}{Y^{(1)}_-+1}}\nonumber\\
 &+ 2 G_{\frac{1}{Y^{(1)}_++1},\frac{1}{Y^{(1)}_++1}}+ 3 G_{\frac{1}{Y^{(1)}_++1},-\frac{1}{\alpha _3+\alpha _2 \left(\alpha _3+1\right)-1}}- 2 G_{\frac{1}{Y^{(1)}_++1},-\frac{1}{\alpha _3 m^2+\alpha _2^2 \left(\alpha _3+1\right) m^2+\alpha _2 \left(2 \alpha _3+1\right) m^2-1}}\nonumber\\
 &+ 3 G_{-\frac{\alpha _2+1}{\left(\alpha _2+1\right) \alpha _3-1},\frac{1}{Y^{(1)}_-+1}}- 3 G_{-\frac{\alpha _2+1}{\left(\alpha _2+1\right) \alpha _3-1},\frac{1}{Y^{(1)}_++1}}- 3 G_{-\frac{1}{\alpha _3+\alpha _2 \left(\alpha _3+1\right)-1},\frac{1}{Y^{(1)}_-+1}}\nonumber\\
 &+ 3 G_{-\frac{1}{\alpha _3+\alpha _2 \left(\alpha _3+1\right)-1},\frac{1}{Y^{(1)}_++1}}+ 2 G_{-\frac{1}{\alpha _3 m^2+\alpha _2^2 \left(\alpha _3+1\right) m^2+\alpha _2 \left(2 \alpha _3+1\right) m^2-1},\frac{1}{Y^{(1)}_-+1}}\nonumber\\
 &- 2 G_{-\frac{1}{\alpha _3 m^2+\alpha _2^2 \left(\alpha _3+1\right) m^2+\alpha _2 \left(2 \alpha _3+1\right) m^2-1},\frac{1}{Y^{(1)}_++1}}\,.
\end{align}

\section{Full result for the non-planar triangle}
\label{app:full results}
In this appendix we provide the full expression for the order $\epsilon^0$ of the off-shell non-planar triangle presented in sec.~\ref{sec:non-planar triangle}. By defining:
\begin{align}
    P_N(\beta_5,s,p_2^2,m^2) &=\left(\beta _5+1\right)^4 m^4-2 \beta _5 \left(\beta _5+1\right)^2 m^2 \left(\beta _5 p_2^2+p_2^2-2 \beta _5 s-s\right)+\beta _5^2 \left(\beta _5 p_2^2+p_2^2-s\right)^2\,,\nonumber\\
    Q_N(\beta_5,s,p_2^2,m^2) &=\left(\beta _5 p_2^2+p_2^2-s\right)^2-4 \left(\beta _5+1\right)^2 m^2 \left(p_2^2-s\right)\,,\nonumber\\
    R_N(\beta_5,s,p_2^2,m^2) &=\beta _5^2 \left(p_2^4-4 m^2 s\right)+2 \beta _5 s \left(p_2^2-4 m^2\right)+s \left(s-4 m^2\right)\,,
\end{align}
and:
\begin{align}
   &a(1)=\frac{m^2 \left(\beta _5+1\right)^2}{\left(p_2^2-s\right) \beta _5},\;a(2)=\frac{\left(\beta _5+1\right)^2}{\beta _5^2+\beta _5+1},\;a(3)=\frac{p_2^2 \beta _5-m^2 \left(\beta _5+1\right)^2}{s \beta _5},\nonumber\\
   &a(4)=\frac{p_2^2 \beta _5-m^2 \left(\beta _5+1\right)^2}{p_2^2 \beta _5},\;a(5)=\frac{p_2^2 \beta _5-m^2 \left(\beta _5+1\right)^2}{\left(p_2^2-s\right) \beta _5},\;a(6)=\frac{m^2 \left(\beta _5+1\right)^2-s \beta _5}{\left(p_2^2-s\right) \beta _5},\nonumber\\
   &a(7)=-\frac{2 \left(m^2 \left(\beta _5+1\right)^2-p_2^2 \beta _5\right)}{\left(\beta _5-1\right) p_2^2+s-\sqrt{Q_N}},\;a(8)=-\frac{2 \left(m^2 \left(\beta _5+1\right)^2-p_2^2 \beta _5\right)}{\left(\beta _5-1\right) p_2^2+s+\sqrt{Q_N}},\nonumber\\
   &a(9)=\frac{2 \left(m^2 \left(\beta _5+1\right)^2-p_2^2 \beta _5\right)}{-\beta _5 p_2^2+s+\sqrt{R_N}},\;a(10)=\frac{m^2 \left(\beta _5+1\right)^2-p_2^2 \beta _5}{\beta _5 m^2+m^2-p_2^2 \beta _5},\;a(11)=-\frac{2 \left(m^2 \left(\beta _5+1\right)^2-p_2^2 \beta _5\right)}{\beta _5 p_2^2-s+\sqrt{R_N}},\nonumber\\
   &a(12)=\frac{2 \left(\beta _5+1\right) \left(m^2 \left(\beta _5+1\right)^2-p_2^2 \beta _5\right)}{\beta _5^2 m^2+2 \beta _5 m^2+m^2-p_2^2 \beta _5^2-p_2^2 \beta _5-s \beta _5+\sqrt{P_N}},\nonumber\\
   &a(13)=-\frac{2 \left(\beta _5+1\right) \left(m^2 \left(\beta _5+1\right)^2-p_2^2 \beta _5\right)}{-\beta _5^2 m^2-2 \beta _5 m^2-m^2+p_2^2 \beta _5^2+p_2^2 \beta _5+s \beta _5+\sqrt{P_N}},\nonumber\\
   &a(14)=-\frac{2 \left(m^2 \left(\beta _5+1\right)^2-p_2^2 \beta _5\right)}{\left(\beta _5-1\right) p_2^2+\sqrt{p_2^2 \left(p_2^2-4 m^2\right)} \left(\beta _5+1\right)},\;a(15)=\frac{2 \left(m^2 \left(\beta _5+1\right)^2-p_2^2 \beta _5\right)}{-\beta _5 p_2^2+p_2^2+\sqrt{p_2^2 \left(p_2^2-4 m^2\right)} \left(\beta _5+1\right)},\nonumber\\
   &a(16)=\frac{s \left(\beta _5+1\right)^2}{-\beta _5 p_2^2+s \beta _5^2+m^2 \left(\beta _5+1\right)^2+s+2 s \beta _5},\;a(17)=\frac{2 \left(\beta _5+1\right) \left(m^2 \left(\beta _5+1\right)^2-p_2^2 \beta _5\right)}{m^2 \left(\beta _5+1\right)^2-\beta _5 \left(\beta _5 p_2^2+p_2^2+s\right)-\sqrt{P_N}},\nonumber\\
   &a(18)=\frac{2 \left(\beta _5+1\right) \left(m^2 \left(\beta _5+1\right)^2-p_2^2 \beta _5\right)}{m^2 \left(\beta _5+1\right)^2-\beta _5 \left(\beta _5 p_2^2+p_2^2+s\right)+\sqrt{P_N}}\,,
\end{align}
we have the following expression for the order $\epsilon^0$ of the non-planar triangle $N_{1,1,1,1,1,1}(s,p_2^2,m^2)$:
\begin{equation}
    N^{(0)}_{1,1,1,1,1,1}(s,p_2^2,m^2)=\int_0^\infty d\beta_5 \frac{1}{\sqrt{P_N(\beta_5,s,p_2^2,m^2)}}N^{(3)}(\beta_5,s,p_2^2,m^2)\,,
\end{equation}
with:
\begin{align}
    N^{(3)}(\beta_5&,s,p_2^2 ,m^2)= -G_{0,a(3),a(12)}+G_{0,a(3),a(13)}+ 2 G_{0,a(4),a(12)}- 2 G_{0,a(4),a(13)}+G_{0,a(6),a(12)}
    \nonumber\\
    &-G_{0,a(6),a(13)}-G_{0,a(7),a(12)}+G_{0,a(7),a(13)}-G_{0,a(8),a(12)}+G_{0,a(8),a(13)}\nonumber\\
    &
    +G_{0,a(12),a(6)}-G_{0,a(13),a(6)}+G_{a(1),a(2),a(17)}-G_{a(1),a(2),a(18)}
    -G_{a(1),a(16),a(17)}\nonumber\\
    &+G_{a(1),a(16),a(18)}+G_{a(1),a(17),a(2)}-G_{a(1),a(17),a(16)}
    -G_{a(1),a(18),a(2)}+G_{a(1),a(18),a(16)}\nonumber\\
    &+G_{a(2),a(1),a(17)}-G_{a(2),a(1),a(18)}
    +G_{a(2),a(3),a(12)}-G_{a(2),a(3),a(13)}- 2 G_{a(2),a(4),a(12)}\nonumber\\
    &+ 2 G_{a(2),a(4),a(13)}
    +G_{a(2),a(7),a(12)}-G_{a(2),a(7),a(13)}+G_{a(2),a(8),a(12)}-G_{a(2),a(8),a(13)}
    \nonumber\\
    &+G_{a(2),a(17),a(1)}-G_{a(2),a(18),a(1)}+G_{a(3),a(2),a(12)}-G_{a(3),a(2),a(13)}
    +G_{a(3),a(12),a(2)}\nonumber\\
    &-G_{a(3),a(12),a(16)}-G_{a(3),a(13),a(2)}+G_{a(3),a(13),a(16)}
    -G_{a(3),a(16),a(12)}+G_{a(3),a(16),a(13)}\nonumber\\
    &- 2 G_{a(4),a(2),a(12)}+ 2 G_{a(4),a(2),a(13)}
    -G_{a(4),a(6),a(12)}+G_{a(4),a(6),a(13)}- 2 G_{a(4),a(12),a(2)}\nonumber\\
    &-G_{a(4),a(12),a(6)}
    + 2 G_{a(4),a(12),a(16)}+ 2 G_{a(4),a(13),a(2)}+G_{a(4),a(13),a(6)}- 2 G_{a(4),a(13),a(16)}\nonumber\\
    &
    + 2 G_{a(4),a(16),a(12)}- 2 G_{a(4),a(16),a(13)}+G_{a(5),0,a(12)}-G_{a(5),0,a(13)}
    -G_{a(5),a(4),a(12)}\nonumber\\
    &+G_{a(5),a(4),a(13)}+G_{a(5),a(7),a(12)}-G_{a(5),a(7),a(13)}
    +G_{a(5),a(8),a(12)}-G_{a(5),a(8),a(13)}\nonumber\\
    &-G_{a(5),a(10),a(12)}+G_{a(5),a(10),a(13)}
    +G_{a(6),0,a(12)}-G_{a(6),0,a(13)}+G_{a(6),a(1),a(17)}\nonumber\\
    &-G_{a(6),a(1),a(18)}
    -G_{a(6),a(4),a(12)}+G_{a(6),a(4),a(13)}+G_{a(6),a(7),a(12)}-G_{a(6),a(7),a(13)}\nonumber\\
    &
    +G_{a(6),a(8),a(12)}-G_{a(6),a(8),a(13)}-G_{a(6),a(10),a(12)}+G_{a(6),a(10),a(13)}
    +G_{a(6),a(17),a(1)}\nonumber\\
    &-G_{a(6),a(18),a(1)}+G_{a(7),a(2),a(12)}-G_{a(7),a(2),a(13)}
    +G_{a(7),a(6),a(12)}-G_{a(7),a(6),a(13)}\nonumber\\
    &+G_{a(7),a(12),a(2)}+G_{a(7),a(12),a(6)}
    -G_{a(7),a(12),a(16)}-G_{a(7),a(13),a(2)}-G_{a(7),a(13),a(6)}\nonumber\\
    &+G_{a(7),a(13),a(16)}
    -G_{a(7),a(16),a(12)}+G_{a(7),a(16),a(13)}+G_{a(8),a(2),a(12)}-G_{a(8),a(2),a(13)}\nonumber\\
    &
    +G_{a(8),a(6),a(12)}-G_{a(8),a(6),a(13)}+G_{a(8),a(12),a(2)}+G_{a(8),a(12),a(6)}
    -G_{a(8),a(12),a(16)}\nonumber\\
    &-G_{a(8),a(13),a(2)}-G_{a(8),a(13),a(6)}+G_{a(8),a(13),a(16)}
    -G_{a(8),a(16),a(12)}+G_{a(8),a(16),a(13)}\nonumber\\
    &+G_{a(9),0,a(12)}-G_{a(9),0,a(13)}
    + 2 G_{a(9),1,a(12)}- 2 G_{a(9),1,a(13)}+G_{a(9),a(3),a(12)}\nonumber\\
    &-G_{a(9),a(3),a(13)}
    -G_{a(9),a(4),a(12)}+G_{a(9),a(4),a(13)}-G_{a(9),a(10),a(12)}+G_{a(9),a(10),a(13)}\nonumber\\
    &
    -G_{a(10),a(3),a(12)}+G_{a(10),a(3),a(13)}+ 2 G_{a(10),a(4),a(12)}- 2 G_{a(10),a(4),a(13)}\nonumber\\
    &
    -G_{a(10),a(6),a(12)}+G_{a(10),a(6),a(13)}-G_{a(10),a(7),a(12)}+G_{a(10),a(7),a(13)}\nonumber\\
    &
    -G_{a(10),a(8),a(12)}+G_{a(10),a(8),a(13)}-G_{a(10),a(12),a(6)}+G_{a(10),a(13),a(6)}
    +G_{a(11),0,a(12)}\nonumber\\
    &-G_{a(11),0,a(13)}+ 2 G_{a(11),1,a(12)}- 2 G_{a(11),1,a(13)}
    +G_{a(11),a(3),a(12)}-G_{a(11),a(3),a(13)}\nonumber\\
    &-G_{a(11),a(4),a(12)}+G_{a(11),a(4),a(13)}
    -G_{a(11),a(10),a(12)}+G_{a(11),a(10),a(13)}-G_{a(14),0,a(12)}\nonumber\\
    &+G_{a(14),0,a(13)}
    - 2 G_{a(14),1,a(12)}+ 2 G_{a(14),1,a(13)}-G_{a(14),a(4),a(12)}+G_{a(14),a(4),a(13)}\nonumber\\
    &
    +G_{a(14),a(7),a(12)}-G_{a(14),a(7),a(13)}+G_{a(14),a(8),a(12)}-G_{a(14),a(8),a(13)}
    +G_{a(14),a(10),a(12)}\nonumber\\
    &-G_{a(14),a(10),a(13)}-G_{a(15),0,a(12)}+G_{a(15),0,a(13)}
    - 2 G_{a(15),1,a(12)}+ 2 G_{a(15),1,a(13)}\nonumber\\
    &-G_{a(15),a(4),a(12)}+G_{a(15),a(4),a(13)}
    +G_{a(15),a(7),a(12)}-G_{a(15),a(7),a(13)}+G_{a(15),a(8),a(12)}\nonumber\\
    &-G_{a(15),a(8),a(13)}
    +G_{a(15),a(10),a(12)}-G_{a(15),a(10),a(13)}-G_{a(16),a(1),a(17)}+G_{a(16),a(1),a(18)}\nonumber\\
    &
    -G_{a(16),a(3),a(12)}+G_{a(16),a(3),a(13)}+ 2 G_{a(16),a(4),a(12)}- 2 G_{a(16),a(4),a(13)}
    -G_{a(16),a(7),a(12)}\nonumber\\
    &+G_{a(16),a(7),a(13)}-G_{a(16),a(8),a(12)}+G_{a(16),a(8),a(13)}
    -G_{a(16),a(17),a(1)}+G_{a(16),a(18),a(1)}\nonumber\\
    &+G_{a(17),a(1),a(2)}-G_{a(17),a(1),a(16)}
    +G_{a(17),a(2),a(1)}+G_{a(17),a(6),a(1)}-G_{a(17),a(16),a(1)}\nonumber\\
    &-G_{a(18),a(1),a(2)}
    +G_{a(18),a(1),a(16)}-G_{a(18),a(2),a(1)}-G_{a(18),a(6),a(1)}+G_{a(18),a(16),a(1)}\,.
\end{align}

\bibliographystyle{JHEP}
\bibliography{mrefs,refs}

\end{document}